\documentclass{atlasnote} 
\usepackage{amsmath}
\usepackage{atlasphysics}

\usepackage{graphicx} 
\usepackage{subfigure}
\usepackage{epstopdf}
\usepackage{lineno}
\usepackage{multirow}
\usepackage{rotating}

\usepackage{cite,./mcite}

\usepackage{hyperref}

\newcommand{\pta}{\ensuremath{p_\mathrm{T}}}
\newcommand{\nev}{\ensuremath{N_{\mathrm{ev}}}}
\newcommand{\nch}{\ensuremath{n_{\mathrm{ch}}}}
\newcommand{\Nch}{\ensuremath{N_{\mathrm{ch}}}}
\newcommand{\meanpt}{\ensuremath{\langle \pta \rangle}}
\newcommand{\nsel}{\ensuremath{n_{\mathrm{sel}}}}
\newcommand{\nselbs}{\ensuremath{n_{\mathrm{sel}}^{\mathrm{BS}}}}

\newcommand{\dzero}{\ensuremath{d_\mathrm{0}}}
\newcommand{\zzero}{\ensuremath{z_\mathrm{0}}}

\newcommand{\dzerobs}{\ensuremath{d_\mathrm{0}^{\mathrm{BS}}}}

\newcommand{\Mchsel}{\ensuremath{M_\mathrm{ch,sel}}}

\newcommand{\trigeff}{\ensuremath{\varepsilon_\mathrm{trig}}}
\newcommand{\vtxeff}{\ensuremath{\varepsilon_\mathrm{vtx}}}
\newcommand{\trkeff}{\ensuremath{\varepsilon_\mathrm{trk}}}

\newcommand{\fsec}{\ensuremath{f_\mathrm{nonp}}}
\newcommand{\fokr}{\ensuremath{f_\mathrm{okr}}}

\newcommand{\py}{{\sc pythia}}
\newcommand{\pho}{{\sc phojet}}

\newcommand{\sqn} {\ensuremath{\sqrt{s} = 0.9}~TeV}
\newcommand{\sqt} {\ensuremath{\sqrt{s} = 2.36}~TeV}
\newcommand{\sqs} {\ensuremath{\sqrt{s} = 7}~TeV}
\newcommand{\sqall} {\ensuremath{\sqrt{s} = 0.9}, 2.36 and 7~TeV}

\title{Charged-particle multiplicities in $pp$ interactions\\ 
 measured with the ATLAS detector at the LHC}

\date{\today}

 \atlasnote{CERN-PH-EP-2010-079\\ (Accepted by New J. Phys.)} 
\abstracttext{

Measurements are presented from proton-proton collisions at centre-of-mass energies of \sqall\ recorded with the ATLAS detector at the LHC.  
Events were collected using a single-arm minimum-bias trigger.
The charged-particle multiplicity, its dependence on transverse momentum and pseudorapidity and the relationship between the mean transverse momentum and charged-particle multiplicity are measured. 
Measurements in different regions of phase-space are shown, providing diffraction-reduced measurements as well as more inclusive ones.
The observed distributions are corrected to well-defined phase-space regions, using model-independent corrections.
The results are compared to each other and to various Monte Carlo models, including a new AMBT1 \py 6 tune.
In all the kinematic regions considered, the particle multiplicities are higher than predicted by the Monte Carlo models.
The central charged-particle multiplicity per event and unit of pseudorapidity, for tracks with $\pt~>~100$~MeV, is measured to be 
3.483 $\pm$ 0.009 (stat) $\pm$ 0.106 (syst) at \sqn\
and
5.630 $\pm$ 0.003 (stat) $\pm$ 0.169 (syst) at \sqs.
}

\begin{document}

%%%%%%%%%%%%%%%%%%%%%%%%%%%%%%%%%%%%
%            Content               % 
%%%%%%%%%%%%%%%%%%%%%%%%%%%%%%%%%%%%

\tableofcontents
\clearpage

%%%%%%%%%%%%%%%%%%%%%%%%%%%%
\section{Introduction}\label{sec:intro}

Inclusive charged-particle distributions have been previously measured in $pp$ and $p{\bar p}$ collisions at a range of different centre-of-mass energies~\cite{MB1, cmsminbias_3, cmsminbias_2,cmsminbias,alice_mb2,alice_mb,Aaltonen:2009ne,Alexopoulos:1994ag,Albajar:1989an,Abe:1989td,Ansorge:1988kn,Ansorge:1988fg,Abe:1988yu,Alner:1987wb,Ansorge:1986xq,Breakstone:1983ns,Arnison:1982ed}.  
These measurements provide insight into the
strong interactions at low energy-scales. Several QCD-inspired
models have been developed to interpret them. These models
are frequently cast into Monte Carlo simulations with free
parameters that can be constrained by measurements such as minimum
bias distributions. 
These measurements contribute to the understanding of soft QCD; 
moreover, they are important to determination of biases on high-\pta\ phenomena due to underlying events and event pileup effects
and are therefore of growing importance for future LHC physics.
The measurements presented in this paper implement a similar strategy to that in~\cite{MB1}. 
A single-arm trigger overlapping with the acceptance of the tracking volume is used.  
Results are presented as inclusive-inelastic distributions, with minimal model-dependence;
a minimum number of charged particles within well-defined \pta\ and $\eta$ selection are required.

This paper reports on measurements of primary charged-particle multiplicity distributions using 
the first $\sim$190~$\mu$b$^{-1}$ of data recorded by the ATLAS experiment at 7~TeV 
and $\sim$7~$\mu$b$^{-1}$ at 0.9~TeV.
At \sqn\ the sample is similar to that used for the first ATLAS minimum-bias publication~\cite{MB1}.
Results are also presented at \sqt\ where the track reconstruction setup differs significantly from that at the other energies, due to the Silicon Tracker (SCT) not being at nominal voltage. 
The integrated luminosity at this energy is estimated to be $\sim$0.1~$\mu$b$^{-1}$.

The following distributions are measured in this paper:
	$$
	\frac{1}{\nev} \cdot  \frac{\mathrm{d} \Nch}{\mathrm{d} \eta}, \ \ \
	\frac{1}{\nev}\cdot \frac{1}{2 \pi  p_\mathrm{T}} \cdot \frac{\mathrm{d}^2 \Nch}{\mathrm{d} \eta \mathrm{d} p_\mathrm{T}}, \ \ \
	\frac{1}{\nev} \cdot \frac{\mathrm{d} \nev}{\mathrm{d} \nch} \ \ \
	{\rm and}
	\ \ \ \langle p_\mathrm{T}\rangle ~ {\mathrm vs.} ~ \nch{\rm ,}
	$$
	where \pta\ is the charged particle momentum component transverse to the beam direction~\footnote[1]{The ATLAS reference system is a Cartesian right-handed co-ordinate system, with
the nominal collision point at the origin. The anti-clockwise beam direction defines the positive $z$-axis, while the positive $x$-axis is defined as pointing from the collision point to the centre of the
LHC ring and the positive $y$-axis  points upwards. 
The azimuthal angle $\phi$ is measured around the beam axis and the polar angle $\theta$ is measured with respect to~the $z$-axis. 
The pseudorapidity
is defined as $\eta =  -\ln \tan (\theta/2) $.},
$\eta$ is the pseudorapidity of the particle, 
\nch\ is the number of charged particles in an event, 
\nev\ is the number of events with a minimum number of charged particles within the selected kinematic range, \Nch\ is the total number of charged particles in the data sample and \meanpt\ is the average \pta\ for a given number of charged particles~\footnote[2]{The factor $2 \pi  p_\mathrm{T}$ in the \pta\ spectrum comes from the Lorentz invariant definition of the cross section in terms of $d^3p$. 
Our results could thus be interpreted as the massless approximation to $d^3p$. }.
Primary charged particles are defined as charged particles with a mean lifetime $\tau > 0.3 \cdot 10^{-10}$~s either directly produced in $pp$ interactions or from subsequent decays of particles with a shorter lifetime.

The charged-particle multiplicity results are compared to particle level Monte Carlo (MC) predictions. 
Three different phase-space regions are considered in this paper, with varying selection both on the \pta\ and the number of charged particles per event; 
all phase-space regions require tracks within $| \eta | < 2.5$.
Diffractive physics is expected to contribute mostly at low numbers of charged particles and at low track momentum.
Therefore varying the selection on \nch\ and \pta\ in effect varies the relative contribution from diffractive events.
Appendix~\ref{sec:more_phase_spaces} shows the results for two additional phase-space regions useful for Monte Carlo tuning.
This measurement, with refined corrections and systematic uncertainty determination supersedes the results presented in~\cite{MB1}.

%%%%%%%%%%%%%%%%%%%%%%%%%%%%
\section{The ATLAS Detector}\label{sec:atlas}

The ATLAS detector~\cite{:2008zzm} at the Large Hadron Collider (LHC)~\cite{Evans:2008zzb}
covers  almost the whole solid angle
around the collision point with 
layers of tracking detectors, calorimeters and muon chambers. It has been designed to study 
a wide range of physics topics at LHC energies. 
For the measurements presented in this
paper,  the tracking devices and the trigger system are of particular importance.

% ID
The ATLAS Inner Detector (ID) has full coverage in $\phi$ and covers the pseudorapidity range $|\eta|~<~2.5$. 
It consists of a silicon pixel detector (Pixel), a silicon microstrip detector (SCT) and a transition radiation tracker 
(TRT). These detectors cover a sensitive radial distance from the interaction point of 
50.5--150~mm, 299--560~mm and 563--1066~mm, respectively, and are immersed in a 2~T  
axial magnetic field.
The inner-detector barrel (end-cap) parts consist of 3 (2$\times$3) Pixel layers, 4 (2$\times$9)
double-layers of single-sided silicon microstrips with a 40~mrad stereo angle,
and 73 (2$\times$160) layers of
TRT straws.  
Typical position resolutions are
10, 17 and 130~$\mu $m for the $R$-$\phi$ co-ordinate and, in case of the Pixel and SCT,  115 and 580~$\mu $m for the second measured co-ordinate.
A track from a charged particle traversing the barrel detector would typically have 11
silicon hits~\footnote[3]{A hit is a measurement point assigned to a track.}(3 pixel clusters and 8 strip clusters) and more than
30 straw hits.

% 2.36 TeV configuration
For the runs at \sqt,  stable beams were not declared by the LHC; 
the high voltage on the SCT detector was thus not turned up to its nominal operating voltage but was left in standby mode. The Pixel detector was at nominal conditions for these runs. 
The hit efficiency in the SCT is thus significantly lower and 
special track reconstruction algorithms are needed; 
the single hit efficiency at nominal voltage in the SCT barrel is above 99.7\%~\cite{atlas_id_perf}, while in standby it drops to $\sim 60\%$ for tracks perpendicular to the silicon surface.

% Trigger
The ATLAS detector has a three-level trigger system:~Level~1~(L1), Level~2 (L2) and Event~Filter~(EF). 
For this measurement, the trigger relies on the 
L1 signals from the
Beam Pickup Timing devices (BPTX) and 
the Minimum Bias Trigger Scintillators (MBTS).
The BPTX stations are composed of electrostatic button pick-up detectors
attached to the beam pipe at $\pm$175~m from the centre of the ATLAS detector.
The coincidence of the BPTX signal between the two sides of the detector is used to determine when bunches are colliding in the centre of the ATLAS detector.
The MBTS are mounted at each end of the detector in front of the liquid-argon end-cap calorimeter cryostats at $z = \pm 3.56$~m.
They are segmented into eight sectors in azimuth 
and two rings in pseudorapidity ($2.09 < |\eta|~<~2.82$ and $2.82 < |\eta|~<~3.84$).   
Data were collected for this analysis using a trigger requiring a BPTX coincidence and MBTS trigger signals.
The MBTS trigger used for this paper is configured to require one hit above threshold from either side of the detector, referred to as a single-arm trigger.  
The efficiency of this trigger is studied with a separate prescaled L1 BPTX trigger, filtered to obtain inelastic interactions by Inner Detector requirements at L2 and EF, the latter only for the 900~GeV data.

%%%%%%%%%%%%%%%%%%%%%%%%%%%%
\section{Monte Carlo Simulation}\label{sec:mc}

Inclusive minimum bias data are modelled using three components in the \py 6~\cite{pythia6} Monte Carlo (MC) event generator: non-diffractive (ND), single- (SD) and double-diffractive (DD).
Non-diffractive processes are modelled from two-to-two processes as described in this section.
Diffractive process modelling is described in Sec.~\ref{sec:diff_models}.

Low-\pta\ scattering processes may be described by 
lowest-order perturbative Quantum Chromodynamics (QCD) two-to-two parton scatters, where the divergence of the cross section at 
\pta\ =~0  is regulated by phenomenological models.
The \py 6 MC event generator 
implements several of these models.
The parameters of these models have been tuned to 
describe charged-hadron production and the underlying event in $pp$ and $p {\bar p}$ data  at centre-of-mass energies between 200~GeV and 1.96~TeV. 

% MC09 as default tune for unfolding
Samples of MC events were produced for single-diffractive, double-diffractive and non-diffractive processes using the \py 6 generator~\footnote[4]{\py\ version 6.4.21}.
The  ATLAS MC09 \py\ tune~\cite{atlasmc09} uses a specific set of optimised parameters;
it employs the MRST LO* parton density functions (PDFs)~\cite{Sherstnev:2007nd} and 
the \pta-ordered parton shower~\cite{py_pt_shower}.
A tune is a particular configuration or set of values of the parameters of the particular Monte Carlo model.
These parameters were derived by tuning to the underlying event (UE) and minimum-bias 
data from the Tevatron at 630~GeV to 1.96~TeV.
The MC samples generated with this tune are used to determine detector acceptances and efficiencies 
and to correct the data.
MC samples were produced at all three centre-of-mass energies considered in this paper. 
The non-diffractive, single-diffractive and double-diffractive contributions in the generated samples are  
mixed according to the generator cross sections.

% detector simulation
All the events are processed through the ATLAS detector simulation program~\cite{bib-ATLAS-simulation-sub}, which is based on {\sc geant}4~\cite{Agostinelli:2002hh}. They are then reconstructed and analysed by the same program chain used for the data.
Particular attention was devoted to the description in the simulation of the size and position of the collision beam spot and of the detailed detector conditions  during data taking.
The MC09 \py 6 samples are used to derive the detector corrections for these measurements. 
The MC samples at 2.36~TeV were generated assuming nominal detector conditions.

% hadron level comparisons
For the purpose of comparing the present measurements 
to different phenomenological models describing minimum-bias events, 
the following additional particle level MC samples were generated:

\begin{itemize}

\item the new ATLAS Minimum Bias Tune 1 (AMBT1) \py 6 tune described in Sec.~\ref{sec:ambt1};

\item the DW~\cite{Albrow:2006rt} \py 6 tune, which uses virtuality-ordered showers and was derived to describe the CDF Run II underlying event and Drell-Yan data;

\item the \py 8 generator~\footnote[5]{\py\ version 8.130}~\cite{py8}, in which the diffraction model produces much harder \pta\ and \nch\ spectra for the single- and double-diffractive contributions than \py 6. The default parton shower model is similar to the one used in \py 6 MC09;

\item the \pho\ generator~\footnote[6]{\pho\ version 1.12.1.35}~\cite{phojet}, which is used as an alternative model to \py-based generators.
\pho\ relies on \py6~\footnote[7]{\py\ version 6.1.15}for the fragmentation of partons.

  \end{itemize}

%----------------------------------------------
\subsection{Diffractive Models}\label{sec:diff_models}
\py 6, \py 8 and \pho\ model the diffractive components very differently. Here we mostly describe the model implemented in \py6.
The \py 6 diffraction is based on a Regge-based pomeron model to generate the cross-section and generate the diffractive mass and momentum transfer~\cite{diff_models_1a,py6_diff}. 
To allow the Regge model to cover the full phase-space, empirical corrections are introduced~\cite{pythia6}. 
These have the effect of enhancing the production of small masses and suppressing production near the kinematic limit. Particle production from low mass states ($M_X < 1$~GeV) is treated as an isotropic two body decay. 
Particle production from high mass states is based on the string model. 
Two string configurations are possible depending on whether the pomeron couples to a quark or gluon~\cite{pythia6}. 

The \py 8 model uses the same model as \py 6 to generate the cross-section and generate the diffractive mass and momentum transfer.
The particle production for low mass states uses the string model but for higher masses ($M_X>10$~GeV) a perturbative element based on pomeron-proton scattering is introduced. 
The non-perturbative string model introduces a mass dependence on the relative probability of the pomeron scattering off a quark to scattering off a gluon, which enhances the gluon probability at high masses. 
The perturbative pomeron-proton scattering uses HERA diffractive PDFs~\cite{hera_pdf} and the standard multiple interactions framework is used to generate the parton-parton scattering. 
The introduction of the perturbative pomeron-proton scattering results in a harder \pta\  and multiplicity spectrum for diffractive events generated with \py 8 compared to those generated with \py 6~\cite{diff_models_3}.
However, it should be noted that relatively little tuning has been made of the diffractive processes in \py 6 and \py 8.

\pho\ is based on the dual parton model.
It generates a harder \pta\ and multiplicity spectrum in diffractive events than \py 6.
The new diffraction model of \py 8 generates distributions quite similar to those from \pho~\cite{diff_models_3}.

%%%%%%%%%%%%%%%%%%%%%%%%%%%%

\subsection{PYTHIA 6 ATLAS Minimum Bias Tune 1}\label{sec:ambt1}

Before the start of the LHC, an ATLAS tune to \py 6 with MRST LO* PDFs using Tevatron underlying event and minimum bias data was produced, the so-called MC09 tune~\cite{atlasmc09}.  
The first ATLAS measurements of charged particle production at the LHC~\cite{MB1} measured the charged particle production at \sqn\ in the central region to be 
5--15\% higher than the Monte Carlo models predict.
In addition, neither the high \nch\ nor
the high \pta\ distributions were well described by this tune and 
the \meanpt\ was overestimated in events with $\nch > 20$.
A new tune, AMBT1, was developed in order to adapt the free parameters of the non-diffractive models to the new experimental data at \sqn\ and \sqs, using the same PDFs and \py 6 model choices as MC09.

The AMBT1 tune is obtained by tuning to ATLAS minimum bias data at both \sqn\ and \sqs\ in a diffraction-reduced phase-space that is presented in this paper: $\nch \geq 6$, $\pta > 500$~MeV, $|\eta| < 2.5$. 
The tune was derived using preliminary versions of these distributions~\cite{MB15T}.
The starting point for this tune is the ATLAS MC09c~\cite{atlasmc09}  \py 6 tune. 
MC09c is an extension of the ATLAS MC09 tune where the strength of the colour reconnection (CR) was tuned to describe the \meanpt\  vs. \nch\ distributions measured by CDF in $p\bar{p}$ collisions at the Tevatron~\cite{Aaltonen:2009ne}.

Charged particle distributions are sensitive to multi-parton interactions (MPI) 
and colour reconnection of the hadronic final state~\cite{skandscolor}; 
the MPI are regulated by a low \pta\ cut-off and the matter overlap distribution of the two protons in which the additional partonic scattering takes place.
These are the main parameters varied for this new tune.
Parameters related to final 
state radiation, hadronisation and fragmentation are not tuned, as these are constrained by many LEP 
results. 
No changes to the diffraction model are made.
The model parameters are adapted in order to best describe these new distributions
over the full range while maintaining consistency with the Tevatron results. 
For the data MC comparisons the {\sc Rivet} \footnote[8]{version 1.2.2a0}~\cite{rivet} package is used;
the tuning is done using the {\sc professor} package \footnote[9]{version 1.0.0a0}~\cite{professor, professor_orig}.
Table~\ref{tab:ambt1_tune} summarizes the parameters varied in this tune;
the meaning of the parameters are given below.

\paragraph{MPI Parameters}
The size of the MPI component in the \py 6 model is regulated by a simple cut-off parameter 
for the $\hat{p}_T$ of two-to-two scattering processes. 
This cut-off parameter is fixed at a reference energy, which is generally taken as 1.8 TeV. 
The cut-off at this reference scale is called PARP(82).
It is then rescaled for other centre-of-mass energies using a parameter PARP(90). 
The rescaling is done according to the following formula:
\begin{equation}
\pta^{min} = \mathrm{PARP}(82) \left( \frac{E}{1.8\,\, \mathrm{TeV}}\right)^{\mathrm{PARP}(90)}.
\end{equation}

The amount of scattering is described by the matter overlap distribution between the two protons,
which regulates how many central, hard scatterings and how many less central, softer scatterings occur.
This distribution is modelled as a double Gaussian probability density function.
The parameter PARP(83) describes the fraction of matter in the narrower of the two Gaussian functions.
The size of this narrower Gaussian is given as a fraction PARP(84) of the wider, main radius.
The optimal value for this parameter was found in a first tuning run.
Further variations of the matter fraction in the narrower cone were found to not have a significant influence on the main distributions used for tuning.

\paragraph{Colour Reconnection Parameters}
The colour reconnection scenario of \py\ used in MC09c minimises the total string length between partons. 
The probability that a given string piece does not participate in the CR
is given by $(1 - \mathrm{PARP}(78))^{n_\mathrm{MI}}$ , where $n_\mathrm{MI}$ is the number of multi-parton interactions~\cite{pythia6};
the larger the parameter, the smaller the probability of the string piece not participating.
In addition to this parameter, an additional parameter PARP(77) is present in \py; 
it is used to describe a
suppression factor for the CR of fast moving string pieces. 
The suppression factor is given
by $1/(1 + \mathrm{PARP}(77)^2 \cdot p^2_\mathrm{avg}$), where $p^2_\mathrm{avg}$ is a measure of the average squared momentum that hadrons
produced by the string piece would have. 

\paragraph{Additional Parameters Investigated}
In an initial study, the cut-off parameter for initial state radiation (PARP(62)) and the cut-off for momentum smearing in primordial $k_\perp$ (PARP(93)) were considered.
The optimal values for these parameters were found in a first tuning run, further
variation of those parameters was not found to have a significant influence on the main distributions used for tuning.

\paragraph{Distributions Used}
The tune described in this paper focuses on the ATLAS minimum bias data.
It primarily attempts to improve the description of the high \pta\ and high \nch\ distributions observed.
For the \pta\ spectrum, only particles above 5 GeV are considered.
For the \nch\ spectrum, only events with 20 or more tracks are used in the tune.
For the \meanpt\ vs. \nch\ distribution, only events with ten or more tracks are considered.
The full $\eta$ distribution is used.
 For  completeness, the preliminary underlying event
results~\cite{ATLAS-UE, ATLAS-UE-paper} are included in the plateau region; however, due to the limited statistics, these data have only very small impact on the tune.
  
Tevatron data in the energy range of 630~GeV to 1.96~TeV are included in the tune, but with a weight which is ten times lower than that of the ATLAS data. 
This weighting allows a check of the consistency of the resulting tune with the Tevatron data while forcing the ATLAS data to drive the tuning process. 
Similar datasets were used for the MC09c tune.
The charged particle multiplicity shown in~\cite{cdf2002} was not included in
the tune as no variation of the tuning parameters considered was able to fit both the ATLAS and the CDF distributions simultaneously.
App.~\ref{sec:ambt1_dist} shows a full list of the distributions and the ranges considered by the tune.

\paragraph{Results}
The final parameter values resulting from the tune are shown in Table~\ref{tab:ambt1_tune}.

\begin{table}[h!]
\begin{center}
\begin{tabular}{llccc}
\hline
Parameter &  Related model & MC09c value & scanning range & AMBT1 value\\
\hline 
PARP(90)  & MPI  (energy extrapolation)  & 0.2487     & $0.18-0.28$  & 0.250\\
PARP(82)  & MPI ($p_T^\mathrm{min}$) & 2.31         & $2.1-2.5$  & 2.292\\
PARP(84)  & MPI matter overlap (core size) & 0.7         & $0.0-1.0$  & 0.651\\
PARP(83)  & MPI matter overlap (fraction in core) & 0.8         & fixed & 0.356\\
PARP(78)  & CR strength & 0.224       & $0.2-0.6$  & 0.538\\ 
PARP(77)  & CR suppression & 0.0         & $0.25-1.15$  & 1.016\\
PARP(93)  & Primordial $k_\perp$ & 5.0         & fixed          & 10.0\\ 
PARP(62)  & ISR cut-off & 1.0         & fixed          & 1.025\\ \hline
\end{tabular}
\caption{
Comparison of MC09c and AMBT1 parameters.
The ranges of the parameter variations scanned are also given.
The parameters declared as `fixed' were fixed to the values obtained after an initial pass of the tuning.
}
\label{tab:ambt1_tune}
\end{center}
\end{table}

%%%%%%%%%%%%%%%%%%%%%%%%%%%%
\section{Data Selection}\label{sec:evt_sel}

Events in which the Inner Detector was fully operational and the solenoid magnet was on are used for this analysis for both \sqn\ and \sqs. 
During this data-taking period, more than $97$\% of the Pixel detector, $99$\% of the SCT and $98$\% of the TRT were operational.
At \sqt\ the requirements are the same, except for the SCT being in standby.

Events were selected from colliding proton bunches
in which the MBTS trigger recorded one or more counters above threshold on either side.  
The maximum instantaneous luminosity is
 approximately $1.9\times10^{27}$~cm$^{-2}$~s$^{-1}$ at 7~TeV.  
The probability of additional interactions in the same bunch crossing is estimated to be of the order of 0.1\%.
In order to perform an inclusive-inelastic measurement, no further requirements beyond the MBTS trigger are applied.

In order to better understand the track reconstruction performance at \sqt , during which time the SCT was in standby, additional data at \sqn\ were taken with the SCT in standby for part of a run. This enables the derivation of data-driven corrections to the track reconstruction efficiency, as described in Sec.~\ref{sec:track_eff_236}.

%------------------------------------------------------
\subsection{Different Phase-Space Regions Considered}

Three separate phase-space regions are considered in the main part of this paper with varying contributions from diffractive events:
\begin{itemize}
\item at least one charged particle in the kinematic range $|\eta|<$~2.5 and $\pta> 500$~MeV,
\item at least two charged particles in the kinematic range $|\eta|<$~2.5 and $\pta> 100$~MeV,
\item at least six charged particles in the kinematic range $|\eta|<$~2.5 and $\pta> 500$~MeV.
\end{itemize}

The first of these phase-space regions is studied at all three centre-of-mass energies. 
This is the region that allows us to best investigate the evolution of charged-multiplicity distributions as a function of centre-of-mass energy and thus constrain the MC parameters that dictate the energy extrapolation of the models.
The second measures the most inclusive charged-particle spectra and is also used as the basis for the model-dependent extrapolation to $\pta = 0$;
in this phase-space region results at $\sqrt{s} =0.9$ and 7~TeV are shown.
The third phase-space region considered is similar to the first but with a higher cut on the number of charged particles, thus reducing the expected contribution from diffractive events in the sample.
These distributions are measured for both 0.9 and 7~TeV. 
This is the phase-space region which was used to produce the new AMBT1 tune.
At 2.36~TeV only the first phase-space region is measured.
Two additional phase-space regions are presented in App.~\ref{sec:more_phase_spaces}.

The relative contribution from diffractive events varies widely between Monte Carlo models and depends strongly on the phase-space region selection applied.
The diffractive contribution is constrained very little by previous data.
Table~\ref{tab:diff_fraction} shows the predicted 
fractions of simulated events originating from diffractive processes, as predicted by \py 6, \py 8 and \pho; 
the values for the different tunes of \py 6 are found to be similar because the acceptances of the different non-diffractive models do not change significantly and the diffractive models are identical.
The large difference in predictions between the models is one of the motivations for not making any model-dependent corrections to the experimental data, 
as such corrections would vary significantly depending on which MC model is used to derive them.

\begin{table}[h!]
 \begin{center}
   \begin{tabular}{|c | c  | c | c | c | c | c | c |}
   \hline\hline
   \multicolumn{2}{|c}{Phase-Space Region} &   \multicolumn{3}{|c}{\sqn} & \multicolumn{3}{|c|}{\sqs} \\
  min \nch & min \pta\ (MeV) & { \py 6} & { \py 8} & { \pho} & { \py 6} & { \py 8} & { \pho}  \\ \hline
2  & 100 & 22\% & 22\% & 20\% & 21\% & 21\% & 14\% \\
1  & 500 & 16\%& 21\% &19\% & 17\% & 21\% & 14\% \\
6  & 500 & 0.4\% &5\% & 8\%& 0.4\% & 10\% & 8\% \\
   \hline\hline
   \end{tabular}
   \caption{
Fraction of simulated events originating from diffractive processes, as
predicted by \py 6, \py 8 and \pho\ in the three phase-space regions measured in this paper at both \sqn\ and \sqs. 
All results are for $|\eta| < 2.5$.
\label{tab:diff_fraction} 
}
 \end{center}
\end{table}

%---------------------------------------------------
\subsection{Event Selection}
To reduce the contribution from background events and non-primary tracks, as well as to minimise the systematic uncertainties, 
the events are required to satisfy the following criteria:

\begin{itemize}
\item to have triggered the single-arm, single-counter level 1 minimum bias trigger scintillators
\item the presence of a primary vertex~\cite{Piacquadio:2008zzb} 
reconstructed using the beam spot information~\cite{beamspot} and at least two tracks, each with:
        \begin{itemize}
	\item \pta~$>$~100~MeV, 
	\item a transverse distance of closest approach with respect to the beam-spot position
	$| \dzerobs | < $ 4~mm;
         \end{itemize}
\item the rejection of events with a second vertex containing four or more tracks, to remove events with more than one interaction per bunch crossing;
\item a minimum number of tracks, depending on the particular phase-space region, as described in Sec.~\ref{sec:track_algos}.
\end{itemize}

%---------------------------------------------------------
\subsection{Track Reconstruction Algorithms}\label{sec:track_algos}

Tracks are reconstructed offline within the full acceptance range $| \eta |<$~2.5 of the Inner Detector~\cite{newt1,newt2}.
Track candidates are reconstructed by requiring a minimum number of silicon hits and 
then extrapolated to include measurements in the TRT.
Due to the SCT being in standby mode at 2.36~TeV, different track reconstruction algorithms are needed;
at 0.9 and 7~TeV, the reconstruction algorithms are collectively referred to as full tracks.
The analysis at \sqt\ has been performed using two complementary methods for reconstructing tracks. 
The first reconstructs tracks using pixel detector information only, denoted Pixel tracks.
The second uses tracks reconstructed from the full Inner Detector information, denoted ID tracks~\footnote[1]{In the context of the other analyses, ID tracks are referred to as track for brevity.}.

\subsubsection{Algorithms for 0.9 and 7~TeV}
For the measurements at 0.9 and 7~TeV, two different track reconstruction algorithms are used. 
The algorithm used for the previous minimum-bias publication~\cite{MB1} is used with a lower \pta\ threshold cut at 100~MeV. 
An additional algorithm configuration is run using only the hits that have not been used by the first algorithm. 
This additional algorithm uses wider initial roads and has a looser requirement on the number of silicon hits. 
This second algorithm contributes around 60\% of the tracks from 100 to 150~MeV, mostly due to the tracks having too low a momentum to go far enough in the SCT detector to satisfy the silicon hit requirement of the original algorithm; 
this fraction decreases rapidly, reaching less than 2\% at 200~MeV.

\begin{table}[h!]
 \begin{center}
   \begin{tabular}{|c | c | c |c |}
   \hline\hline
   Criteria & $\sqrt{s} = $ 0.9 and 7~TeV & \multicolumn{2}{|c|}{ $\sqrt{s} = $ 2.36 TeV} \\
& Full Tracks &ID Tracks & Pixel Tracks \\
   \hline
 $\pta>100$ or 500~MeV & YES & YES & YES \\
$|\eta | < 2.5$  & YES & YES & YES \\
layer-0 hit if expected  & YES & YES & YES(*) \\
$>1$ Pixel hit  & YES & YES & YES \\
$> 2$, 4 or 6 SCT hits for tracks (**) & YES & NO &NO \\
$| \dzero |< 1.5$~mm and $| \zzero | \cdot \sin \theta < 1.5$~mm & YES & YES & YES(***) \\
$\chi^2$ probability $>  0.01$ for $\pta~>~10$~GeV & YES & N/A & N/A \\
     \hline\hline
   \end{tabular}
   \caption{
Selection criteria applied to tracks for the full reconstruction, ID tracks and Pixel tracks.
The transverse momentum cut applied depends on the phase-space region in question.
(*) For the Pixel track method the layer-0 is required even if not expected.
(**) The SCT hit selection are for $\pta < 200$, $200< \pta < 300$ or $\pta~>~300$~MeV, respectively.
(***) For the Pixel track method, the \dzero\ and \zzero\ selection are after the track refitting is performed (see Sec.~\ref{sec:2.36config}).
}\label{tab:track_sel}
 \end{center}
\end{table}
Tracks are required to pass the selection criteria shown in Table~\ref{tab:track_sel}; 
the column labelled Full Tracks refers to the algorithms used at 0.9 and 7~TeV. 
The transverse, \dzero, and longitudinal, \zzero,  impact parameters are calculated with respect to the event primary vertex.  
The layer-0 selection requires a hit in the inner-most layer of the Pixel detector if a hit is expected~\footnote[2]{A hit is expected if the extrapolated track crosses an active region of a Pixel module that has not been disabled.}.
The track-fit $\chi^2$ probability~\footnote[3]{This probability function is computed as $1-P(n_\mathrm{dof}/2,\chi^2/2)$, where $P(n_\mathrm{dof}/2,\chi^2/2)$ is the incomplete gamma function and $n_\mathrm{dof}$ is the number of degrees of freedom of the fit. It represents the probability that an observed $\chi^2$ exceeds the observed value for a correct model.} cut is applied to remove tracks with mis-measured \pta\ due to mis-alignment or nuclear interactions.

These tracks are used to produce the corrected distributions and will be referred to as selected tracks. 
The multiplicity of selected tracks within an event is denoted by \nsel.
The tracks used by the vertex reconstruction algorithm are very similar to those used for the analysis; 
the \pta\ threshold is also 100~MeV.
Due to the requirement that the vertex be made from a minimum of two such tracks and the fact that we do not wish to correct our measurement outside of the observed phase-space region, the minimum number of particles per event for the phase-space region with $\pta>100$~MeV also needs to be set at two.
Table~\ref{tab:nevents} shows the total number of selected events and tracks for all phase-space regions considered.

Trigger and vertex reconstruction efficiencies are parameterised as a function of \nselbs . 
\nselbs\ is defined as the number of tracks passing all of the track selection requirements except for the constraints with respect to the primary vertex;  
instead, the unsigned transverse impact parameter with respect to the beam spot, $|\dzerobs|$, is required to be less than 1.8~mm.

\begin{table}[h!]
 \begin{center}
   \begin{tabular}{|c | c | c | c | c | c | c | c | c |}
   \hline\hline
   \multicolumn{2}{|c}{Phase-Space Region} &   \multicolumn{2}{|c}{ { \sqn}} & \multicolumn{2}{|c}{ { \sqs}} & \multicolumn{2}{|c|}{ { \sqt}}  \\
 \nch  &  min \pta & \multicolumn{2}{c}{ { Full Tracks}} & \multicolumn{2}{|c}{ { Full Tracks}} &  \multicolumn{2}{|c|}{ { ID Tracks (Pixel Tracks)}}   \\
& { (MeV)} & { Events} & {Tracks} & { Events} & {Tracks} & {Events} & {Tracks} \\ \hline
2  & 100 & 357,523 & 4,532,663 
		& 10,066,072 & 209,809,430 
		& - &-  \\
1  & 500 	& 334,411 & 1,854,930
		& 9,619,049 & 97,224,268
		& 5,929 (5,983) & 38,983 (44,788) \\
6  & 500 	& 124,782 & 1,287,898
		& 5,395,381 & 85,587,104
		& - &  -  \\
   \hline\hline
   \end{tabular}
   \caption{\label{tab:nevents} Number of events and tracks in the three phase-space regions at each centre-of-mass energy considered in this paper.
}
 \end{center}
\end{table}

%%%%%%%%%%%%%%%%%%%%%%%%%%%%%%%%%%%
\subsubsection{Track Reconstruction Algorithms at 2.36~TeV}\label{sec:2.36config}

Operation of the SCT at standby voltage during 2.36~TeV data taking led to reduced SCT hit efficiency. 
Consequently, ID tracks are reconstructed at this centre-of-mass energy using looser requirements on the numbers of hits and holes~\footnote[4]{A hole is defined as an absence of a hit when it is expected given the track trajectory.}~\cite{newt1, newt2}. 
There are no simulation samples that fully describe the SCT operating at reduced voltage. 
A technique to emulate the impact of operating the SCT in standby was developed in simulation; 
this corrects the Monte Carlo without re-simulation by modifying the silicon clusterisation algorithm used to study the tracking performance. 
However, the final ID track efficiency at \sqt\ was determined using a correction to the track reconstruction efficiency derived from data at \sqn. 

Pixel tracks were reconstructed using the standard track reconstruction algorithms limited to Pixel hits and with different track requirements. 
There is little redundant information, because at least three measurement points are needed to obtain a momentum measurement and the average number of Pixel hits per track is three in the barrel. 
Therefore the Pixel track reconstruction efficiency is very sensitive to the location of inactive Pixel modules. 
The total distance between the first and the last measurement point in the pixel detector, as well as the limited number of measurement points per track, limit the momentum resolution of the tracks;
therefore the Pixel tracks were refit using the reconstructed primary vertex as an additional measurement point. 
The refitting improves the momentum resolution by almost a factor of two.
However, the Pixel track momentum resolution remains a factor of three worse than the resolution of ID tracks.

The selection criteria used to define good Pixel and ID tracks are shown in Table~\ref{tab:track_sel}. The total number of accepted events and tracks at this energy are shown in Table~\ref{tab:nevents}.
These two track reconstruction methods have different limitations; 
the method with the best possible measurement for a given variable is chosen when producing the final plots.
The Pixel track method is used for the \nch\ and $\eta$ distributions,
while the ID track method is used for the \pta\ spectrum measurement;
the \meanpt\ distribution is not produced for this energy as neither method is able to describe both the number of particles and their \pta\ accurately.

%%%%%%%%%%%%%%%%%%%%%%%%%%%%%%%%%%%
\section{Background Contribution}\label{sec:background}

\subsection{Event Backgrounds}
There are three possible sources of background events that can contaminate the selected sample:  cosmic rays, beam-induced background and the presence of another collision inside the same bunch crossing.
The fraction of cosmic ray background events was estimated in~\cite{MB1}, where it was found to be smaller than $10^{-6}$.
Beam-induced backgrounds are estimated from non-colliding empty bunches using the same method as described in~\cite{MB1}; 
after final event selection, fewer than 0.1\% of events are predicted to originate from beam-induced backgrounds. 
The reconstructed primary vertex requirement is particularly useful in suppressing the beam-induced background.
The instantaneous luminosity at \sqs\ is high enough that the effect of multiple collisions inside the same bunch crossing cannot be ignored. 
Events are rejected if they have a second vertex with four or more tracks~\footnote[5]{Events with two vertices with fewer than four tracks are dominated by events where a secondary interaction is reconstructed as another primary vertex and are thus not removed from our data samples.}. 
After this cut, the fraction of events with more than one interaction in the same bunch crossing is measured to be about 0.1\%; 
the residual effect is thus neglected. At the lower centre-of-mass energies, the rate of multiple interactions is lower and thus also neglected.

%%%%%%%%%%
\subsection{Backgrounds to Primary Tracks}
\begin{figure}[htbp]
\begin{center}
\includegraphics[width=0.49\textwidth]{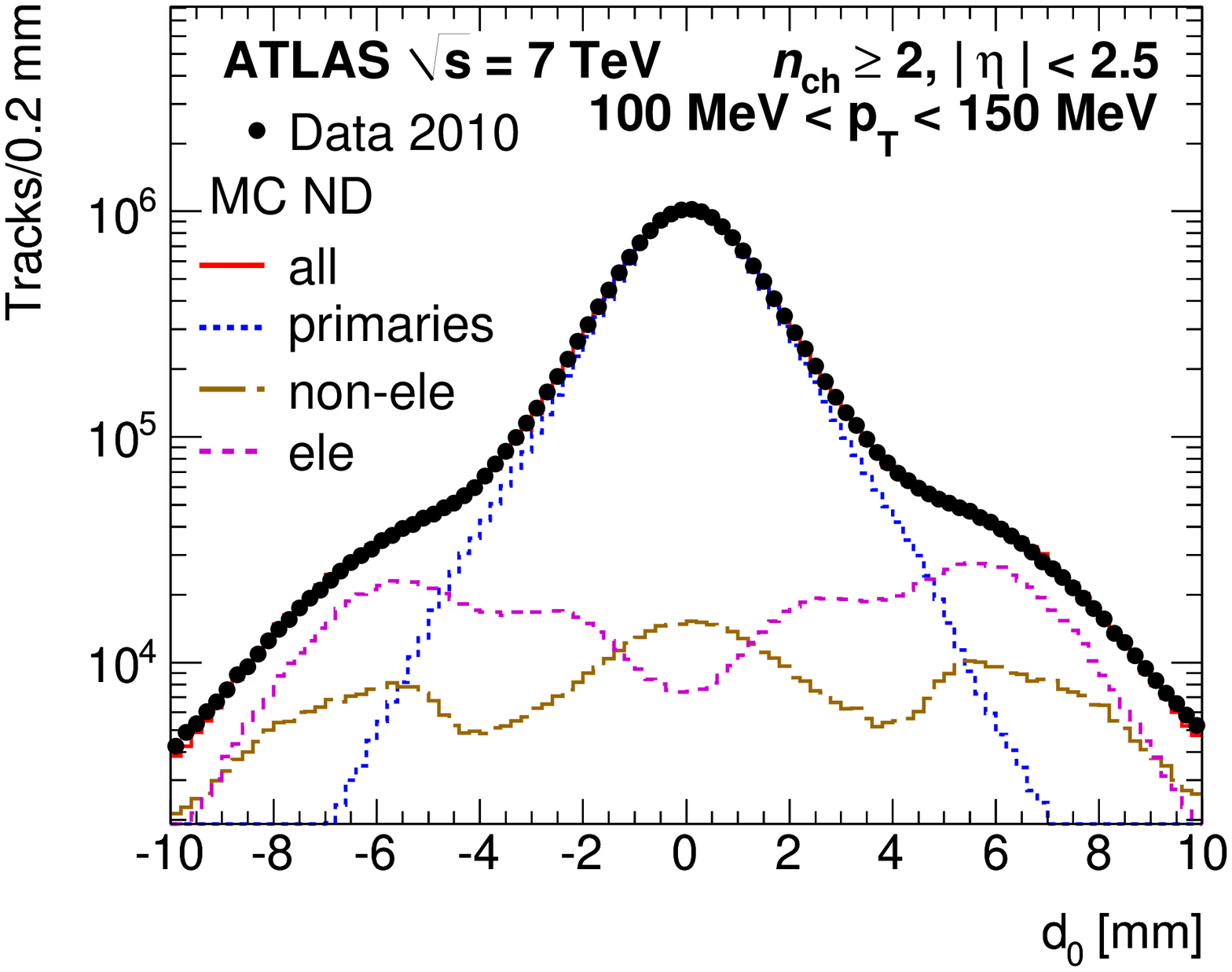}
\end{center}
\vspace{-1cm}
\caption{Transverse impact parameter, \dzero , distribution at \sqs\ for primary (blue short dashed) and non-primary particles after scaling them to the best fit value for $100 < \pta < 150$~MeV.
The non-primary particles are split into electrons (pink long-dashed) and non-electrons (green dot-dashed).
The full red curve shows the non-diffractive (ND) MC prediction for the sum over the three components which agrees well with the data (black points). 
}
 \label{fig:d0_fit}
\end{figure}

Primary charged-particle multiplicities are measured from selected-track distributions after correcting for the fraction of non-primary particles in the sample. 
Non-primary tracks are mostly due to hadronic interactions, photon conversions and decays of long-lived particles, as well as a small fraction of fake tracks.
Their contribution is estimated using MC predictions for the shape of the \dzero\ distribution for primaries, non-primaries from electrons and other non-primaries. 
The separation between non-primaries from electrons and non-electrons is needed as the electrons are mostly from conversions in the detector material and would thus be sensitive to a mis-modeling of the detector material, whereas the non-electron non-primary tracks are mostly from long-lived particles and this fraction is thus also sensitive to the underlying physics.
The Gaussian peak of the \dzero\ distribution, shown in Fig.~\ref{fig:d0_fit} for $100<\pta< 150$~GeV, is dominated by the primary tracks and their resolution. 
The non-primary tracks populate the tails. 
The dominant contribution to non-primary tracks inside the acceptance cut on $|\dzero|$ comes from non-electrons.

The primary, electron non-primary and non-electron non-primary \dzero\ distributions are obtained from MC and used as templates to extract the relative fractions in data.
A fit is performed in the side-bands of the distribution, i.e. outside the range in \dzero\ used for selecting tracks.
The fractions of primary, electron non-primary and non-electron non-primary tracks are all allowed to float with the total number of events constrained to that of the data. 
The contribution of non-primaries from electrons within the analysis acceptance of 1.5~mm is small, while it dominates at high values of $|\dzero|$. 
The requirement on having a hit on layer-0 suppresses this contribution enough to allow the fit to be performed down to the lowest \pta\ region. 
The fit is performed in bins of 50~MeV in \pta\ from 100 to 500~MeV. 
A single fit is used for all tracks with $\pta~>~500$~MeV; in this bin the distinction is not made between the two sources of non-primary tracks. 
The fraction of non-primary tracks varies from 3.4\% for $100 < \pta < 150$~MeV to 1.6\% above 500~MeV at \sqs. 
Figure~\ref{fig:d0_fit} shows the observed \dzero\ distribution for the bin $100 < \pta < 150$~MeV compared to the MC predictions after the fit.

%%%%%%%%%%%%
\paragraph {Systematic Uncertainties}
The full difference between the non-primary fraction in MC and that in data obtained using the fit is taken as a systematic uncertainty.  
The largest difference is found to be an increase of non-primaries in data by 25\% relative to the MC for $\pta~>~500$~MeV. This conservative estimate is taken to be constant as a function of \pta\ and results in only a small effect, up to 0.9\%, on the final corrected distributions.
In order to estimate the effect of the choice of the variable used to obtain the fit, the fraction of primary and non-primary track contributions are obtained by fitting the \zzero\ distributions.
The difference is measured to be 12\% in the first bin, 8\% in the last bin and less than 4\% in all other bins; this difference is taken as a source of systematic uncertainty.
The estimated number of non-primary tracks in $|\dzero| < 1.5$~mm is found to be stable with respect to a change in the fit range of 1~mm in all \pta\ bins except the first one ($100 < \pta <150$~MeV), where a 10\% difference is observed;  this difference is taken as a systematic uncertainty.
The fraction of non-primary tracks is found to be independent of \nsel , but shows a small dependence on $\eta$, taken as a small systematic uncertainty of 0.1\%.

The total uncertainty on the fraction of non-primary tracks is taken as the sum in quadrature of all these effects.
The total relative uncertainty on the measured distributions at \sqn\ and \sqs\ is 1.0\% for the first \pta\ bin, decreasing to 0.5\% above 500~MeV.
At  \sqt\ this uncertainty for the Pixel track method is 0.6\%.

%%%%%%%%%%%%%%%%%%%%%%%%%%%%

\section{Selection Efficiency}\label{sec:eff-corr}

The data are corrected to obtain inclusive spectra for charged primary particles satisfying the different phase-space region requirements. 
These corrections include inefficiencies due to trigger selection, 
vertex and track reconstruction. They also account for effects due to the momentum scale and resolution and for the residual background from non-primary tracks.

In the following sections the methods used to obtain these efficiencies, as well as the systematic uncertainties associated with them are described. 
Plots are shown for the phase-space region $\nch\geq2$, $\pta>100$~MeV, $|\eta|~<~2.5$ at \sqs, but similar conclusions can be drawn at the other energies and phase-space regions.

%%%%%%%%%%%%%%
\subsection{Trigger Efficiency}\label{sec:trigeff}

The trigger efficiency, \trigeff , is measured from a data sample selected using a control trigger.  
The control trigger used for this analysis selects events from random filled bunch crossings which are then filtered at L2.
At \sqn\ the L2 filter requires a minimum of 
seven pixel clusters and seven SCT hits and the EF requires at least one track with $\pta~>~200$~MeV.
At \sqs\ the L2 requirement is loosened to four pixel clusters and four SCT hits. No EF requirements are made at this energy.
The vertex requirement for selected tracks is removed for these trigger studies, to account for correlations between the trigger and vertex reconstruction efficiencies.
The trigger efficiency is determined by taking the ratio of events from the control trigger in which the L1 MBTS also accepted the event, over the total number of events in the control sample.
For \sqt\ there is not sufficient data to measure the trigger efficiency and thus the \sqn\ parametrisation is used to correct the 2.36~TeV data.

\begin{figure}[htbp]
\begin{center}
\includegraphics[width=0.49\textwidth]{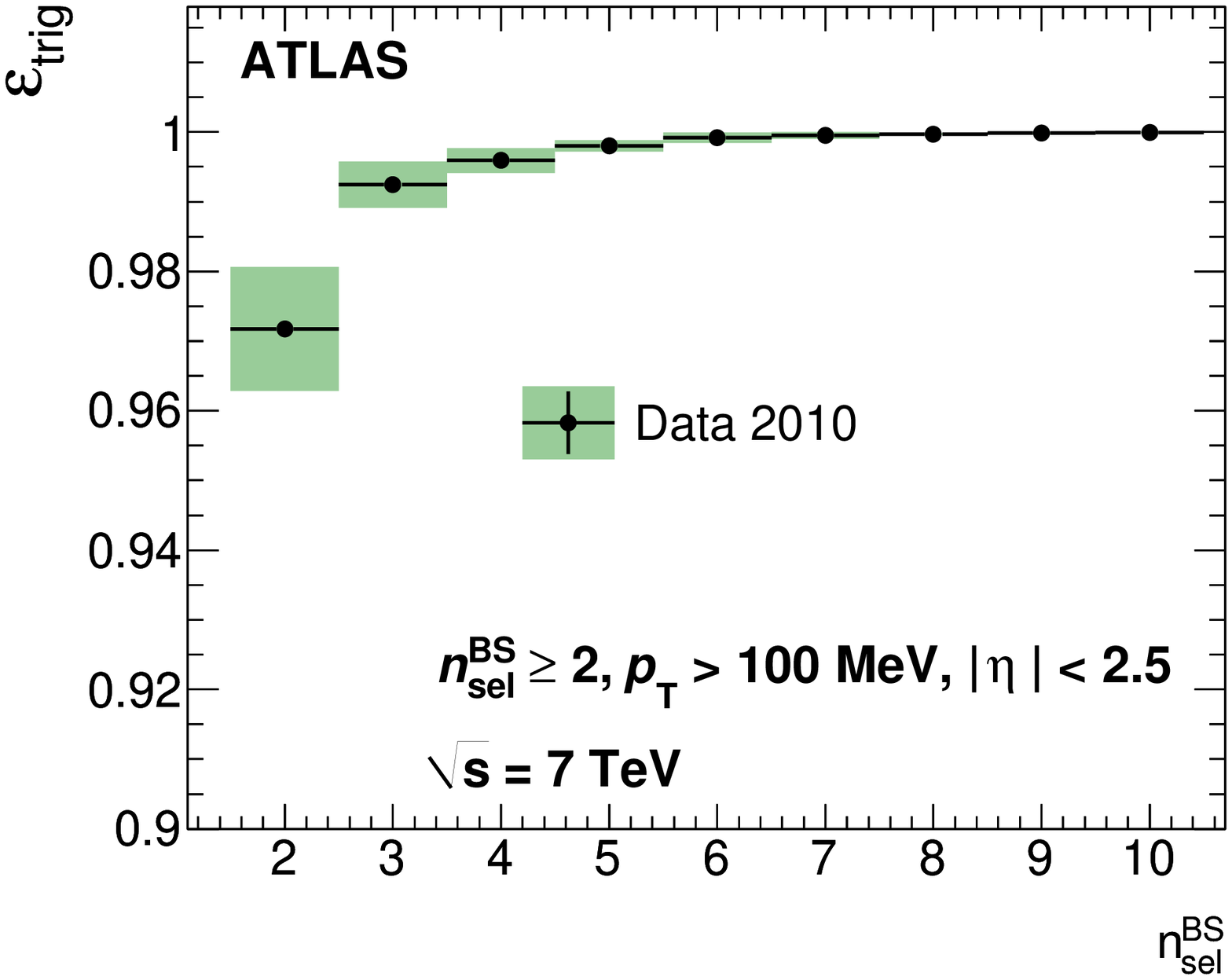}
\includegraphics[width=0.49\textwidth]{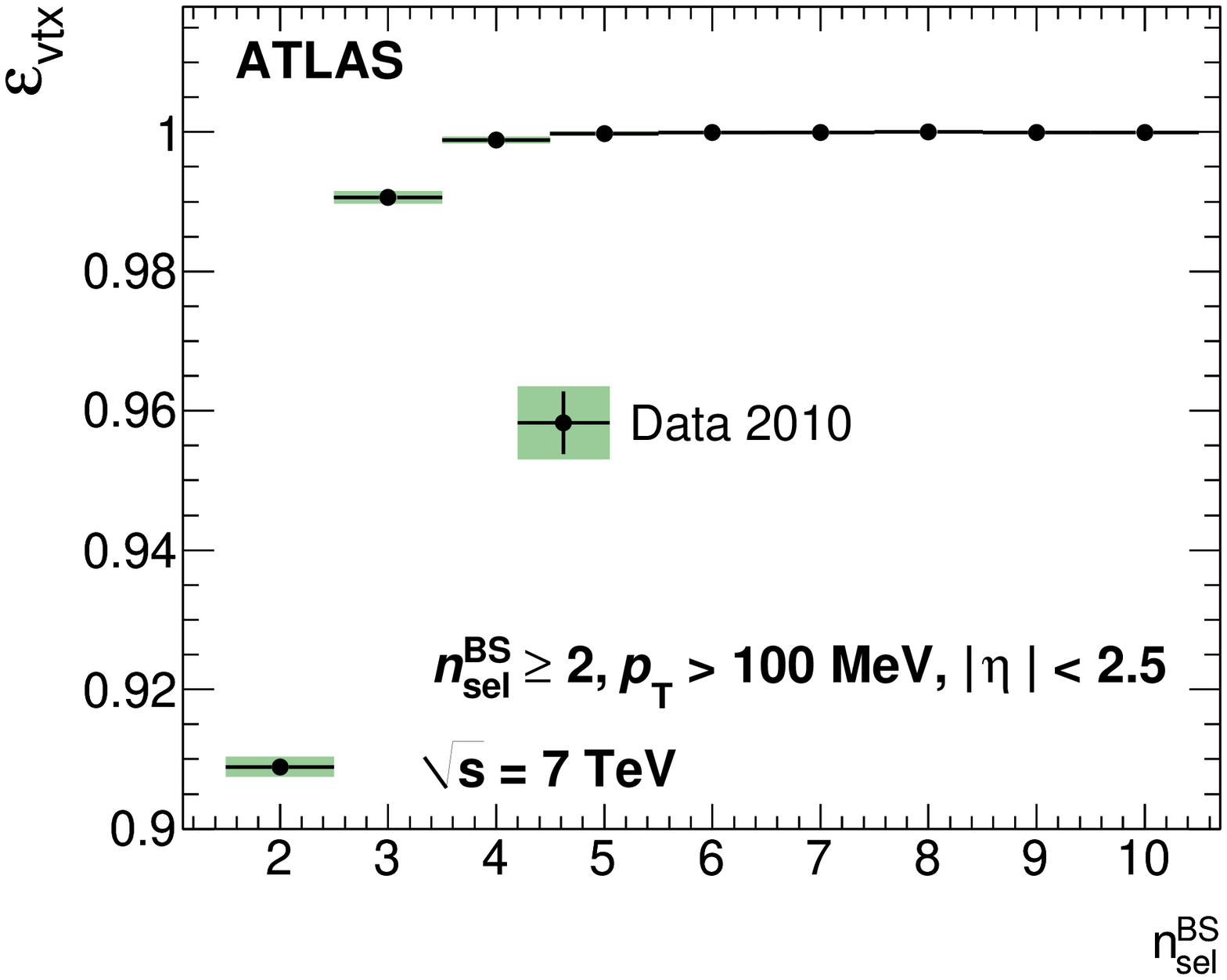}
\includegraphics[width=0.49\textwidth]{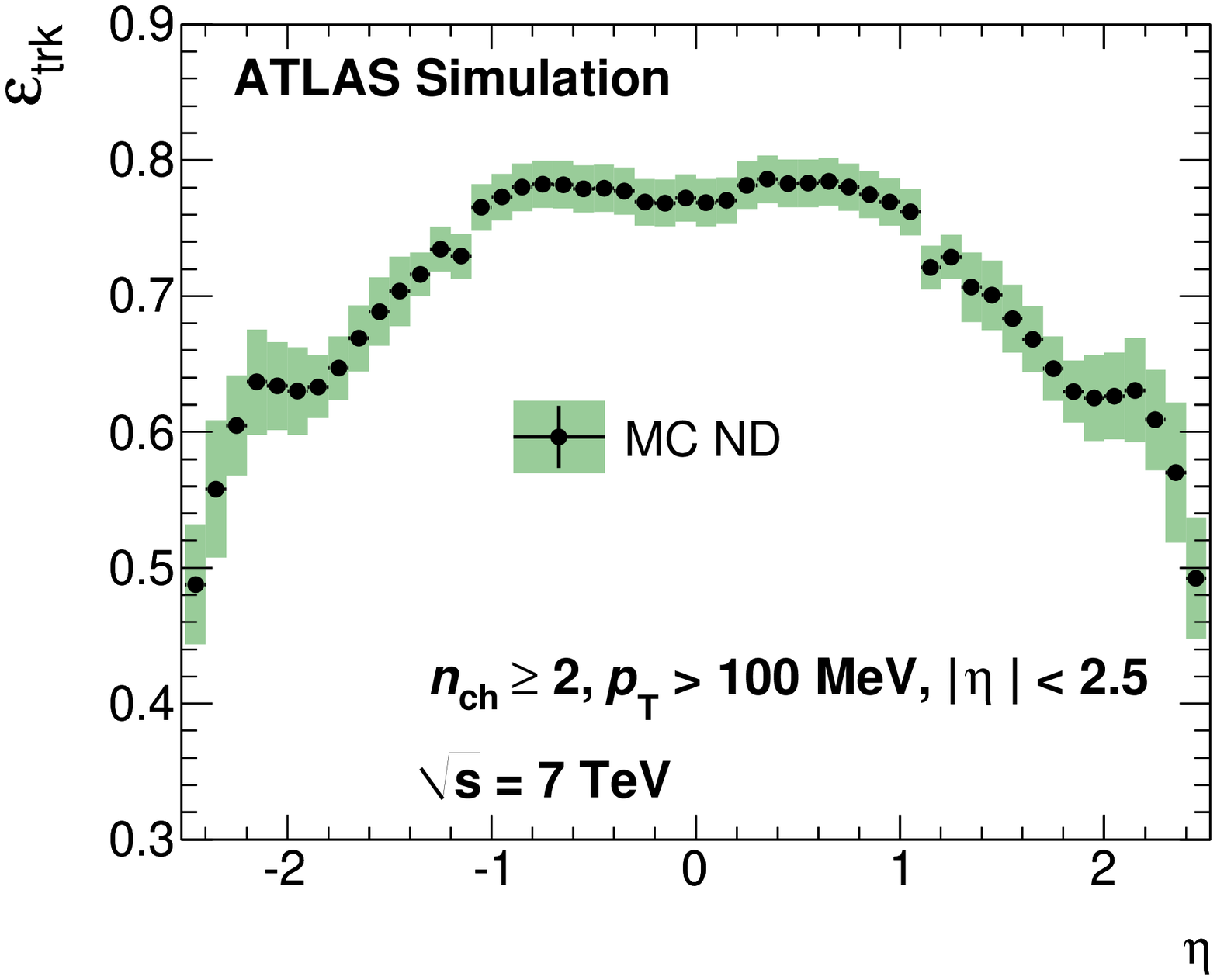}
\includegraphics[width=0.49\textwidth]{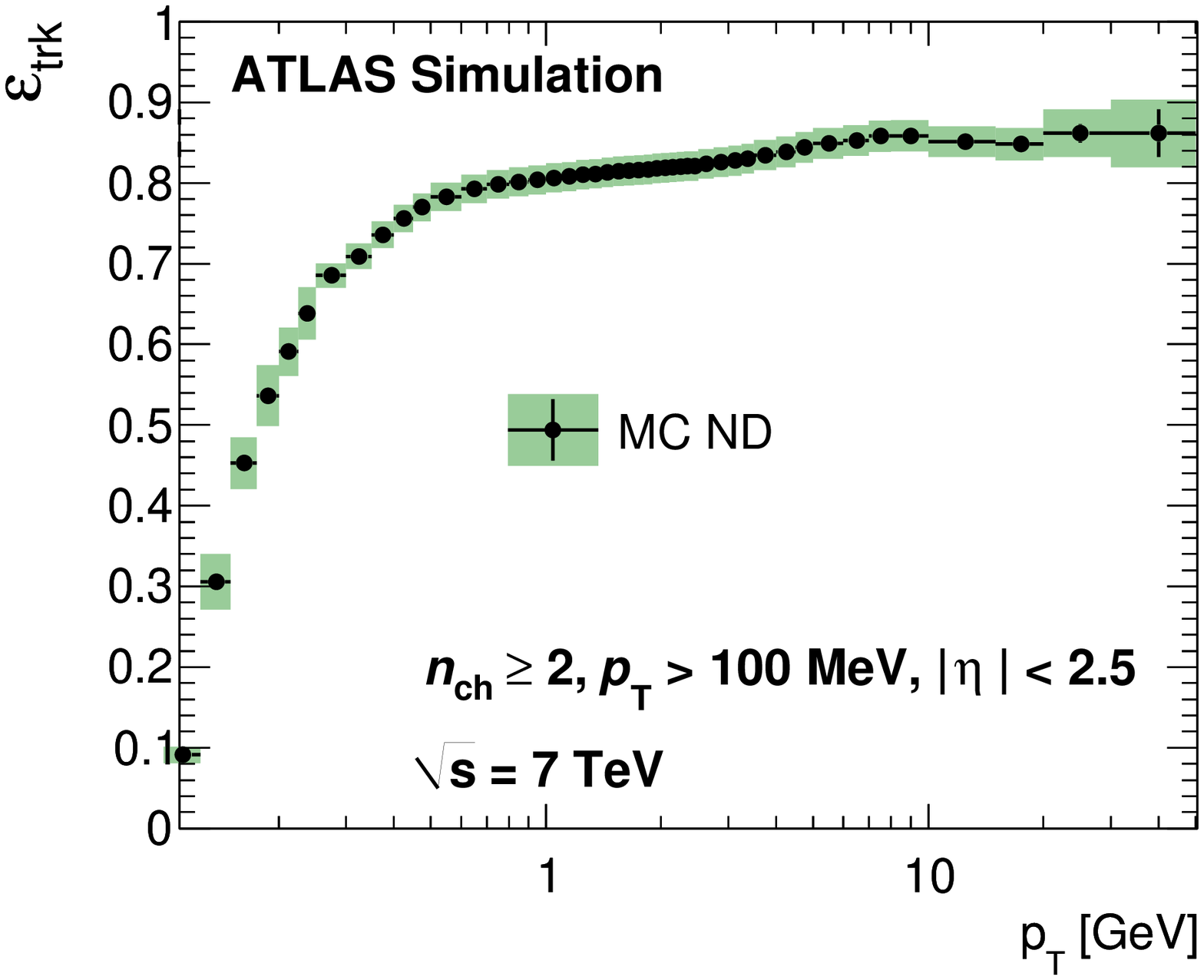}
\end{center}
 \begin{picture} (0.,0.)
    \setlength{\unitlength}{1.0cm}
    \put ( 0.8,13.2){(a)}
    \put (8.7,13.2){(b)}
    \put ( 0.8,7.1){(c)}
    \put (8.7,7.1){(d)}
 \end{picture}
 \vspace{-1cm}
\caption{Trigger efficiency (a) and vertex reconstruction efficiency (b) with respect to the event selection, 
as a function of the number of reconstructed tracks before the vertex requirement (\nselbs).
The track reconstruction efficiency as a function of $\eta$ (c) and \pta\ (d) is derived from non-diffractive (ND) MC.
The statistical errors are shown as black lines, the total errors as green shaded areas.
All distributions are shown at \sqs\ for $\nch~\geq~2$, $\pta~>~100$~MeV, $|\eta|< 2.5$. 
For the vertex and trigger efficiencies, the selection requires $\nselbs~\geq~2$.
  \label{fig:eff}}
\end{figure}

The trigger efficiency is parametrised as a function of \nselbs ; it is 97\% (99\%) in the first \nselbs\ bin and rapidly increases to nearly 100\% for $\nselbs \geq 2$, $\pta~>~100$~MeV ($\nselbs \geq 1$, $\pta~>~500$~MeV).
The trigger requirement is found to introduce no observable bias in the \pta\ and $\eta$ distributions of selected tracks within the statistical uncertainties of the the data recorded with the control trigger.
The resulting trigger efficiency is shown in  Fig.~\ref{fig:eff}a for the phase-space region with $\nselbs \geq 2$, $\pta~>~100$~MeV at \sqs.

\paragraph{Systematic Uncertainties}
Since there is no vertex requirement in the data sample used to measure the trigger efficiency, it is not possible to make the same impact-parameter selection as is made on the final selected tracks. 
In order to study potential effects due to this, the trigger efficiency is measured after applying the impact-parameter constraints with respect to the primary vertex if available or with respect to the beam spot if not. 
The difference in the efficiency obtained this way and in the nominal way is considered as a systematic uncertainty.
This variation provides a conservative estimate of the effect of beam-induced background and non-primary tracks on the trigger efficiency at low values of \nselbs. 
The systematic uncertainty arising from possible correlation of the MBTS trigger with the control trigger is studied using simulation, and the effect of correlations on the trigger efficiency is found to be less than 0.1\%. 
The total systematic uncertainty on the trigger efficiency determination, which also includes the statistical uncertainty on the control sample, is of the order of 1\% in first \nselbs\ bin, decreasing rapidly as \nselbs\ increases.

%%%%%%%%%%%%%%%%%%%%%%%
\subsection{Vertex Reconstruction Efficiency}
 \label{sec:vertexeff}

The vertex reconstruction efficiency, \vtxeff , is determined from data by taking the ratio of triggered events with a reconstructed vertex to the total number of triggered events, after removing the expected contribution from beam background events. 
The efficiency is measured to be 90-92\% in the first \nselbs\ bin for the different energies and phase-space regions;
it rapidly rises to 100\% at higher track multiplicities.
The vertex reconstruction efficiency at \sqs\ for $\nselbs \geq 2$, $\pta~>~100$~MeV is shown in Fig.~\ref{fig:eff}b as a function of \nselbs.

The dependence of the vertex reconstruction efficiency on the $\eta$ and \pta\ of the selected tracks is studied as well as the dependence on the projection along the beam-axis of the separation between the perigees~\footnote[6]{The perigee of a track is here the point of closest approach of the track and the coordinate origin (0,0,0).}of the tracks ($\Delta z$), for events with more than one track. 
For all phase-space regions, only the dominant effect is corrected for as the other effect is always found to be significantly smaller and would thus not affect the final result.

For the lower \pta\ threshold selection, a strong dependence is observed as a function of $\Delta z$ for events with two tracks; 
this bias is corrected for in the analysis using two different parametrisations depending on the \pta\ of the lowest \pta\ track: one for tracks below 200~MeV and one for those above that threshold.
The dependence on the vertex reconstruction efficiency due to the $\eta$ of the tracks is found to be smaller than the $\Delta z$ correction and is neglected for this phase-space region.
For the 500~MeV \pta\ threshold selection, the $\eta$ dependence is corrected for events with $\nselbs = 1$.
For events with higher multiplicities the $\Delta z$ dependence is found to be very small and is neglected.

\paragraph{Systematic Uncertainties}
The difference between the vertex reconstruction efficiency measured with beam background removal and the vertex reconstruction efficiency measured without beam background removal is assigned as the systematic uncertainty on the vertex reconstruction efficiency. 
For determination of this difference, the contribution of beam-related backgrounds is estimated using non-colliding bunches, as in~\cite{MB1}. 
The highest rate of beam-related background is found in the phase-space region with $\pta~>~100$~MeV at 900~GeV, 
where it is 0.8\% without vertex selection and 0.2\% with vertex selection, 
although it is found to decrease rapidly at higher multiplicities. 
(This beam-related background contribution is larger than that given if Sec.~\ref{sec:background} 
where a reconstructed primary vertex was required.)
The total uncertainty due to the vertex reconstruction efficiency is significantly below 1\% for all phase-space regions at all energies.
Fig~\ref{fig:eff}b shows the total error for the phase-space region with  $\pta~>~100$~MeV at \sqs.

%%%%%%%%%%%%%%%%%%%%%%
\subsection{Track Reconstruction Efficiency for the 0.9 and 7~TeV Data Samples}\label{sec:trkeff}

\begin{figure}[htb!]
\begin{center}
\includegraphics[width=0.49\textwidth]{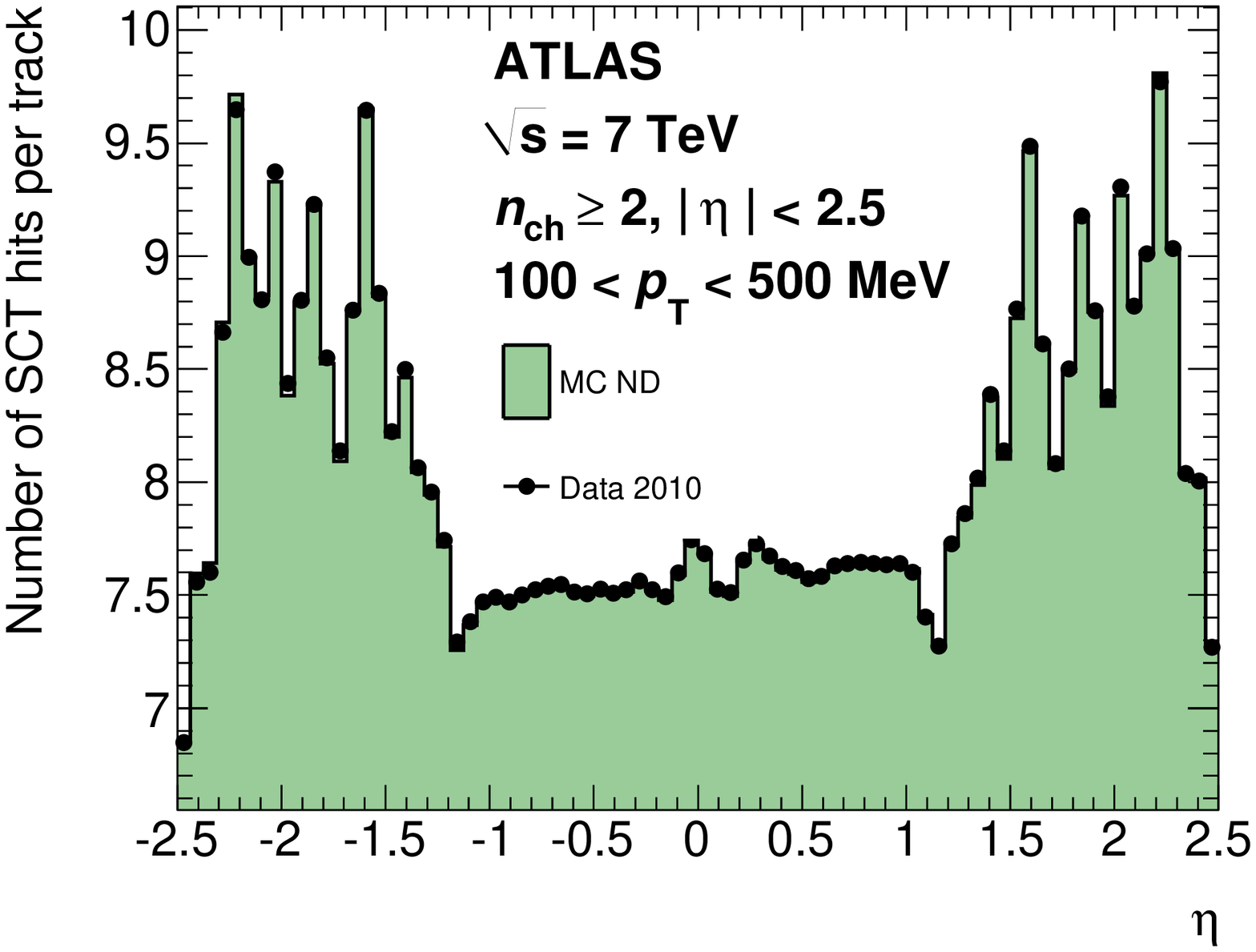}
\includegraphics[width=0.49\textwidth]{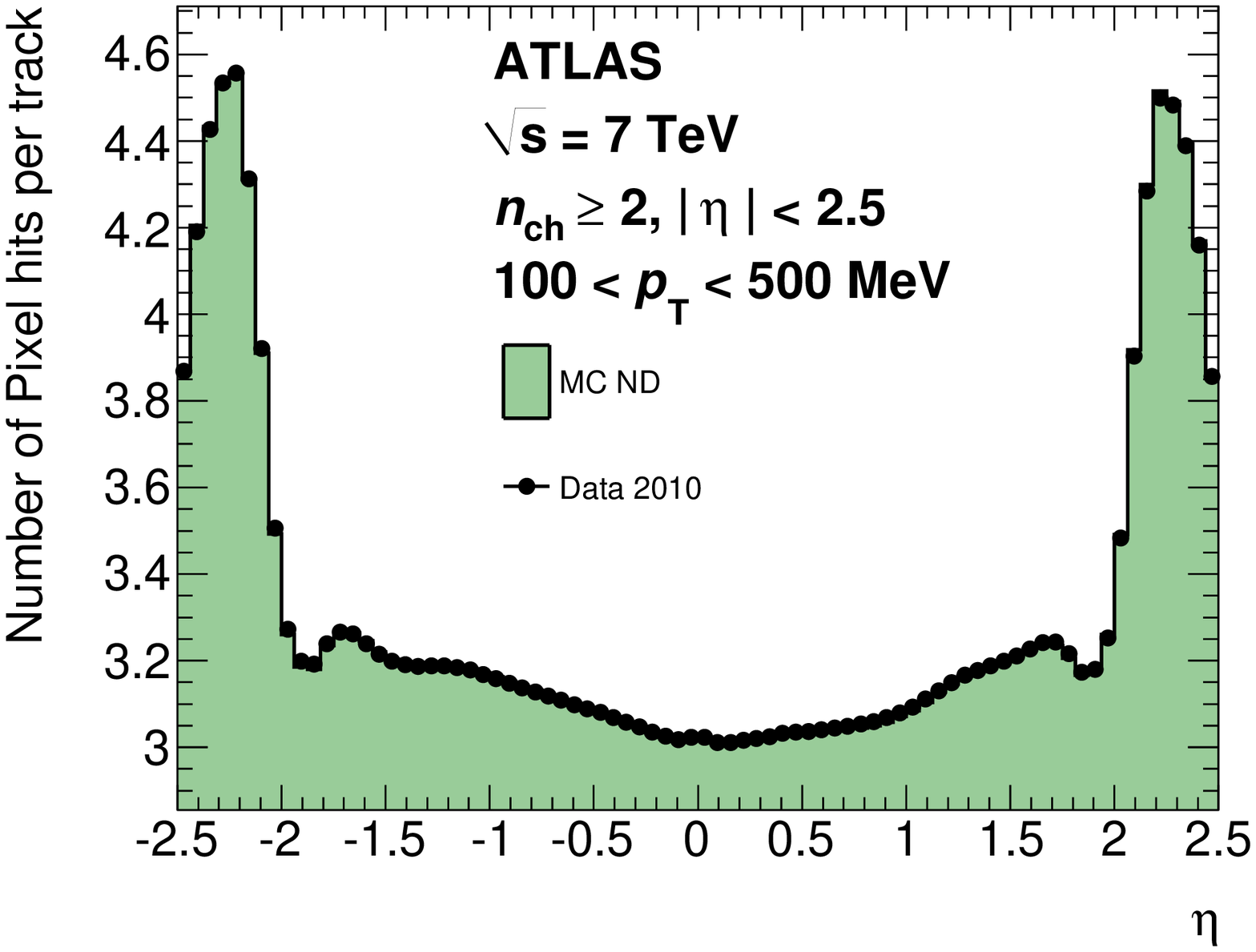}
\includegraphics[width=0.49\textwidth]{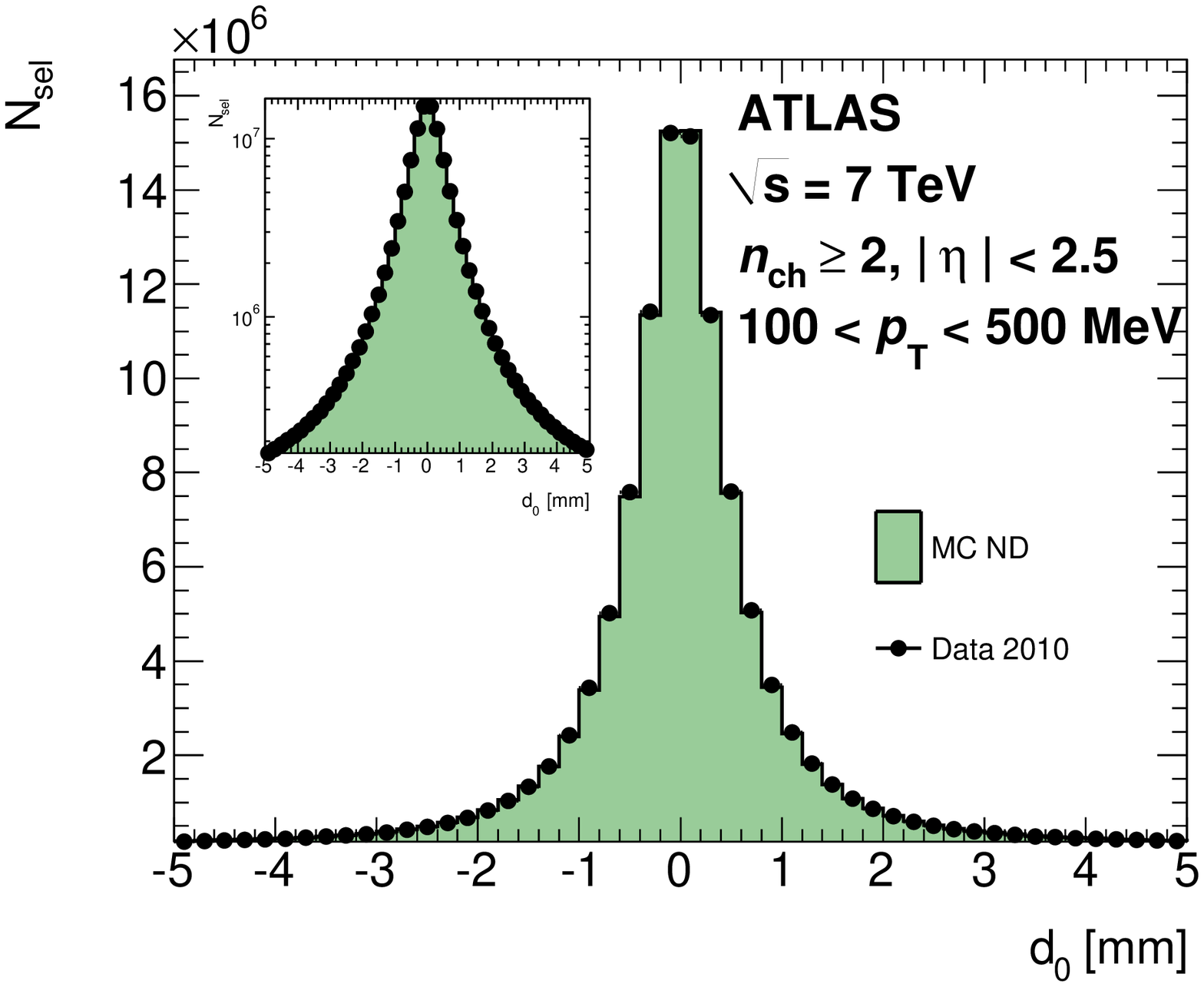}
\includegraphics[width=0.49\textwidth]{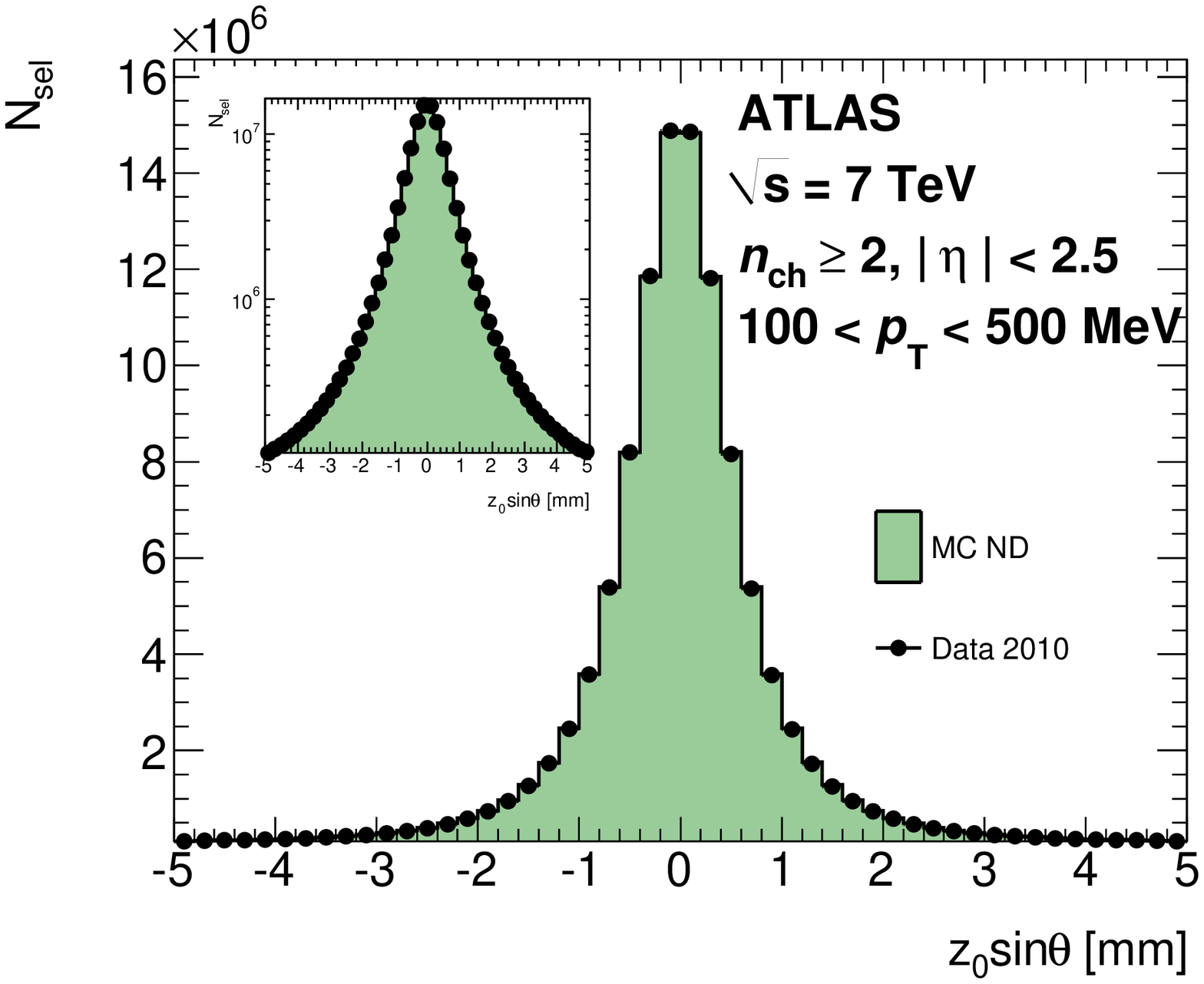}
\end{center}
 \begin{picture} (0.,0.)
    \setlength{\unitlength}{1.0cm}
    \put ( 0.8,13.0){(a)}
    \put (8.7,13.0){(b)}
    \put ( 0.8,6.8){(c)}
    \put (8.7,6.8){(d)}
 \end{picture}
 \vspace{-1cm}

\caption{Comparison between data and simulation at $\sqrt{s}=7$~TeV for tracks with transverse momentum between 100 and 500~MeV: 
the average number of silicon hits on reconstructed track as a function of $\eta$ in the  SCT (a) and Pixel (b) detectors, 
the transverse impact parameter (c) and longitudinal impact parameter multiplied by $\sin{\theta}$ (d). 
The inserts for the impact parameter plots show the log-scale plots. The \pta\ distribution of the tracks in non-diffractive (ND) MC is re-weighted to match the data and the number of events is scaled to the data.\label{fig:trk_val}}
\end{figure}

The track reconstruction efficiency, \trkeff , determined from MC, is parametrised in bins of \pta\ and $\eta$.
The excellent agreement between data and MC of basic track quantities for tracks above 500~MeV was previously demonstrated~\cite{MB1}. 
Figure~\ref{fig:trk_val} highlights the agreement for tracks in the additional range covered in this paper, $100 < \pta < 500$~MeV.

The track reconstruction efficiency is defined as:
$$
\trkeff (p_\mathrm{T},\eta) \ = \ \frac{N^\mathrm{matched}_\mathrm{rec}(p_\mathrm{T},\eta)}
                                   {N_\mathrm{gen}(p_\mathrm{T},\eta)},
$$
where  \pta\ and $\eta$ are generated particle properties,
 $N^\mathrm{matched}_\mathrm{rec}(p_\mathrm{T},\eta)$ is the number of reconstructed tracks matched to a generated charged particle  
and $N_\mathrm{gen}(p_\mathrm{T},\eta)$ is the number of generated charged particles in that bin. 
The matching between a generated particle and a reconstructed track uses a cone-matching algorithm in the $\eta$--$\phi$ plane, associating the particle to the track with the smallest $\Delta R \ = \ \sqrt {(\Delta\phi)^2+(\Delta \eta )^2}$
within a cone
of radius 0.15. 
In addition, the particle trajectory must be compatible with the position of one of the pixel hits of the track.
The larger cone size than in~\cite{MB1} is needed to account for the degraded resolution at lower track \pta.

The resulting reconstruction efficiency as a function of $\eta$ integrated over \pta\ is
shown in Fig.~\ref{fig:eff}c at \sqs\ for the phase-space region with the lowest \pta\ threshold. 
The track reconstruction efficiency is lower in the region $|\eta| > 1$ due to particles passing through more material in that region.
Figure~\ref{fig:eff}d shows the efficiency as a function of \pta\ integrated over $\eta$.
The initial rise with \pta\ is due to the requirement on the minimum number of silicon hits required in the analysis, 
which indirectly constrains the tracks to pass through a minimum number of detector layers and thus have a minimum \pta.

%-------------------------------------------------------------------
\paragraph{Systematic Uncertainties}\label{sec:tracking_sys}
As the track reconstruction efficiency is determined from MC, the main systematic uncertainties result from the level of agreement between data and MC. 
The overwhelming majority of particles in the selected events are hadrons. 
These are known to suffer from hadronic interactions with the material in the detector. 
Thus a good description of the material in the detector is needed to get a good description of the track reconstruction efficiency.
To quantify the influence of an imperfect description of the detector description, in particular the material in the simulation, two different data-driven methods are used.
The first reconstructs the invariant mass of $K_s^0$ mesons decaying to two charged pions; 
the second compares the track lengths in data and simulation. 
The $K_s^0$ mass method studies the mass as a function of the decay radius of the meson; it has greatest sensitivity to small radii, while the track length study probes the material description in the simulation in terms of nuclear interaction length ($\lambda$) in the SCT detector. 
The combination of both methods provides good sensitivity throughout the silicon detectors.
They allow us to constrain the material to better than 10\% in the central barrel region and better than 30\% at the highest $|\eta|$ measured.
The material uncertainty is the largest uncertainty in almost all regions of all distributions plotted in this paper.
In the barrel region, the total uncertainty due to the material is 8\% at low \pta , going down to 2\% above 500~MeV.
The uncertainty increases with increasing $|\eta|$; 
the largest uncertainties are in the region $2.3 < |\eta | < 2.5$: 15\% in the first \pta\ bin decreasing to 7\% above 500 MeV.

The track-fit $\chi^2$ probability cut has been found to offer powerful discrimination against tracks with mis-measured momenta. 
These are mostly very low momentum particles that are reconstructed with much higher momentum due to mis-alignment or nuclear interactions~\footnote[7]{
Note that the momentum spectrum falls by many orders of magnitude in the measured range.
}.
Mis-measured tracks are seen predominantly at the edges of the $\eta$ acceptance where
the distance between consecutive measurement points of the outer layer of the Pixel and the first layer of
 the SCT can reach up to  $\sim$~1~m.
The fraction of mis-measured tracks is observed to be significantly more in data than in Monte Carlo even after this cut is applied. 
Two different methods are used to estimate the fraction of mis-measured tracks in data. 
The first compares the momentum obtained from the tracks reconstructed using only the SCT hit information with that obtained for fully reconstructed tracks. 
After normalising the number of well-measured tracks in MC to data,
the scaling of the MC high-\pta\ tails needed to model the data is obtained.
The second method uses the difference between data and MC seen in the tails of the \dzero\ distributions at high \pta\ because mis-measured tracks tend to have poorly reconstructed \dzero. Again a scaling factor is obtained to scale the MC tails in order to describe the data.
These two methods give very similar results.
Both methods are used to obtain the systematic uncertainty for all but the outer-most regions in $\eta$ where the effect is the most significant. 
In this region an additional method is used that compares the $\eta$ distributions, normalised in the central region, in bins of \pta. 
The variation with \pta\ of the $\eta$ distribution due to physics is small compared to the differences observed due to mis-measured tracks. 
The additional tracks at high $|\eta|$, high \pta\ are considered to be due to mis-measured tracks and the fraction of mis-measured tracks in data is obtained.
This third method gives the systematic uncertainty for the outer-most $\eta$ bins.
Averaged over the whole $\eta$ region, the fraction of mis-measured tracks in data is found to be negligible for $\pta < 10$~GeV, 3\% for $10 < \pta <15$~GeV and increases to 30\% for $30 < \pta < 50$~GeV.
An additional systematic on the track reconstruction efficiency of 10\% is taken for all tracks with $\pta~>~10$~GeV due to different efficiencies of the $\chi^2$ probability cut in data and MC.
All systematic uncertainties on the mis-measured high-\pta\ tracks are taken as single-sided errors.

Studies using $Z \rightarrow \mu \mu$ events show that the resolution in data is about 10\% worse than the nominal MC resolution above 10~GeV. The impact of a 10\% Gaussian smearing of the reconstructed track \pta\ in MC is performed and found to have a 7\% effect for the binning used in this paper. This effect is taken as a systematic uncertainty on tracks above 10~GeV. 
This systematic uncertainty is single-sided and added linearly with the systematic uncertainty due to the mis-measured high-\pta\ tracks. 
The effect on tracks below 10~GeV is found to be negligible.

The \pta\ cut applied at various stages of the pattern recognition inside the track reconstruction algorithm introduces an inefficiency due to the momentum resolution. 
A different momentum resolution or a bias in the momentum estimation in data compared to MC can result in a change in the migration out of the first bin in \pta\ ($100<\pta <150$~ MeV) and thus a gain or loss of observed tracks. 
The default migration correction is derived using the resolution in Monte Carlo.
The track \pta\ resolution at the seed finding stage in Monte Carlo is increased by a very conservative 10~MeV, making the \pta\ resolution effectively 15~MeV instead of 10~MeV. The effect of this shift on the track reconstruction efficiency in the first \pta\ bin is found to be about 5\%; this difference is assigned as a systematic uncertainty.

A detailed comparison of track properties in data and simulation is performed by varying the track selection criteria. 
The largest deviations between data and MC are observed at high $\eta$ and are found to be $\sim 1\%$.
For simplicity, a constant 1\% uncertainty is assigned over the whole range.

A summary of the track reconstruction systematic uncertainties is shown in Table \ref{tab:trk_sys}.
The total uncertainty due to the track reconstruction efficiency determination is obtained by adding all effects in quadrature except for tracks above 10~GeV where the resolution and mis-measured track effects are added linearly; asymmetric errors are considered for these effects.

  \begin{table}
	\centering
\begin{tabular}{ | c | c | c  |}
\hline\hline
{Systematic Uncertainty} & {Size} & {Region}  \\

\hline
Material & $\pm2-15\%$  & decreases with \pta, increases with $|\eta|$ \\
\hline
$\chi^2$ prob. cut & $\pm 10\%$ &  flat, only for $\pta~>~10$~GeV \\
\hline
 \multirow{3}{*}{Resolution} & $\pm 5\%$ & $ 100 < \pta < 150$~MeV \\
                     & negligible &  $0.15 < \pta < 10$~GeV \\
                     & $-7\%$ & $\pta~>~10$~GeV \\                 
\hline
Track Selection& $\pm1$\%  & flat in \pta\ and $\eta$ \\
\hline
Truth Matching & $\pm 1\%$ & only for \sqt\ Pixel Tracks \\
\hline
Efficiency correction factor & $\pm 4\%$ & only for \sqt\ ID Track \\
\hline
 \multirow{2}{*}{Alignment and other high \pta\ } &  \multirow{2}{*}{-3\% to -30\%} &   only for $\pta~>~10$~GeV \\
  & &   averaged over $\eta$ , increases with increasing \pta\ \\
\hline
\hline
\end{tabular}
\caption{The systematic uncertainties on the track reconstruction efficiency for \sqn,  \sqs\ and \sqt\ Pixel Track and ID Track methods.  
Unless otherwise stated, the systematic is similar for all energies and phase-space regions.
All uncertainties are quoted relative to the track reconstruction efficiency.
\label{tab:trk_sys}}
\end{table}

%%%%%%%%%%%%%%%%%%%%%%%%%%%%%%%%%%%
\subsection{Track-Reconstruction Efficiency for the 2.36~TeV Data Sample}\label{sec:track_eff_236}

Both the Pixel track and the ID track methods apply a data-driven correction to the primary track reconstruction efficiency, $\varepsilon_{\textrm{MC}}$
\begin{equation}
	\varepsilon (x) \ = \varepsilon_{\textrm{MC}} (x) \cdot \varepsilon_{\textrm{corr}} (\eta) ,
\end{equation}
where $\varepsilon_{\textrm{MC}}$ is derived from nominal simulation at \sqt.
Here $x$ is either both \pta\ and $\eta$ for the ID track or only $\eta$ for the Pixel track method, as those are the parameters that the correction factors were found to depend on.

The correction, $\varepsilon_{\textrm{corr}}$, is derived from the reference dataset taken at \sqn\ where the high voltage on the SCT was lowered for part of the run. 

For the Pixel track method, $\varepsilon_{\textrm{corr}}$ is the ratio of the relative Pixel track reconstruction efficiency, $\varepsilon_{\textrm{rel}}$, in data to simulation. The relative Pixel track efficiency is the efficiency to reconstruct a Pixel track if a track has been reconstructed using hits in the SCT and TRT detectors only. 
\begin{equation}
\varepsilon_{\textrm{corr}} (\eta) = \frac{\varepsilon^{\textrm{Data}}_{\textrm{rel}} (\eta)}{\varepsilon^{\textrm{MC}}_{\textrm{rel}} (\eta)}
\label{erelDef}
\end{equation}

\begin{figure}[htb!]
\begin{center}
\subfigure[]{\includegraphics[width=0.49\textwidth]{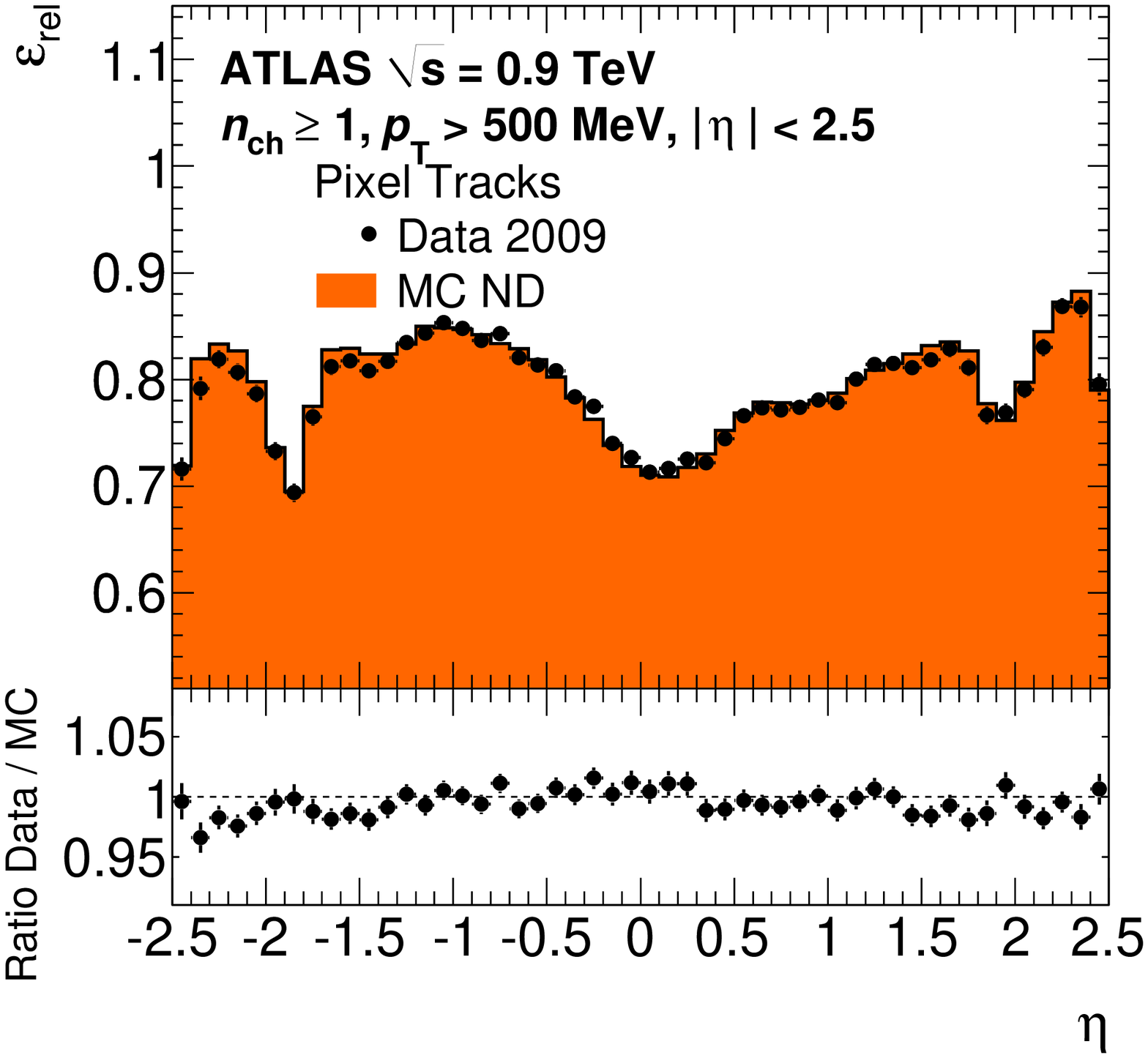}}
\subfigure[]{\includegraphics[width=0.49\textwidth]{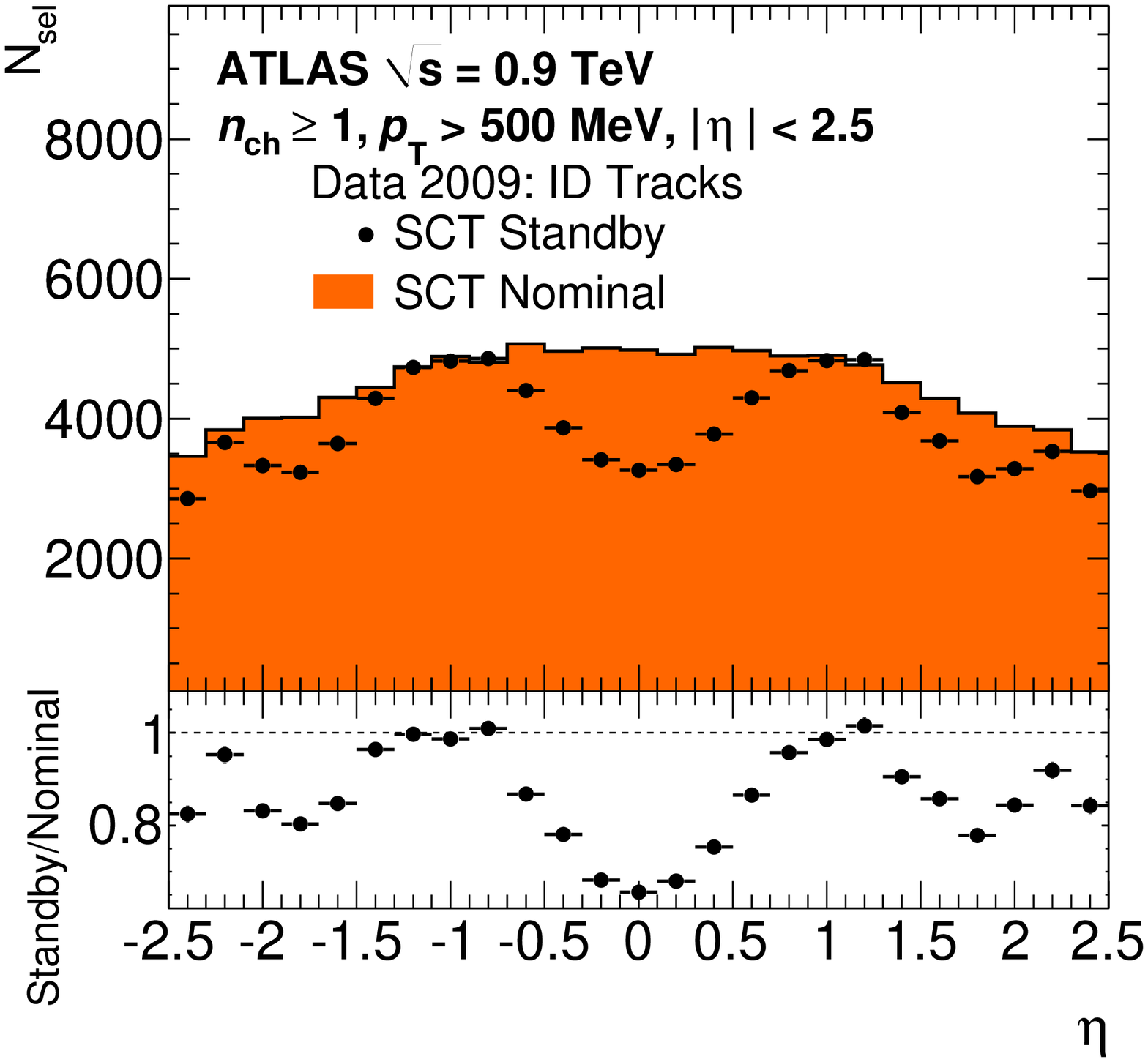}}
\caption{
Relative efficiency of Pixel tracks in
data and non-diffractive (ND) MC simulation at \sqn\ (a).
Both Pixel track distributions are re-weighted to have the same beam spot distribution as the \sqt\ data.
The number of reconstructed ID tracks in data at \sqn\ as a function of $\eta$ with the SCT in nominal and standby (b).  
The ID track distributions are normalised to the number of events passing the trigger requirement.
\label{fig:tracksVsEta}}
\end{center}
\end{figure}

Figure~\ref{fig:tracksVsEta}a shows the relative Pixel track efficiency in data and simulation. The ratio of the two distributions, shown in the insert, is used to correct the track reconstruction efficiency for the Pixel track method at \sqt.

For the ID track method the efficiency derived from simulation with nominal conditions is corrected by $\varepsilon_{\textrm{corr}}$ to account for the lower SCT efficiency in standby mode. Figure~\ref{fig:tracksVsEta}b shows the distribution of the number of reconstructed tracks in data in both SCT configurations at \sqn\ normalised to the same number of events satisfying the trigger requirement. The ratio of the number of reconstructed tracks with the SCT in standby, $N_{\textrm{tr}}^{\textrm{sb}}$, to the number of reconstructed tracks with the SCT at nominal, $N_{\textrm{tr}}^{\textrm{nom}}$, shown in the inset, is used to correct the track reconstruction efficiency for the ID track method at \sqt:

\begin{equation}
\varepsilon_{\textrm{corr}} (\eta) = \frac{N_{\textrm{tr}}^{\textrm{sb}} (\eta)}{N_{\textrm{tr}}^{\textrm{nom}} (\eta)}.
\label{etrackDef}
\end{equation}

\paragraph{Systematic Uncertainties}

Most systematic uncertainties on the ID track reconstruction efficiency are similar to the full tracking at other energies. 
The major additional systematic uncertainty is due to the efficiency correction factor for the SCT configuration.  
The uncertainty due to the statistical limitations of the reference dataset is 2\%. An additional 3\% uncertainty accounts for the extrapolation from \sqn\ to \sqt, which was estimated by comparing the distributions of the number of ID tracks between \sqn\ and \sqt. 
The total uncertainty on the efficiency correction factor adds those two effects in quadrature to obtain a total uncertainty of 4\%.

The material uncertainty is estimated using a similar method as for the other energies; the absolute uncertainty is found to be 2\% (3\%) for the Pixel (ID) track reconstruction efficiency. 
The uncertainty is larger for ID tracks, because such tracks are sensitive to the material throughout the whole silicon detector. The uncertainty due to the momentum resolution is negligible because the phase-space cuts are sufficiently far from the track algorithm cuts.

There is an additional 1\% uncertainty on the Pixel track method due to the matching procedure.
The relative Pixel track reconstruction efficiency differs from the primary efficiency due to material effects and contributions from non-primary tracks. There is an additional discrepancy of 4\% in for $2.4 < |\eta| < 2.5$  that is assigned as a systematic uncertainty for those bins. 
At central $\eta$ the total uncertainty on the Pixel (ID) track reconstruction efficiency is estimated to be 3.4\% (6\%). 
Table~\ref{tab:trk_sys} shows the track reconstruction systematics at \sqt\ and the differences with respect to the uncertainties at other centre-of-mass energies are indicated.

%%%%%%%%%%%%%%%%%%%%%%%%%%%%%%%%%%%
\section{Correction Procedure}\label{sec:corr}

The effect of events lost due to the trigger and vertex requirements is corrected using an event-by-event weight:

$$
w_\mathrm{ev}(\nselbs) =  \frac{1}{\trigeff (\nselbs)} \cdot \frac{1}{\vtxeff (\nselbs,x)},
$$
where $x$ is either the $\Delta z$ between tracks or the $\eta$ of the tracks, as described in Sec.~\ref{sec:vertexeff}.

The \pta\ and $\eta$ distributions of selected tracks are corrected for using a track-by-track weight:
$$
w_\mathrm{trk}(\pta, \eta) =  \frac{1}{\trkeff( \pta, \eta)}\cdot(1-\fsec( \pta ))\cdot(1-\fokr(\pta , \eta)), 
$$
where \fsec\ is the fraction of non-primary tracks determined as described in Sec.~\ref{sec:background}. 

The fraction of selected tracks passing the kinematic selection for which the corresponding primary particle is outside the kinematic range, $\fokr(\pta, \eta)$, originates from resolution effects and has been estimated from MC.
The uncertainty on \fokr\ is mostly due to the resolution difference between data and MC. 
This uncertainty is negligible for all cases except at \sqt\ for the Pixel track method where the uncertainty is estimated to be 1\%, due to the poor momentum resolution of the Pixel tracks.
No additional corrections are needed for the $\eta$ distribution; 
the additional corrections needed for the other distributions are described in the following sections.

For all distributions in all phase-space regions considered, closure tests are carried out.
These are tests carried out on MC where the reconstructed samples are corrected according to the same procedure as used on the data; 
the resulting difference between the corrected distribution and the known particle level distribution is defined as the amount of non-closure; 
if the correction procedure were perfect, the non-closure would be zero. 
For this analysis, closure tests are carried out on all distributions in all phases-space regions and unless explicitly mentioned in the text the level of non-closure is less than 1\%.

%-----------------
\subsection{Correction to $\frac{\mathrm{d} \nev}{\mathrm{d} \nch}$}\label{sec:nch_unfolding}

First, the observed \nsel\ distribution is corrected for the trigger and vertex reconstruction efficiencies.
Then, an event-level correction is applied using Bayesian unfolding~\cite{D'Agostini:1994zf} to correct the observed track multiplicity to the distribution of the number of primary charged particles, as follows. 
An unfolding matrix, \Mchsel, is defined that expresses the probability that a given selected track multiplicity, after all other event-level corrections are applied, \nsel , is due to \nch\ primary particles.
This matrix is normalised such that the number of events does not change except for the rare cases where $\nsel > \nch$ and \nch\ is below our acceptance selection.
This matrix is populated from MC09 MC and 
applied to data to obtain the observed \nch\ distribution. 
The resulting distribution is then used to re-populate  
the matrix and the correction is re-applied. 
This procedure is repeated without a regularisation term and converges after four iterations in data;
convergence is defined as the first iteration in which the $\chi^2$ difference between the result of the unfolding and the input distribution for that iteration is less than the number of bins used in the unfolding.

After the \nsel\ distribution has been unfolded, the resulting charged particle multiplicity distribution is corrected for events migrating out of the selected kinematic range ($\nch~\geq~X$), which the matrix does not account for. This is achieved by adding an additional term to the correction. The correction terms for the phase-space regions with $\nch~\geq~2$ is
\begin{equation}
\label{eq:warsinsky_equation}
1/ ( 1 - (1 - \trkeff)^{\nch} - \nch \cdot \trkeff \cdot (1- \trkeff)^{(\nch-1)})  
\end{equation}
where $\trkeff$ is the mean effective track reconstruction efficiency for a given \nch\ bin.
Corresponding terms are used for the other phase-space regions.
This track reconstruction efficiency can in principle be different for each \nch\ bin,
but the difference is found to be small and thus the mean effective track reconstruction efficiency for lowest \nch\ bin is used.

\paragraph{Systematic Uncertainties}
The systematic uncertainties on the unfolding procedure are obtained by modifying the input distributions as described below, applying the unfolding procedure and comparing the output to that obtained when using the nominal input; 
the matrix and the correction factors are not modified. 

There are two sources of systematic uncertainties considered. One of them is due to the track reconstruction efficiency uncertainties while the second one accounts for the different \pta\ spectrum reconstructed in data and MC.
The first source of uncertainty is estimated by starting from the observed \nsel\ spectrum in data; 
tracks are randomly removed from the distribution according to the mean \pta\ and $\eta$ of the tracks for each value of \nsel\ and the uncertainty on the track reconstruction efficiency for those \pta\ and $\eta$ values. 
A new input distribution is obtained, put through the unfolding procedure and the difference with respect to the nominal \nch\ distribution is taken as a systematic uncertainty. 
The uncertainty is then symmetrised.
The uncertainty on \nch\ due to the uncertainty on the track reconstruction efficiency is found to be 
$\sim 3\%$ to $\sim$~25\% 
at \sqs\ in the most inclusive phase-space region, $\nch~\geq~2$, $\pta~>~100$~MeV, $|\eta|< 2.5$.

The other source of uncertainty originates from the unfolding method that is carried out in a single dimension at a time, in this case \nch. 
There is some dependency on the \pta\ spectrum of the MC sample used to populate the matrix, due to the strong dependence of the track reconstruction efficiency on \pta. 
To investigate this effect, the average track reconstruction efficiency derived using the \pta\ spectrum in data and that obtained from MC are compared.
The difference in these two mean efficiencies is then treated in the same way as the uncertainty on track reconstruction efficiency, described in the previous paragraph.
This uncertainty is taken as being asymmetric; only the contribution from a shift of the spectrum in the direction of the data is taken.
The mean value is kept as that given by the nominal \pta\ spectrum in MC.
The uncertainty varies with increasing \nch\ from $-2\%$ to $+40\%$ at \sqs\ in the most inclusive phase-space region.

The only additional systematic uncertainty due to the tuning of the track reconstruction efficiency is due to the difference between the bias introduced by the vertex correction in MC and data.
The estimation of this error is done by comparing the $\Delta$\zzero\ distribution in \nselbs=2 between data and MC.
The $\Delta$\zzero\ distribution is a very good probe of the correlation between \nsel/\nch\ and \nselbs\ as
events with high \nsel\ tend to have small $\Delta$\zzero\ values while events with \nsel$<$2 tend to have large $\Delta$\zzero.
Very good agreement is found between data and MC.
Re-weighting the $\Delta$\zzero\ distribution in MC to match data or applying the vertex correction extracted from data to the MC closure test leads to a systematic uncertainty of the order of 0.1\% for \nch=2 where this effect is most pronounced.
As this error is much smaller than other systematic uncertainties considered, it is neglected.
The systematic uncertainty due to track-track correlation in a single event is small and is neglected everywhere in this analysis.

%-----------------
\subsection{Corrections to \nev}
The total number of events, \nev, used to normalise the final distributions, is defined as the integral of the \nch\ distributions, after all corrections are applied.

\paragraph{Systematic Uncertainties}
The systematic uncertainties on \nev\ are obtained in the same way as for the \nch\ distributions.
Only those systematics affecting the events entering or leaving the phase-space region have an impact on \nev.
The total uncertainty on \nev\ at \sqs\ for the most inclusive phase-space region is 0.3\%, due mostly to the track reconstruction efficiency.
At \sqt\ the total uncertainty on \nev\ is 1.4\% for the Pixel track and 2.6\% for the ID track methods.

%-----------------
\subsection{Corrections to  $\frac{1}{ p_\mathrm{T}} \cdot \frac{\mathrm{d} \Nch}{ \mathrm{d} p_\mathrm{T}}$}
The tracks are first corrected for the event level inefficiencies of the trigger and the vertex reconstruction.
Then the tracks are corrected for the track reconstruction inefficiencies, non-primary track contamination and out of kinematic range factors.
Finally,
a similar unfolding method to that used on the \nch\ distribution is used to correct the measured track \pta\ to the primary particle momentum. 
More bins are used for the unfolding than are shown in the final distributions; 
this is necessary in order to avoid amplification of 
small data MC differences with successive iterations, causing large fluctuations.
For this distribution four iterations are required before convergence is reached;
convergence is defined as for the \nch\ distribution.

\paragraph{Systematic Uncertainties}
In order to estimate the effect on the final \pta\ distributions of the uncertainties affecting the correction steps prior to the unfolding, 
the unfolding procedure is re-run on the corrected \pta\ distribution shifting the distribution used as input to the unfolding procedure by the systematic uncertainties.
This new \pta\ distribution is put through the unfolding procedure and the difference with respect to the nominal corrected \pta\ spectrum is taken as a systematic uncertainty.

The high-\pta\ systematic uncertainties are obtained using the MC samples.
The systematic uncertainty associated to the mis-measured high-\pta\ tracks is obtained by scaling the number of mis-measured tracks in MC 
to match those found in data.
This new input distribution is put through the unfolding procedure and the final difference with respect to the nominal MC is taken as a systematic uncertainty.
The systematic uncertainty associated to the resolution is obtained by smearing the well-measured tracks, in MC, by the resolution uncertainty obtained in Sec.~\ref{sec:tracking_sys}. The effect on the final unfolded distribution is taken as a systematic uncertainty.
Those two high-\pta\ systematics are added linearly. Both cause only single-sided variations.
This combined uncertainty is measured to be from -10\% for \pta\ = 10 GeV to -30\% for the last \pta\ bin ($30~<~\pta~<~50
$~GeV) at \sqs\  for the $\nch~\geq~2$, $\pta~>~100$~MeV phase-space region.
The variations for other phase-space regions at this energy are similar. 
At \sqn\ this uncertainty is found to be -20\% for all three bins above \pta\ of 10~GeV.

In order to assess the stability of the results under varying starting hypotheses for the MC spectrum used to fill the matrix, a flat initial prior is used as an input. While convergence is only typically reached after seven iterations, instead of three for the nominal prior, the final difference in the unfolded spectra is small. 
The difference between the resulting distribution obtained with a flat prior and that obtained with the MC \pta\ spectrum as a prior is taken as a systematic uncertainty.
At \sqs\ this uncertainty is less than 2\% for nearly all pT bins, with the exception of a couple of bins around changes in bin width, where the effect is 3-5\%.
At \sqn , due to more limited statistics in the MC, the largest change seen is 7\% with a few others around 3-4\%.

%----------------------
\subsection{Mean \pta\ versus \nch}

The correction procedure for the \meanpt\ vs. \nch\ distribution is designed to correct separately two components: 
$\sum_i \pta(i)$ vs. \nch\ and 
$\sum_i 1$ vs. \nch\ and take the ratio only after all corrections are applied. 
The sum is over all tracks and all events; 
the first sum is the total \pta\ of all tracks in that bin in \nch;
the second sum represents the total number of tracks in that bin.
The sums will be referred to as the numerator and denominator, respectively.
Each of these distributions, $\sum_i \pta(i)$ and $\sum_i 1$, is corrected in two steps.

First the two distributions as a function of \nsel\ are corrected on a track-by-track basis by applying the appropriate track weights; 
this track-by-track correction is applied to the data distribution and thus no longer relies on the \pta\ spectrum of the MC.
Second, the matrix obtained after the final iteration of the \nch\ unfolding described in Sec.~\ref{sec:nch_unfolding} is applied to each of the distributions to unfold \nsel\ to \nch.
Finally, the ratio of the two distributions is taken to obtain the corrected \meanpt\ vs. \nch\ distribution.
For this distribution we exclude tracks with $\pta~>~\sqrt{s}/2$ as they are clearly un-physical; 
this removes 1 track at \sqn\ and 1 track at \sqs.

This unfolding procedure assumes that
the tracking efficiency depends only on 
\pta\ and $\eta$ and is independent of the track particle multiplicity, 
and that the \pta\ spectrum of the tracks in events that migrate back from a given \nsel\ bin to a given \nch\ bin
is the same as the \pta\ spectrum of tracks in events in the corresponding \nsel\ bin. 
The fact that these assumptions are not completely valid is taken as a systematic uncertainty.
This uncertainty is obtained by looking at the non-closure of the corrected distribution in the MC. 
This residual non-closure is, we believe, a consequence of the two main assumptions.
A full parametrisation of the track reconstruction efficiency in terms of \pta, $\eta$ and \nch\ would remove the need for the first assumption,
while a  full two-dimensional unfolding as a single step where the two dimensions were \pta\ and \nch\ would remove the need for the second. 
Both of these are beyond the scope of the current paper.
In order to understand if the amount of non-closure is a realistic estimate of the uncertainty on the method  when applied to data, 
in particular to investigate its dependence on the \pta\ spectrum, the whole unfolding procedure is carried out using \py6 DW tune samples and the \py8 samples;
we varied both the input distribution and the matrix used to do the unfolding.
The level of non-closure is found to be similar to that obtained with the MC09 \py6 samples. 
We thus conclude that the level of non-closure is not strongly dependent on the \pta\ spectrum. 
This allows us to use the residual non-closure as a systematic uncertainty on the unfolding method as described in the next section.

\paragraph{Systematic and Statistical Uncertainties}
For the calculation of the statistical uncertainty, the full correlation between the tracks inside the same event was not computed. The statistical uncertainty in the numerator and denominator are computed separately then added in quadrature after taking the ratio. This is found to be a conservative estimate of the uncertainty.

Systematic uncertainties considered for the \meanpt\ vs. \nch\ distribution are either due to assumptions made during the correction 
procedure or to uncertainties on quantities taken from the MC and used during the correction procedure.

The first category refers to the assumptions on the method, the effects of which are visible in the closure test.
To account for these imperfections, we apply a systematic uncertainty of 2\%, which covers the non-closure in MC, 
except for the highest \nch\ bin and the first few \nch\ bins in some of the phase-space regions. 
For these cases a larger systematic uncertainty is applied to cover the non-closure. 
For the analyses with $\pta>500$~MeV, where the size of a non-closure
is larger, a 3\% systematic error is applied in the \nch =1 bin. 
This systematic uncertainty also covers the difference in the 
non-closure between samples created using MC09 (default) and those with DW tune of \py6 and \py8.
In the correction procedure we use the approximation that \nsel = \nselbs .
The effect of such an approximation is studied on simulation and found to be negligible with respect 
to the other sources of uncertainty. 

The second category comprises uncertainties on the track correction weights $w_\mathrm{ev}(\nselbs)$ and $w_\mathrm{trk}( \pta, \eta)$ 
and on the migration probabilities obtained from the unfolding matrix.
The dominant systematic uncertainties that affects both the track corrections weights and the migration probabilities 
are the same as those affecting the \nch\ distribution unfolding:
the uncertainty on the track reconstruction efficiency
and the effect of the difference in the \pta\ spectra between data and MC. 
These uncertainties are propagated by varying the input distribution for both the $\sum_i \pta(i)$ vs. \nsel\ and $\sum_i 1$ vs. \nsel . 

Smaller effects are also studied, for example the uncertainty on the rate of non-primary tracks and 
the effect of the systematic uncertainties affecting the high-\pta\ tracks mentioned in Sec.~\ref{sec:tracking_sys}.
Excluding the systematic uncertainties due to the assumptions made during the correction procedure, 
the systematic uncertainties are between 0.5\% and 2\% for all bins in \nch, all energies and all phase-space regions.

 %-------------------
\subsection{Correction for Different Minimum \nch\ Requirements}\label{sec:nch_6}
The only difference in the correction procedure from track to particle level for $\nch~\geq~6$ with respect to $\nch~\geq~1$
is the need for an additional correction that takes into account the effect on the tracks due to the tighter cut on both the number of tracks and number of particles.

The \nch\ distribution and the number of events \nev\ are obtained by correcting and unfolding the multiplicity distribution of the whole spectrum and then applying the higher \nch\ cut on the final distribution. 
For the \pta\ and $\eta$ track distributions an extra correction is needed.
For events with $\nsel \geq 6$, the tracks are added to the distribution as for all other phase-space regions; a weight corresponding to the product of the track ($w_{trk}$) and event weights ($w_{ev}$) is applied.
For events with $\nsel < 6$ the tracks are added to the distribution with an additional weighting factor, $w_{\nch < 6}$
that represents the probability that a track from an event with \nsel\ tracks is from an event with $\nch~\geq~6$.
This additional weight is taken from the final \nch\ unfolding matrix, after the final iteration;
each column in the matrix represents the probability that an event with  \nsel\ tracks has \nch\ particles.
The total probability ($p(\nch~\geq~6~|~\nsel) $) for a given $\nsel~<~6$ is therfor the sum over the matrix elements for $\nch~\geq~6$

\begin{equation*}
w_{\nch < 6} = p(\nch~\geq~6 \, | \, \nsel) = \sum_{\nch~\geq~6} M_\mathrm{\nch,\nsel},
\end{equation*}
where $M_\mathrm{\nch,\nsel}$ is the entry in the unfolding matrix for \nch\ and \nsel.
This weight is about 65\% for $\nsel = 5$ and rapidly drops to 1\% for $\nsel =2$.

 \paragraph{Systematic Uncertainties}

All uncertainties related to the distributions with the lower \nch\ cut are taken into account.
In addition, an extra systematic uncertainty due to the uncertainty on the track reconstruction efficiency is needed for the correction to higher \nch\ selection. 
By varying the track reconstruction efficiency down by its uncertainty, different $w_{\nch < 6}$ weights are obtained. 
The shift in the resulting \nch\ distribution is symmetrised and taken as an additional systematic uncertainty.

 %-------------------
\subsection{Extrapolation to $\pta = 0$}\label{sec:pt_extrap}
Comparing the results in our well-defined phase-space regions to other inclusive measurements from other experiments requires additional model-dependent corrections. 
One such correction is described here, but applied only for comparative purposes. 
This particular correction is derived to extrapolate the average multiplicity in the phase-space region with the lowest measured \pta\ to the multiplicity for all $\pta > 0$. 
No attempt is made to correct for the $\nch \geq 2$ requirement. 
Results are quoted for the average multiplicity in the rapidity interval $|\eta| < 2.5$ and are not considered to be the main results of this paper.
This correction is obtained using three independent methods: fitting the \pta\ spectrum to a given functional form, assuming a flat distribution at low \pta\ in the observed fully corrected $\frac{1}{ p_\mathrm{T}} \cdot \frac{\mathrm{d} \Nch}{ \mathrm{d} p_\mathrm{T}}$ distribution and obtaining the correction factor from the AMBT1 \py6 MC.

In the first method, the corrected \pta\ spectrum is fit with a two-component
Tsallis~\cite{tsallis, tsallis2} distribution 
\begin{eqnarray*}
f(\pta) = \frac{1}{2 \pi \eta'} \sum_{i=\pi,p} 
& \frac{d \Nch}{d y} \biggm|_{y=0,i} 
 \frac{(n_i-1)(n_i-2)}{(n_i T_i +m_{0,i}(n_i -1))(n_i T_ i +m_{0,i})}  \\
& \cdot 
 \left[ \frac{n_i T_i +m_T (\pta)_ i}{n_ i T_ i +m_{0,i}} \right]^{-n_ i } 
 \tanh^{-1}\left(\frac{\pta \sinh \eta'}{\sqrt{m_{0,i}^{2} + \pta^{2} \cosh^{2} \eta'}} \right) \Biggm|_{\eta'=2.5},
\label{eqn:ptfit}
\end{eqnarray*}
where $m_{T}(\pta)$ is the transverse mass $m_{T} = \sqrt{\pta^{2}+m_{0}^{2}}$ 
and $m_{0}$ is the particle rest mass 
$m_{0}=\{m_{\pi},m_{p}$\} and $d \Nch / d y|_{y=0,i}$, $T_{i}$ and $n_{i}$ are the six parameters of the fit. 
$\eta'$ represents the pseudorapidity at the edge of our acceptance, $\eta = 2.5$.
$d \Nch / d y |_{y=0} $ represents the integrated yield of the particle production at mid-rapidity, but is left here as a free parameter of the fit. Mesons (pions and kaons) are merged into a single Tsallis function since there is insufficient information in the measured distribution to fit three independent shapes. 
The $ \tanh^{-1}$ factor accounts for the variation in $E/p$ of each track over the entire measured pseudorapidity range. It is derived by integrating $\frac{d y}{d \eta} d\eta$ over $|\eta| < 2.5$.

From this functional form and using the parameters obtained from the fit, the fraction of particles with $ \pta < 100$~MeV is extracted.
This procedure gives the correction factor to be applied to the mean charged-particle multiplicity per unit $\eta$ , averaged over $|\eta|~<~2.5$, in order to get the inclusive multiplicity.
The correction factor from $\pta> 100$~MeV to $\pta~>~0$~MeV is found to be 1.065 at \sqn\ and 1.063 at \sqs.

The second method assumes that the $\frac{1}{ p_\mathrm{T}} \cdot \frac{\mathrm{d} \Nch}{ \mathrm{d} p_\mathrm{T}}$ distribution is flat at low \pta. One can thus use the value of this distribution in the lowest \pta\ bin ($100<\pta < 150$~MeV) to extract the value for tracks below 100~MeV. 
From this assumption, the fraction of particles below 100~MeV and the scale factor used to correct our observed distributions are derived. 
The scale factors are found to be  1.068 at \sqn\ and 1.065 at \sqs. 
The third and final method simply obtains the correction factor using one of the MC models. AMBT1 \py6 is chosen; the correction factors are found to be 
1.055 at \sqn\ and 1.051 at \sqs. 
We chose to use the scale factor obtained from the functional form fit as the central value and consider the difference between this and the other two methods as a systematic uncertainty.

\paragraph{Systematic Uncertainties}
Several sources of systematic uncertainty on the calculated scale factor are considered.
The dominant uncertainty comes from the difference in the scale factors obtained from the three different extrapolation methods. 
The largest difference between the value obtained from the fit and the values from the MC and from the flat extrapolation is considered as the uncertainty and then symmetrised. This uncertainty is found to be $0.007$ at \sqn\ and $0.012$ at \sqs.

The other sources of uncertainty are related to the fitting procedure such as 
the variation within the uncertainty on the fit parameters and
the variation due to a change of the the fit range.
All sources of uncertainty are assumed to be uncorrelated and thus added in quadrature.
The final scale factors, with total uncertainty, are then $1.063 \pm 0.014_{\textrm{tot}}$ at \sqs\ and $1.065 \pm  0.011_{\textrm{tot}}$ at \sqn.

%%%%%%%%%%%%%%%%%%%%%%%%%%%%

\section{Total Systematic Uncertainties} \label{sec:sys}

The individual sources of systematic uncertainties have already been discussed in previous sections.
The effect on the final distribution from each source is treated independently and propagated to the final distributions; 
the total error is the sum in quadrature from the different sources, unless explicitly mentioned in the text.
In most bins of all distributions the largest uncertainty comes from the track reconstruction efficiency.
The uncertainties at \sqt\ are larger than at the other two energies due to the uncertainties related to the operation of the SCT at reduced bias voltage during 2.36~TeV data taking.
The total uncertainties are shown as shaded bands in the final distributions presented in the next section.

%%%%%%%%%%%%%%%%%%%%%%%%%%%%
\section{Results and Discussion}\label{sec:results}

The corrected distributions for primary charged particles for events in three separate phase-space regions are shown in Fig.~\ref{fig:dndeta_1} to~\ref{fig:sqrts}. 
The results are compared to predictions of models tuned to a wide range of measurements. 
The measured distributions are presented as inclusive-inelastic distributions within a given phase-space region with minimal model-dependent
corrections to facilitate the comparison with models. 

%-------------------------
\subsection{Charged-Particle Multiplicities as a Function of the Pseudorapidity}
\begin{figure}[htb!]
\begin{center}
	\subfigure[\label{dndeta_900_pt500}]{\includegraphics[width=0.43\textwidth]{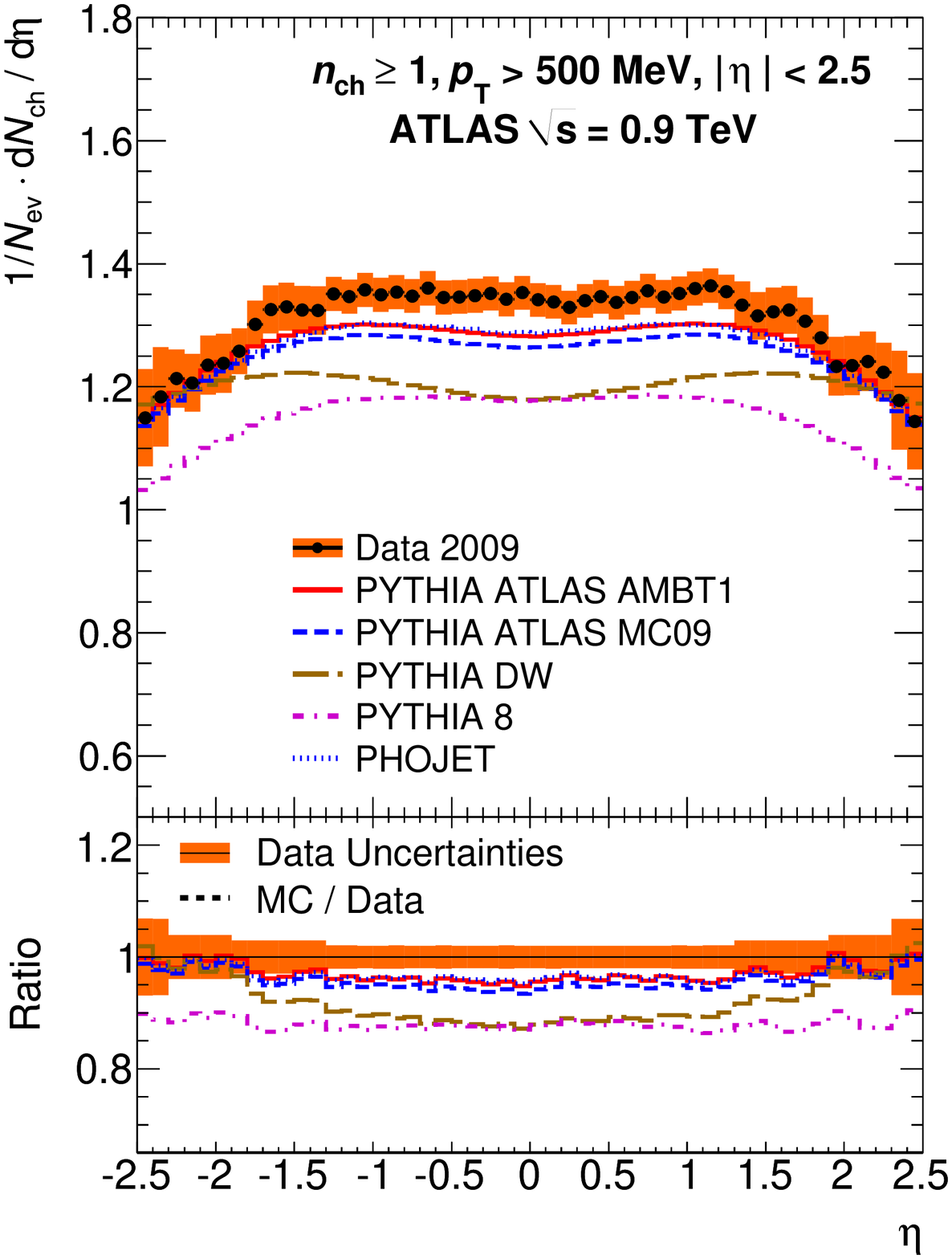}}	
	\subfigure[\label{dndeta_236_pt500}]{\includegraphics[width=0.43\textwidth]{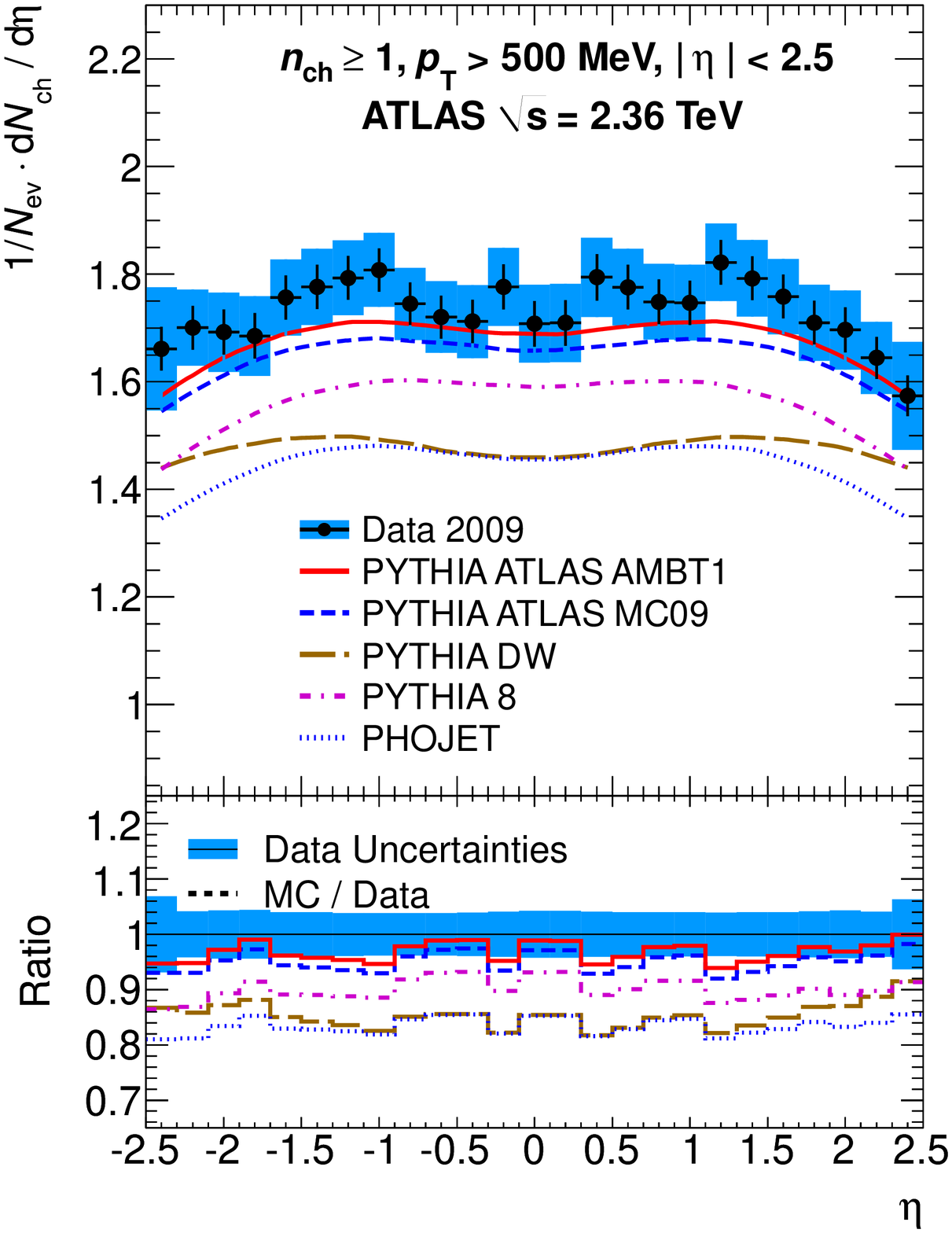}}	
	\subfigure[\label{dndeta_7_pt500}]{\includegraphics[width=0.43\textwidth]{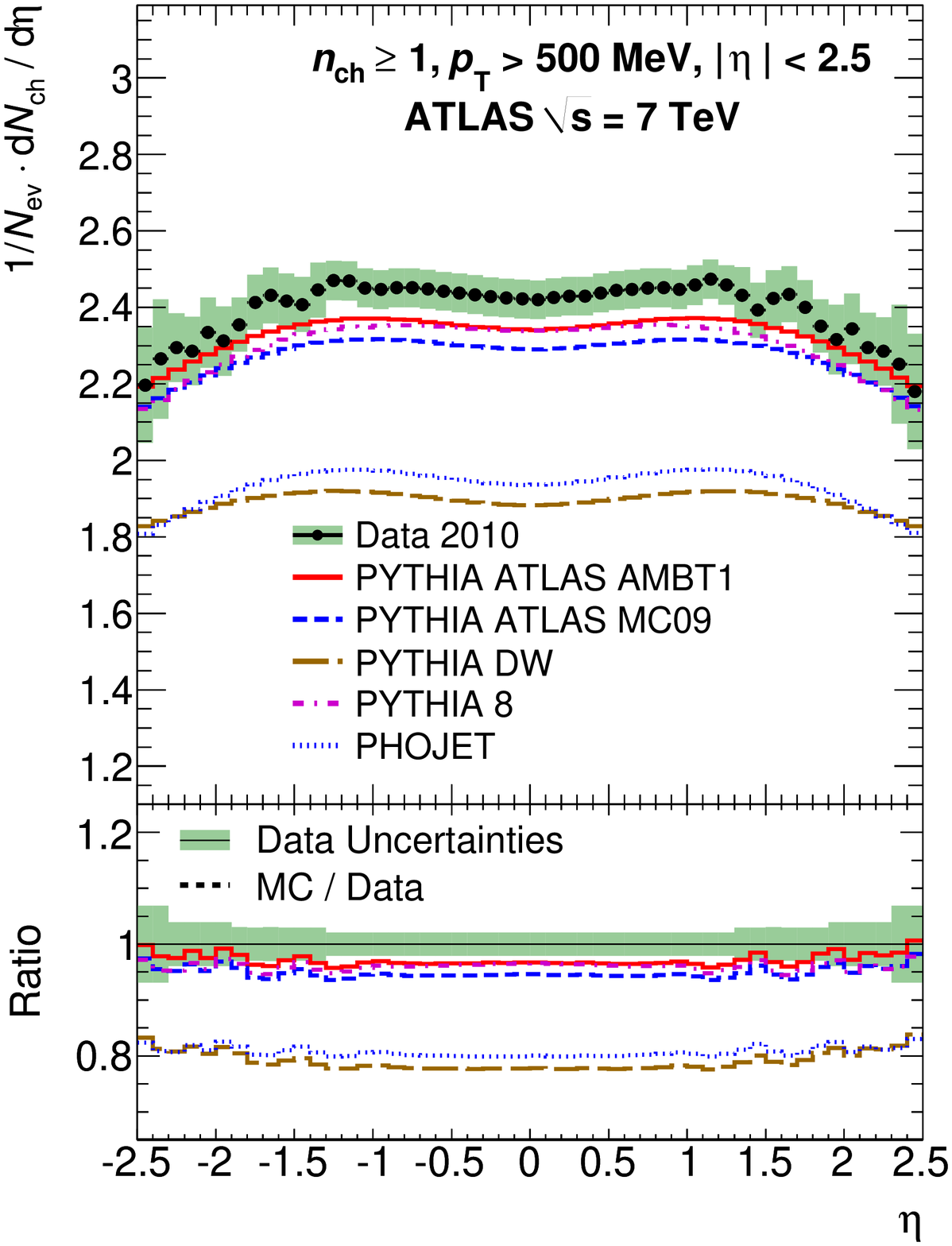}}	
\caption{Charged-particle multiplicities as a function of the pseudorapidity
for events with $\nch~\geq~1$, $\pta~>~500$~MeV and $|\eta|~<~2.5$ at \sqn\ (a), \sqt\ (b) and \sqs\ (c). The dots represent the data and the curves the predictions from different MC models. The vertical bars represent the statistical uncertainties,
while the shaded areas show statistical and systematic uncertainties added in quadrature.
The bottom inserts show the ratio of the MC over the data. The values of the ratio histograms refer to the bin centroids.}
\label{fig:dndeta_1}
\end{center}
\end{figure}

\begin{figure}[htb!]
\begin{center}
	\subfigure[\label{dndeta_900_pt100}]{\includegraphics[width=0.43\textwidth]{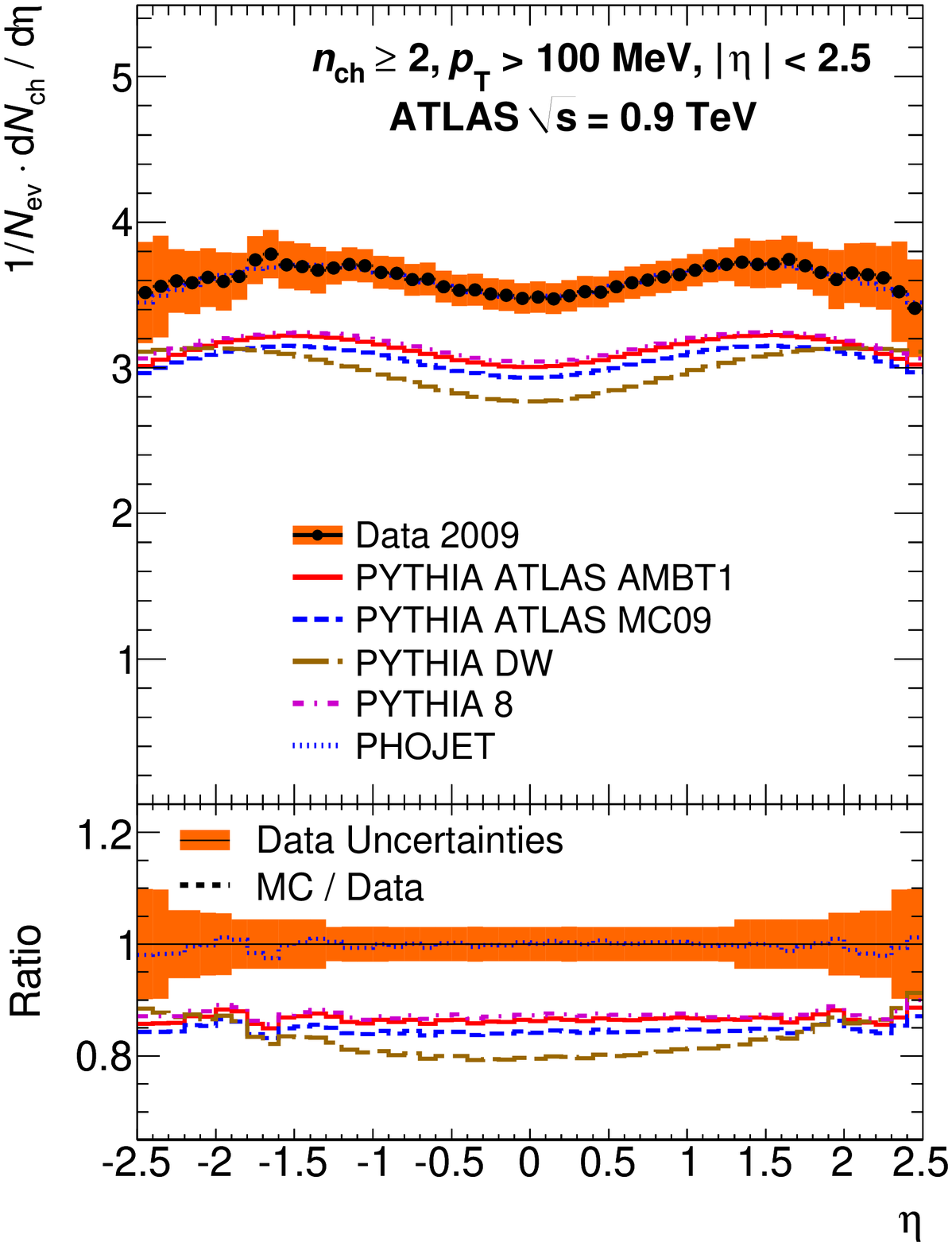}}	
	\subfigure[\label{dndeta_7_pt100}]{\includegraphics[width=0.43\textwidth]{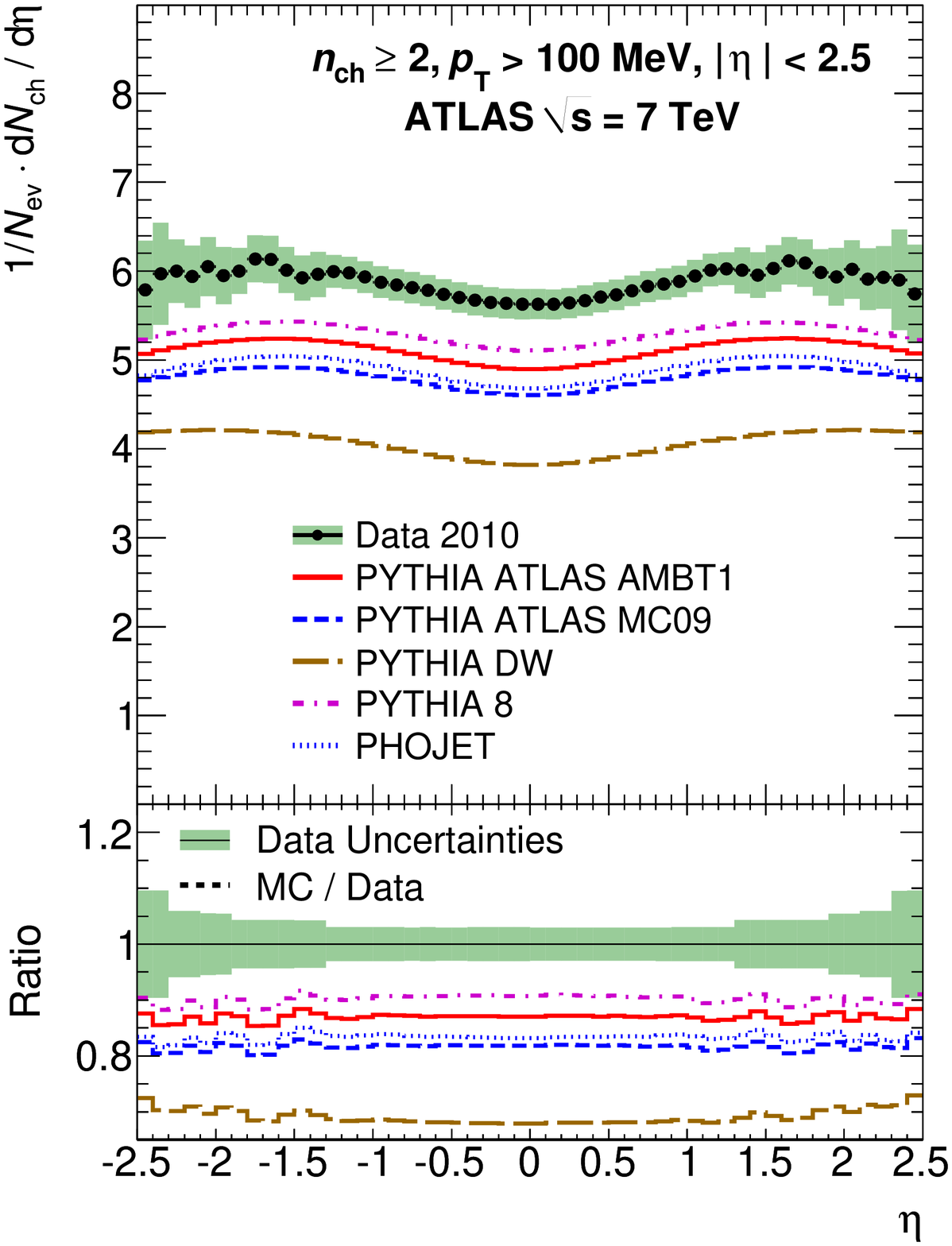}}	
	\subfigure[\label{dndeta_900_pt500_nch6}]{\includegraphics[width=0.43\textwidth]{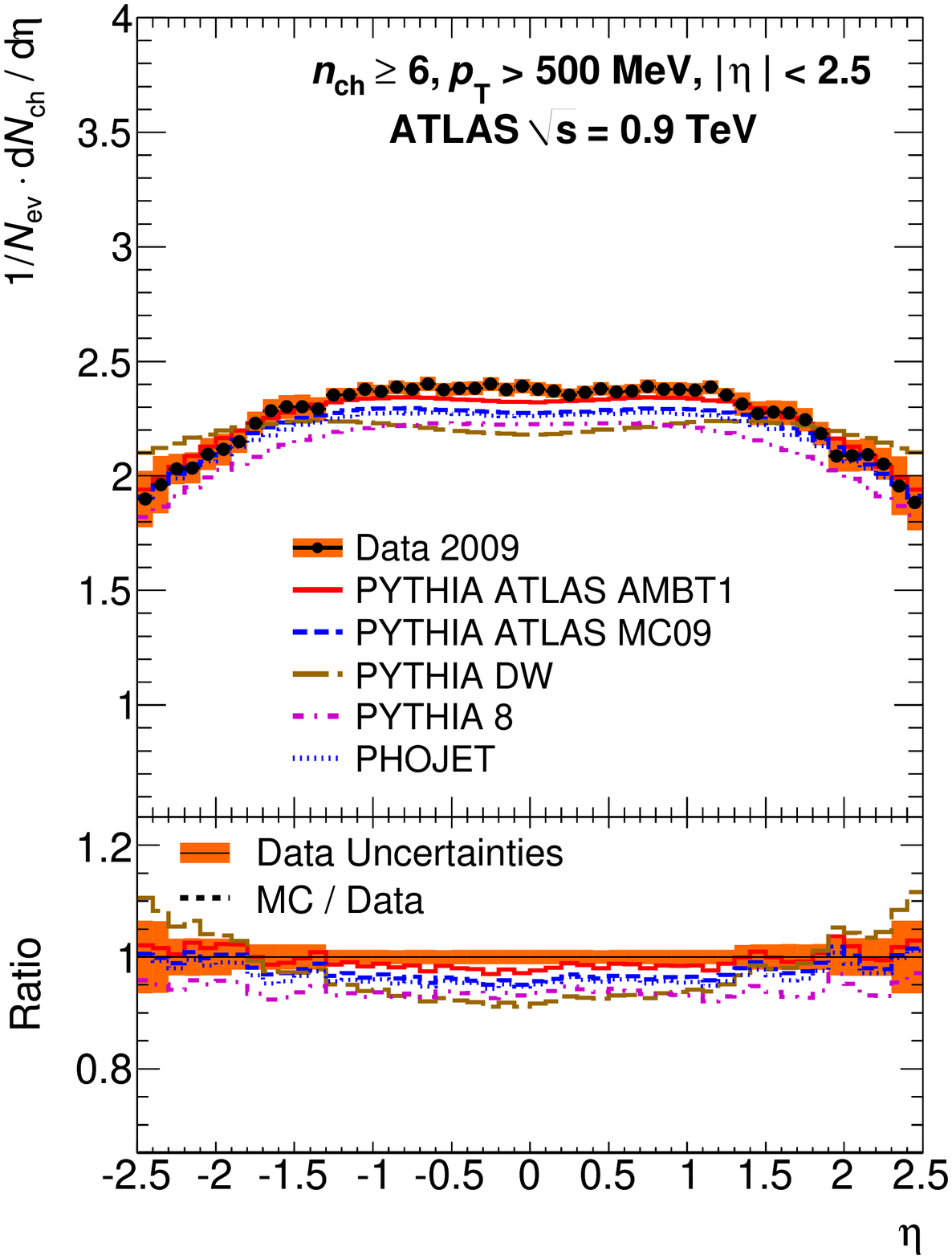}}	
	\subfigure[\label{dndeta_7_pt500_nch6}]{\includegraphics[width=0.43\textwidth]{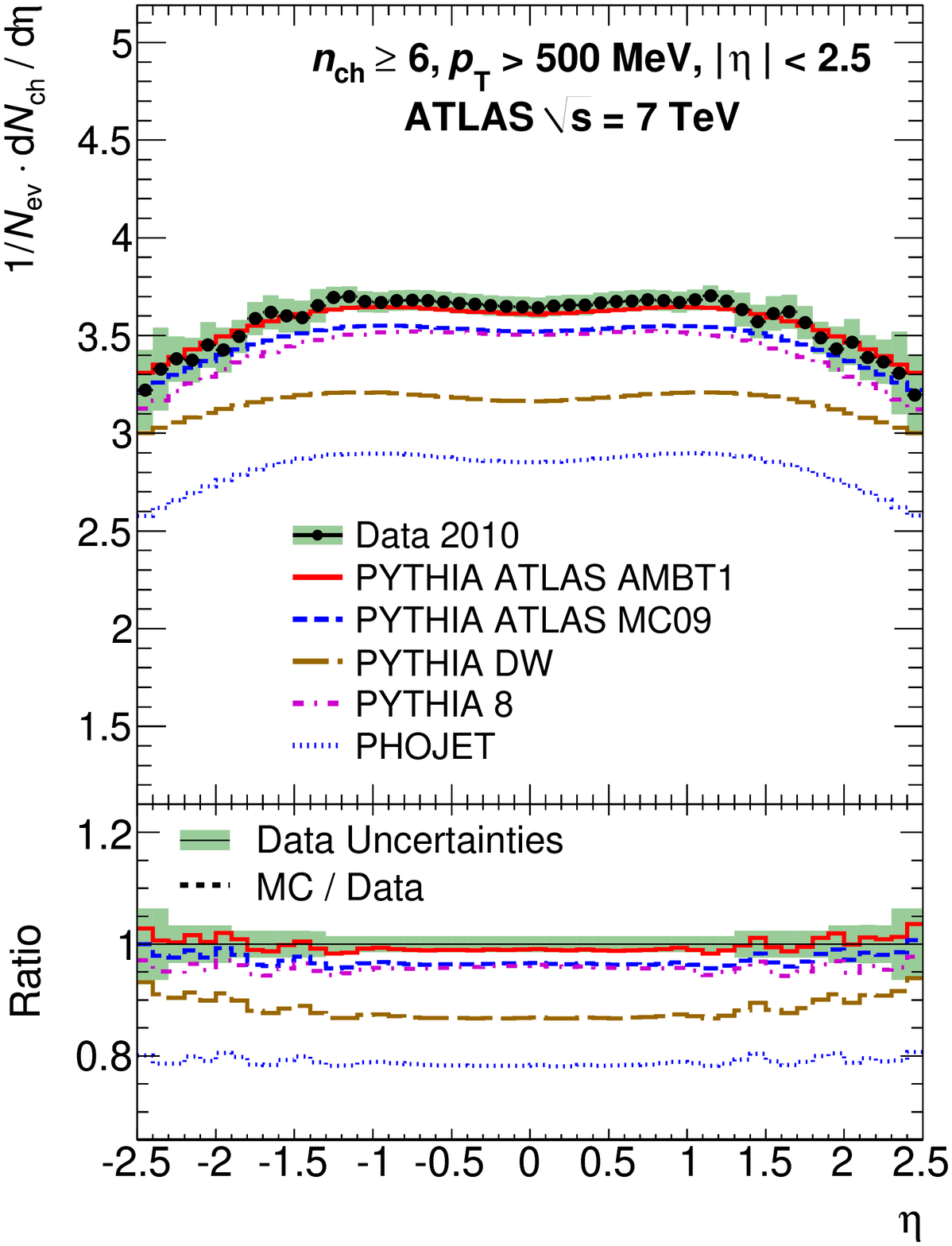}}	

\caption{Charged-particle multiplicities as a function of the pseudorapidity
for events with $\nch~\geq~2$, $\pta~>~100$~MeV (a,b) and $\nch~\geq~6$, $\pta~>~500$~MeV (c,d) and $|\eta|~<~2.5$ at \sqn\ (a,c) and \sqs\ (b,d). 
The dots represent the data and the curves the predictions from different MC models. 
The vertical bars represent the statistical uncertainties,
while the shaded areas show statistical and systematic uncertainties added in quadrature.
The bottom inserts show the ratio of the MC over the data. The values of the ratio histograms refer to the bin centroids.}
\label{fig:dndeta_2}
\end{center}
\end{figure}

Figures~\ref{fig:dndeta_1} and~\ref{fig:dndeta_2} show the charged-particle multiplicity as a function of pseudorapidity. 
Figure~\ref{fig:dndeta_1} shows the distribution at all three centre-of-mass energies in the phase-space region, $\nch~\geq~1$, $\pta~>~500$~MeV, $|\eta|< 2.5$. 
The mean particle density is roughly constant for $|\eta|~<~1.0$ and decreases at higher values of $|\eta|$. 
There is little shape variation between the models except for the DW \py 6 tune which has a flatter spectrum and a more pronounced dip at central $\eta$ , especially at low~$\sqrt{s}$.
At all three energies the AMBT1 \py 6 tune gives the best shape and normalisation description of the data, although it was tuned for $\nch \geq 6$.

Figure~\ref{fig:dndeta_2}a and b show the $\eta$ distributions for the most inclusive phase-space region, $\nch~\geq~2$, $\pta~>~100$~MeV, $|\eta|< 2.5$. 
There is less $\eta$ variation than in the previous figure.
%Diffractive physics is expected to be roughly flat in $\eta$, partially due to the lower \pta\ spectrum, and thus a larger fraction of diffractive events leads to a flatter distribution.
At 900~GeV there is very little difference between the models both in shape and normalisation with the exception of \pho\ which shows an excellent agreement with the data; 
the other models show on average too few particles.
The shape of the distribution is reasonably well described by all models.
At 7~TeV again the shapes seem to all model reasonably well the observed spectrum, 
but at this energy the difference in normalisation among the models varies more widely and no model reproduces the data.

Figure~\ref{fig:dndeta_2}c and d show the $\eta$ distributions for the phase-space region with the least amount of diffraction, $\nch~\geq~6$, $\pta~>~ 500$~MeV, $|\eta|< 2.5$. 
The distributions in this phase-space region have the largest drop at high $|\eta|$. 
All but \py 6 DW and \pho\ at \sqs\ show reasonable agreement in both shape and normalisation at both energies.

%%%%%%%%%%%%%%%%%%%
% dn/deta:
% pt > 500

% pt > 100

\subsection{Charged-Particle Multiplicities as a Function of the Transverse Momentum}
\begin{figure}[htb!]
\begin{center}
	\subfigure[\label{dndpt_900_pt500}]{\includegraphics[width=0.43\textwidth]{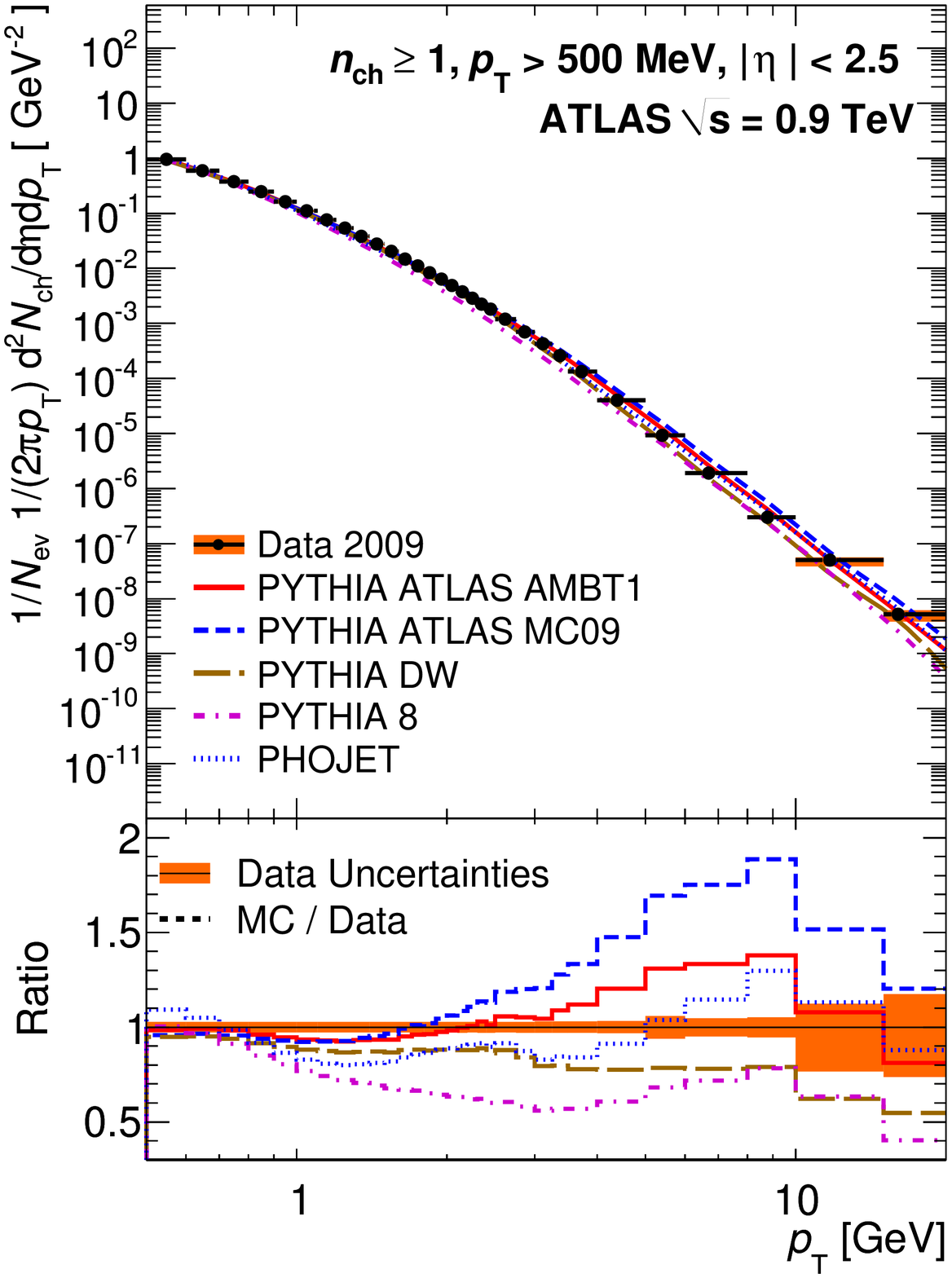}}	
	\subfigure[\label{dndpt_236_pt500}]{\includegraphics[width=0.43\textwidth]{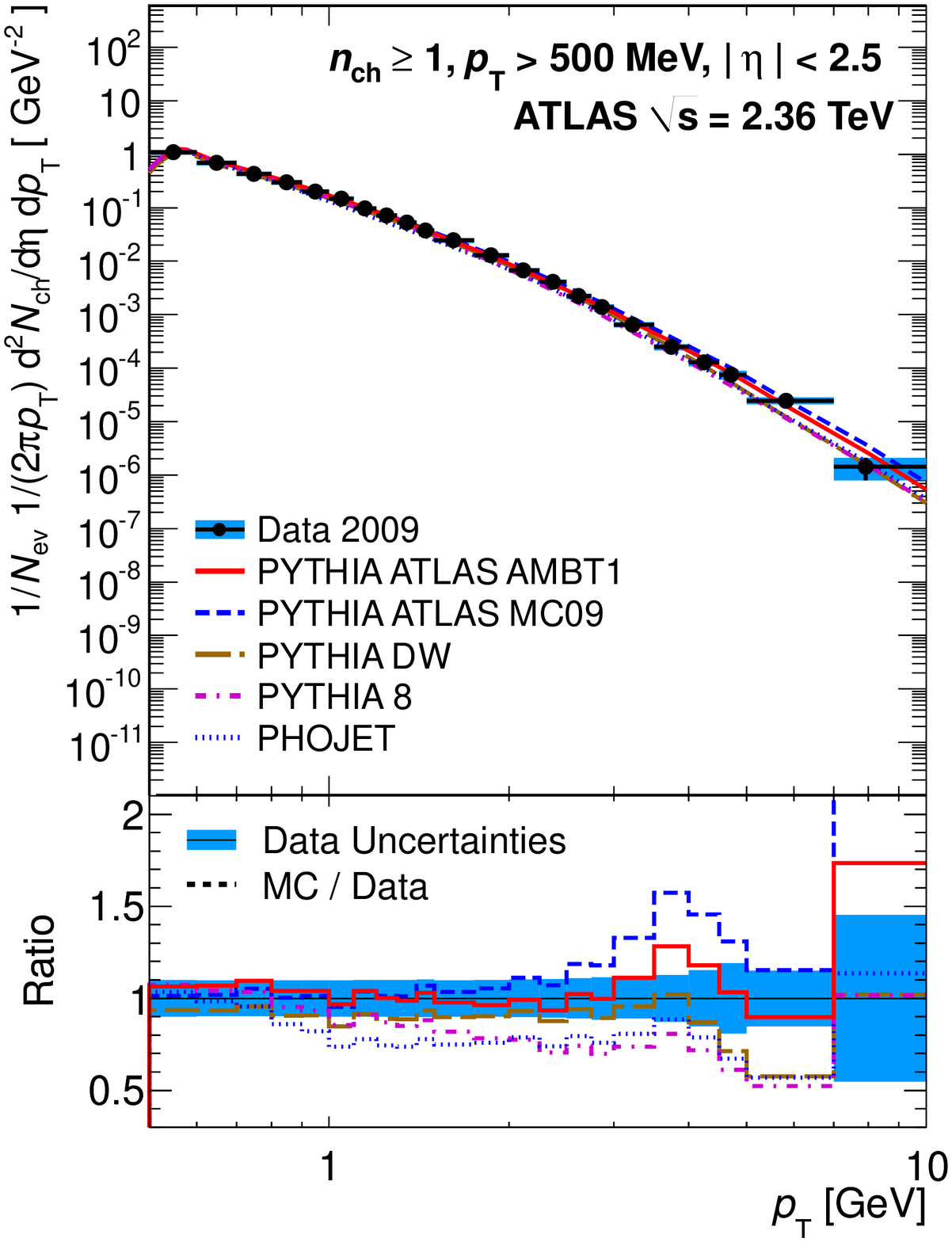}}	
	\subfigure[\label{dndpt_7_pt500}]{\includegraphics[width=0.43\textwidth]{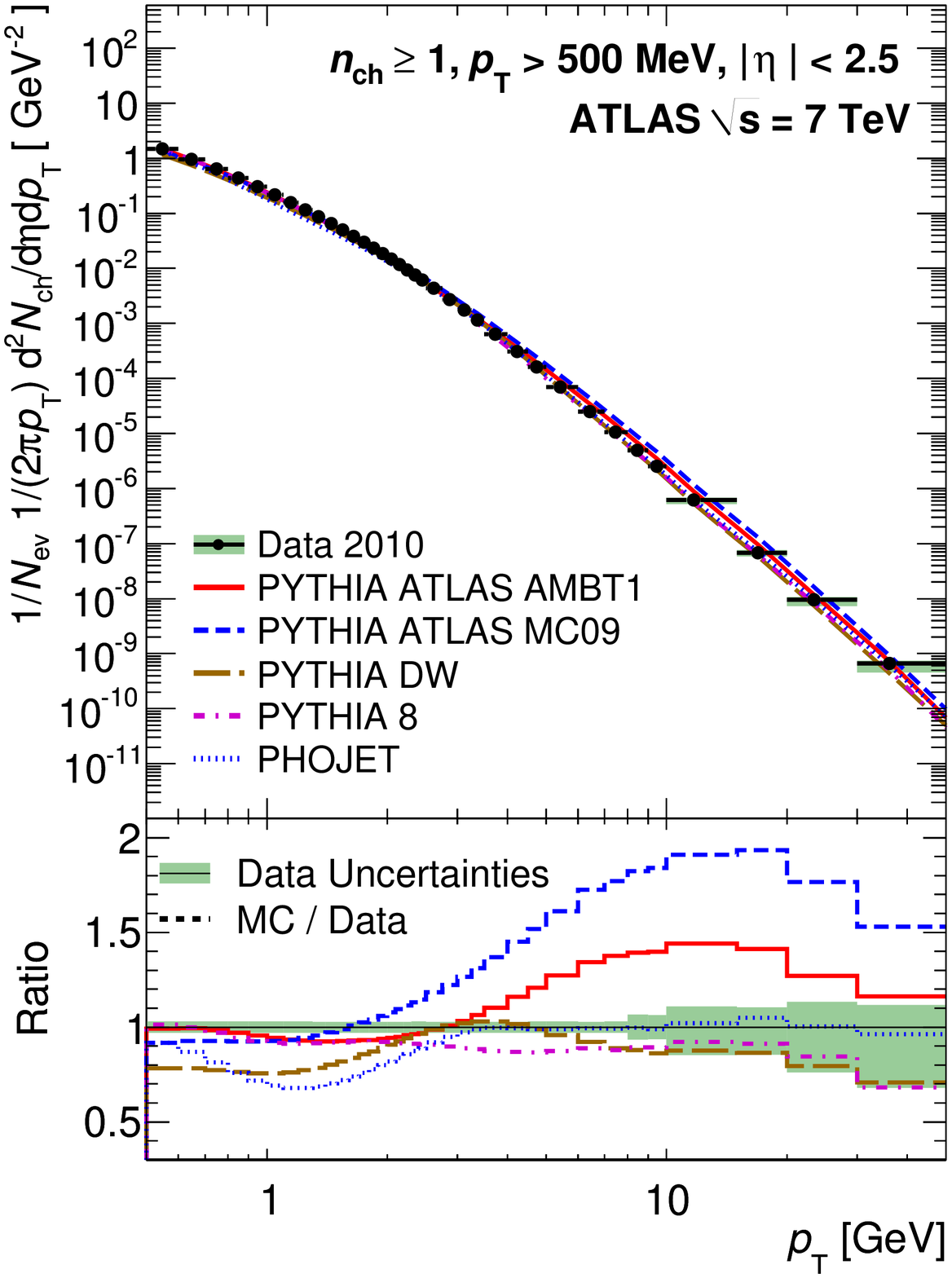}}	

\caption{Charged-particle multiplicities as a function of the transverse momentum
for events with $\nch~\geq~1$, $\pta~>~500$~MeV and $|\eta|<2.5$ at \sqn (a), \sqt (b) and \sqs (c). 
The dots represent the data and the curves the predictions from different MC models. 
The vertical bars represent the statistical uncertainties,
while the shaded areas show statistical and systematic uncertainties added in quadrature.
The bottom inserts show the ratio of the MC over the data. 
The values of the ratio histograms refer to the bin centroids.}
\label{fig:dndpt_1}
\end{center}
\end{figure}

\begin{figure}[htb!]
\begin{center}
	\subfigure[\label{dndpt_900_pt100}]{\includegraphics[width=0.43\textwidth]{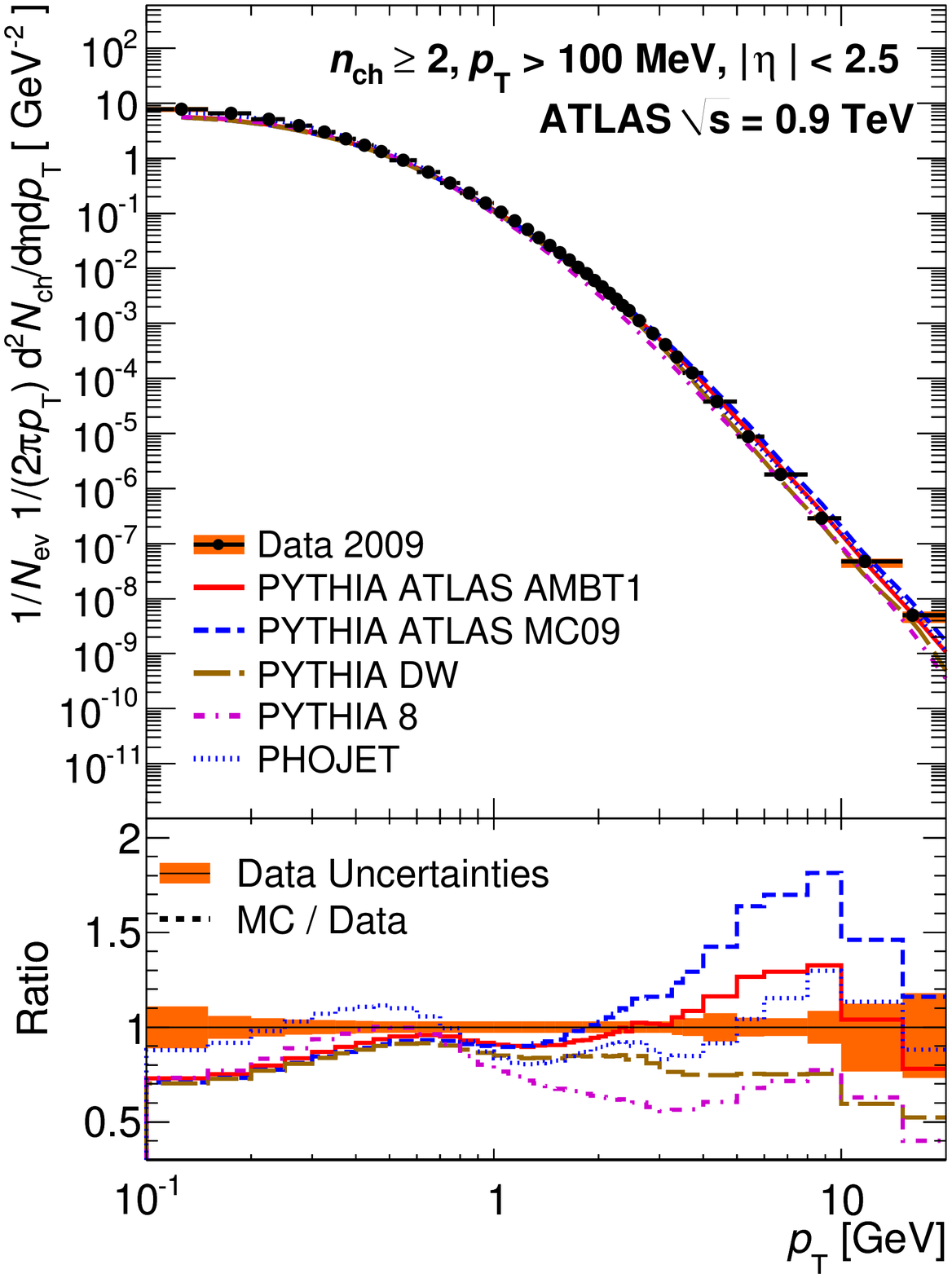}}	
	\subfigure[\label{dndpt_7_pt100}]{\includegraphics[width=0.43\textwidth]{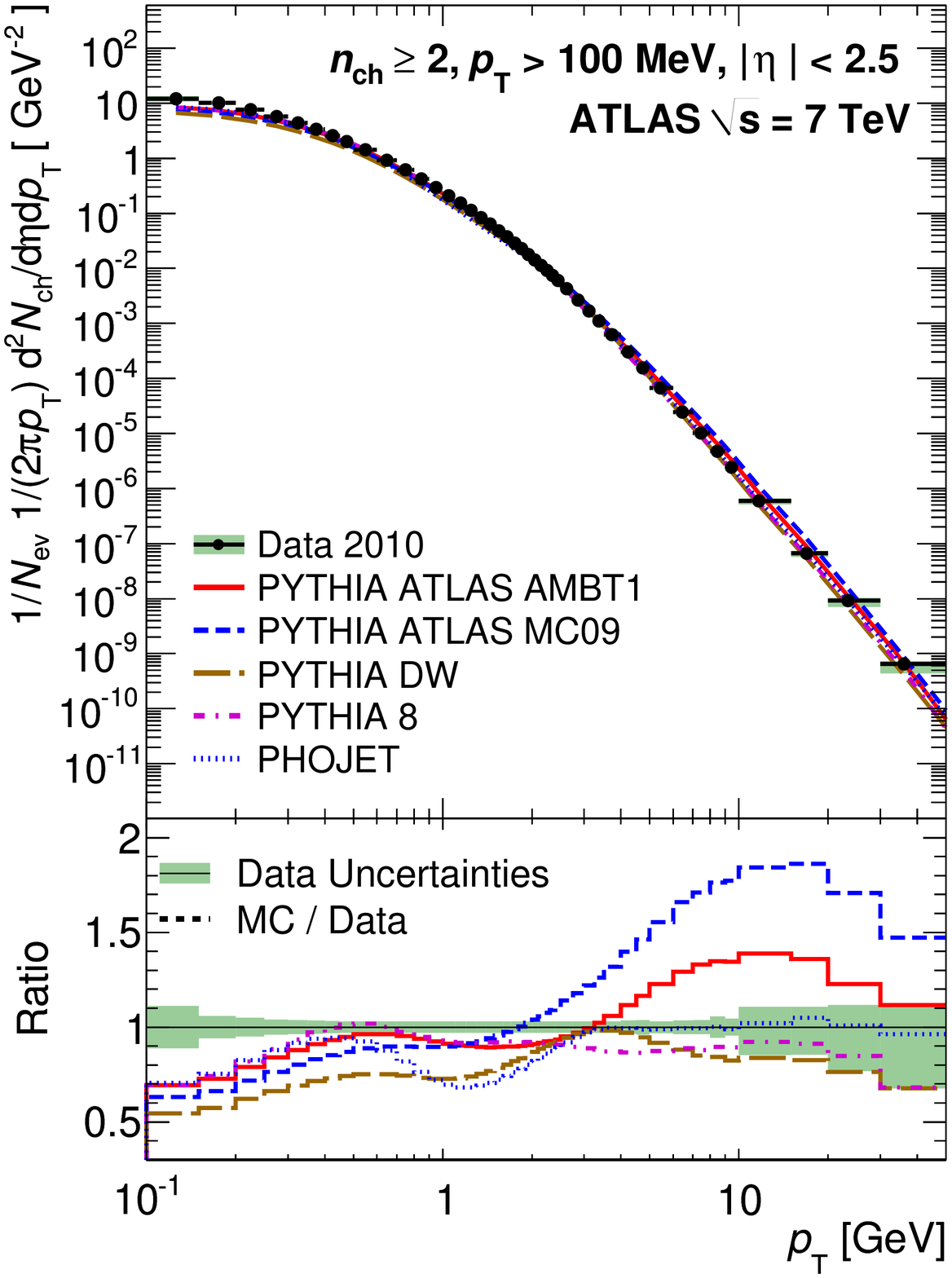}}	
	\subfigure[\label{dndpt_900_pt500_nch6}]{\includegraphics[width=0.43\textwidth]{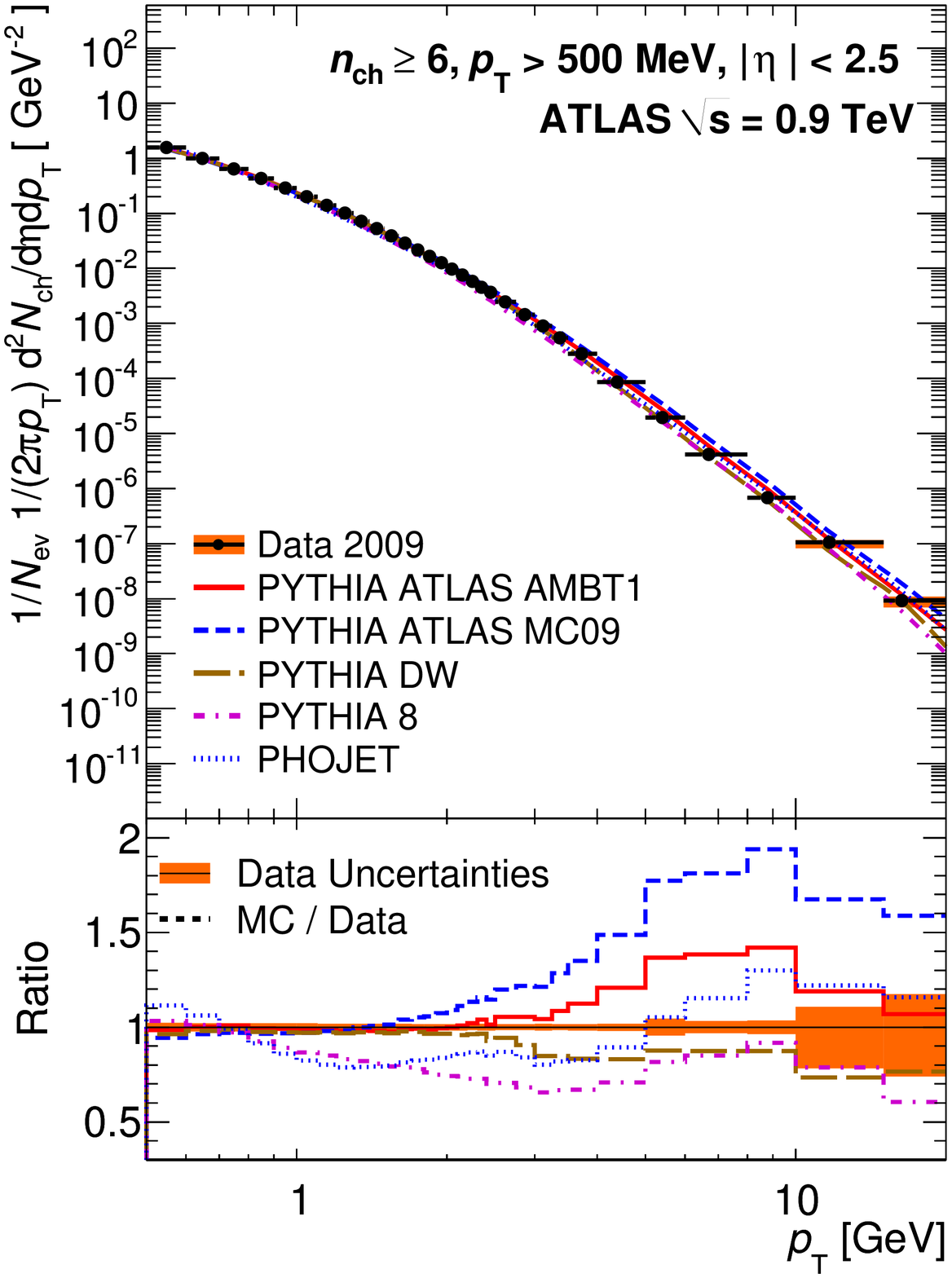}}
	\subfigure[\label{dndpt_7_pt500_nch6}]{\includegraphics[width=0.43\textwidth]{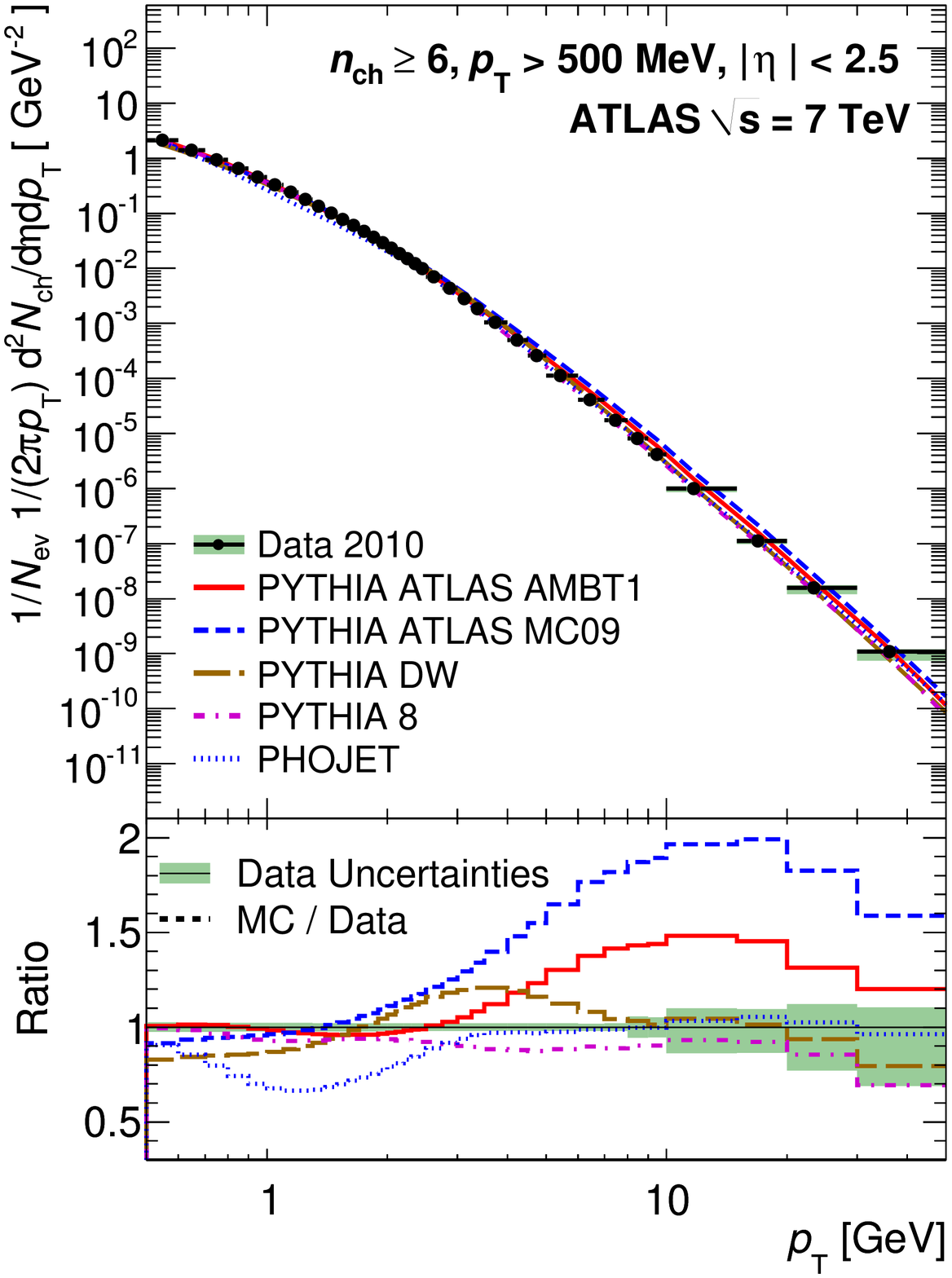}}	

\caption{Charged-particle multiplicities as a function of the transverse momentum
for events with $\nch~\geq~2$, $\pta~>~100$~MeV (a,b) and $\nch~\geq~6$, $\pta~>~500$~MeV (c,d) and $|\eta|~<~2.5$ at \sqn\ (a,c) and \sqs\ (b,d). 
The dots represent the data and the curves the predictions from different MC models. The vertical bars represent the statistical uncertainties,
while the shaded areas show statistical and systematic uncertainties added in quadrature.
The bottom inserts show the ratio of the MC over the data. 
The values of the ratio histograms refer to the bin centroids.}
\label{fig:dndpt_2}
\end{center}
\end{figure}

Figures~\ref{fig:dndpt_1} and~\ref{fig:dndpt_2} show the charged-particle multiplicities as a function of the transverse momentum.
The first of these figures shows all three centre-of-mass energies considered in the phase-space region $\nch~\geq~1$, $\pta~>~500$~MeV and $|\eta|~<~2.5$.
The observed \pta\ spectrum is not described by any of the models over the whole range. 
The region that the models have the most difficulty describing is the region above 1~GeV.

Figures~\ref{fig:dndpt_2}a and b show the charged-particle multiplicities in the most-inclusive phase-space region.
At 900~GeV \pho\ describes the data best over the whole range even though the agreement is still not excellent. 
The other models tend to under-predict the number of low \pta\ particles
while at higher \pta\ the models vary widely.
At 7~TeV the effect at low \pta\ is more pronounced, while at high \pta\ the agreement of \py 8 and \pho\ with the data is quite good.
The AMBT1 and MC09 tunes of \py 6 predict too many particles at higher \pta.

Figures~\ref{fig:dndpt_2}c and d show the charged-particle multiplicities with the smallest contribution from diffractive events. 
This distribution carried the most weight in the AMBT1 tune. 
Considerable improvement in the agreement with data is seen between the older MC09 and the newly tuned AMBT1 but the parameters varied in this tune were not sufficient to describe the full spectrum.

% dn\dpt2
% pt > 500

% pt > 100

\subsection{Charged-Particle Multiplicity Distribution}

\begin{figure}[htb!]
\begin{center}
	\subfigure[\label{dndnch_900_pt500}]{\includegraphics[width=0.43\textwidth]{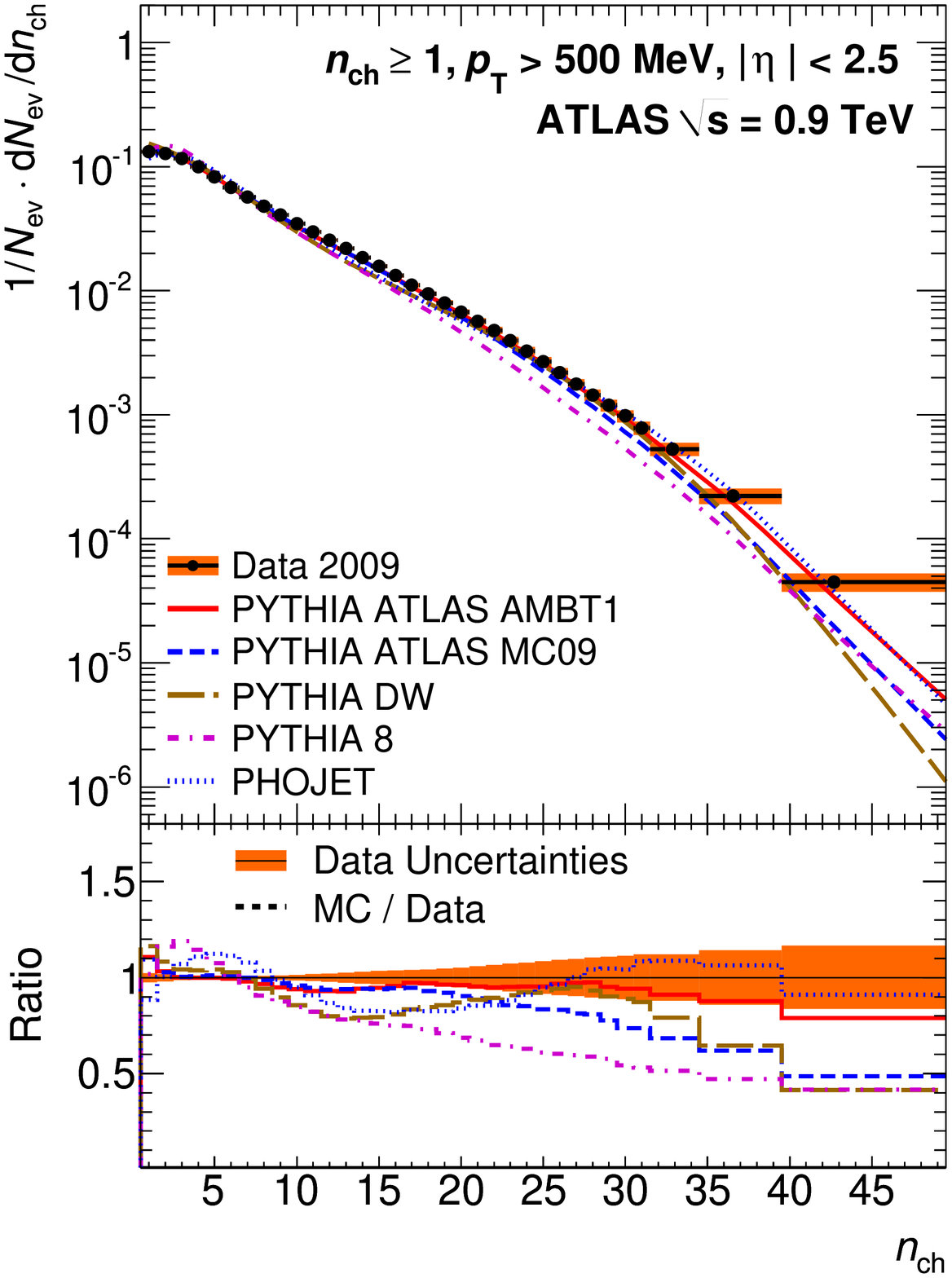}}	
	\subfigure[\label{dndnch_236_pt500}]{\includegraphics[width=0.43\textwidth]{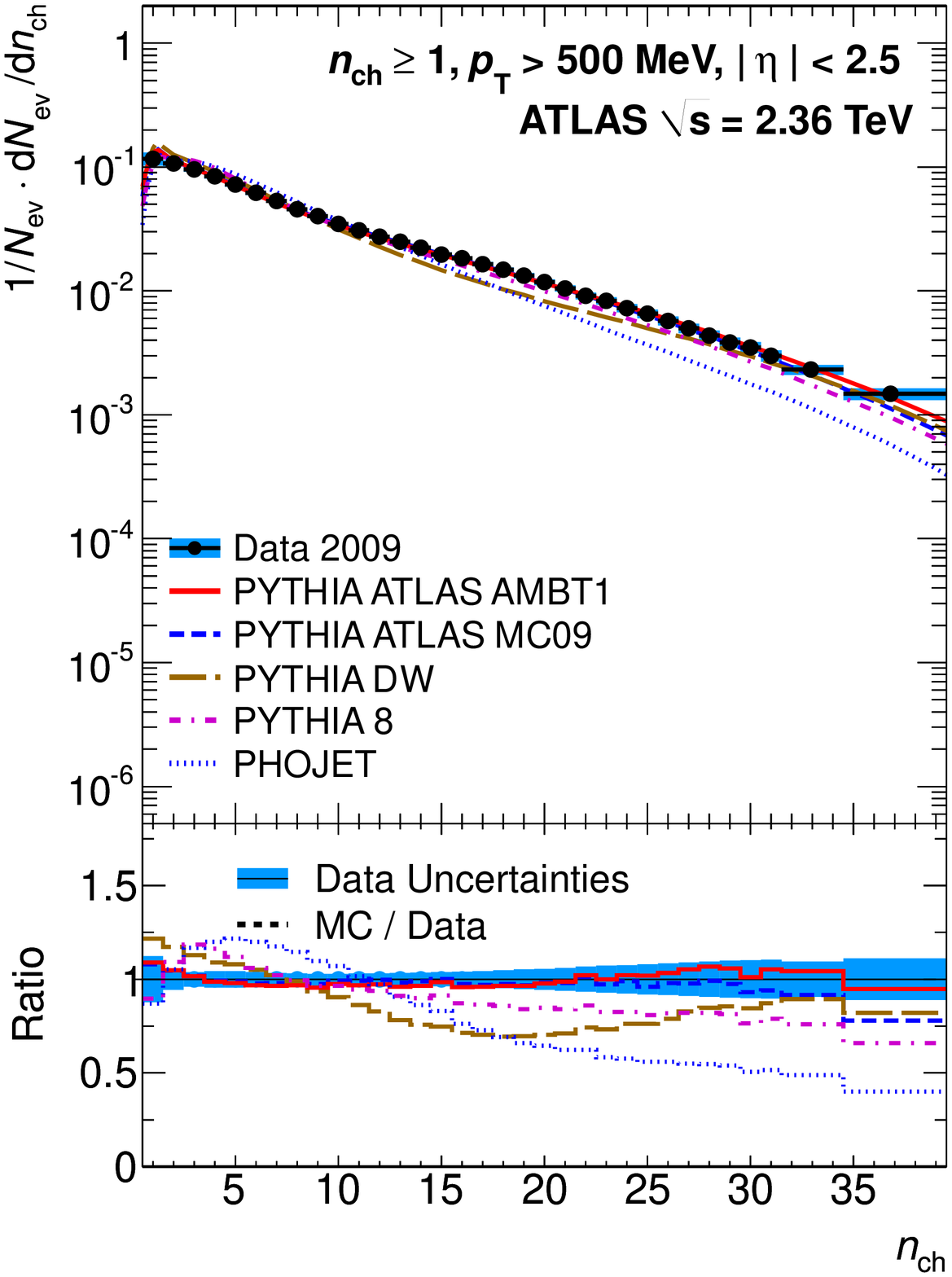}}	
	\subfigure[\label{dndnch_7_pt500}]{\includegraphics[width=0.43\textwidth]{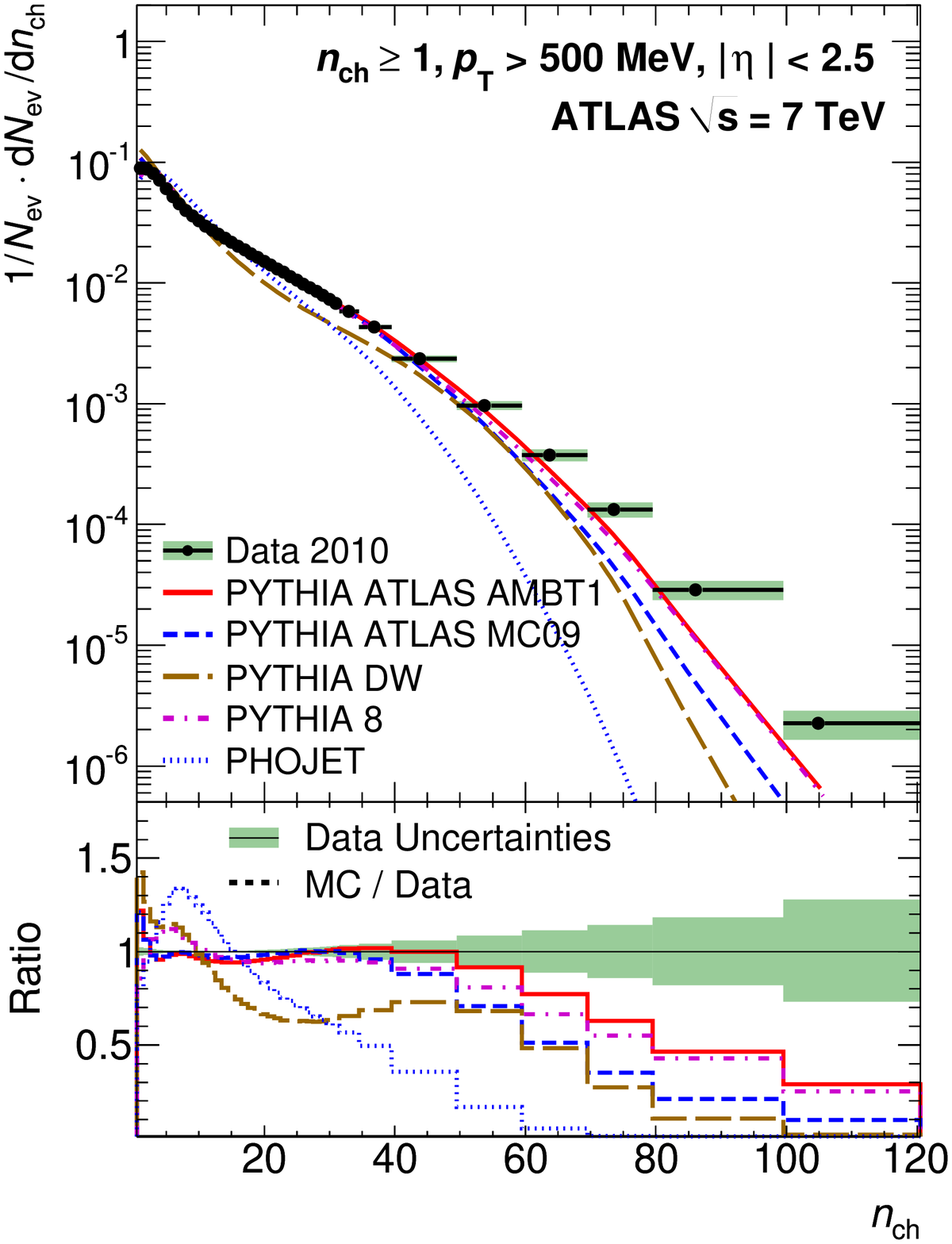}}	

\caption{Charged-particle multiplicity distributions
for events with $\nch~\geq~1$, $\pta~>~500$~MeV and $|\eta|~<~2.5$ at \sqn (a), \sqt (b) and \sqs (c). 
The dots represent the data and the curves the predictions from different MC models. The vertical bars represent the statistical uncertainties,
while the shaded areas show statistical and systematic uncertainties added in quadrature.
The bottom inserts show the ratio of the MC over the data. The values of the ratio histograms refer to the bin centroids.}
\label{fig:dndnch_1}
\end{center}
\end{figure}

Figure~\ref{fig:dndnch_1} shows the charged-particle multiplicity distributions for $\nch~\geq~1$, $\pta~>~500$~MeV and $|\eta|~<~2.5$ at all three centre-of-mass energies.
At low number of charged particles, all models predict more events than observed in data, which is compensated by an under-prediction in the tails of the distributions.
It should be noted that due to the normalisation, $1/\nev$, a deviation observed in one region needs to be compensated for by one in the other direction somewhere else.
Although the predictions of \pho\ at 0.9~TeV model the data reasonably well, at 2.36~TeV and 7~TeV they do not model the observed spectrum.
The new AMBT1 \py 6 tune seems to provide the best agreement with data.

\begin{figure}[htb!]
\begin{center}
	\subfigure[\label{dndnch_900_pt100}]{\includegraphics[width=0.43\textwidth]{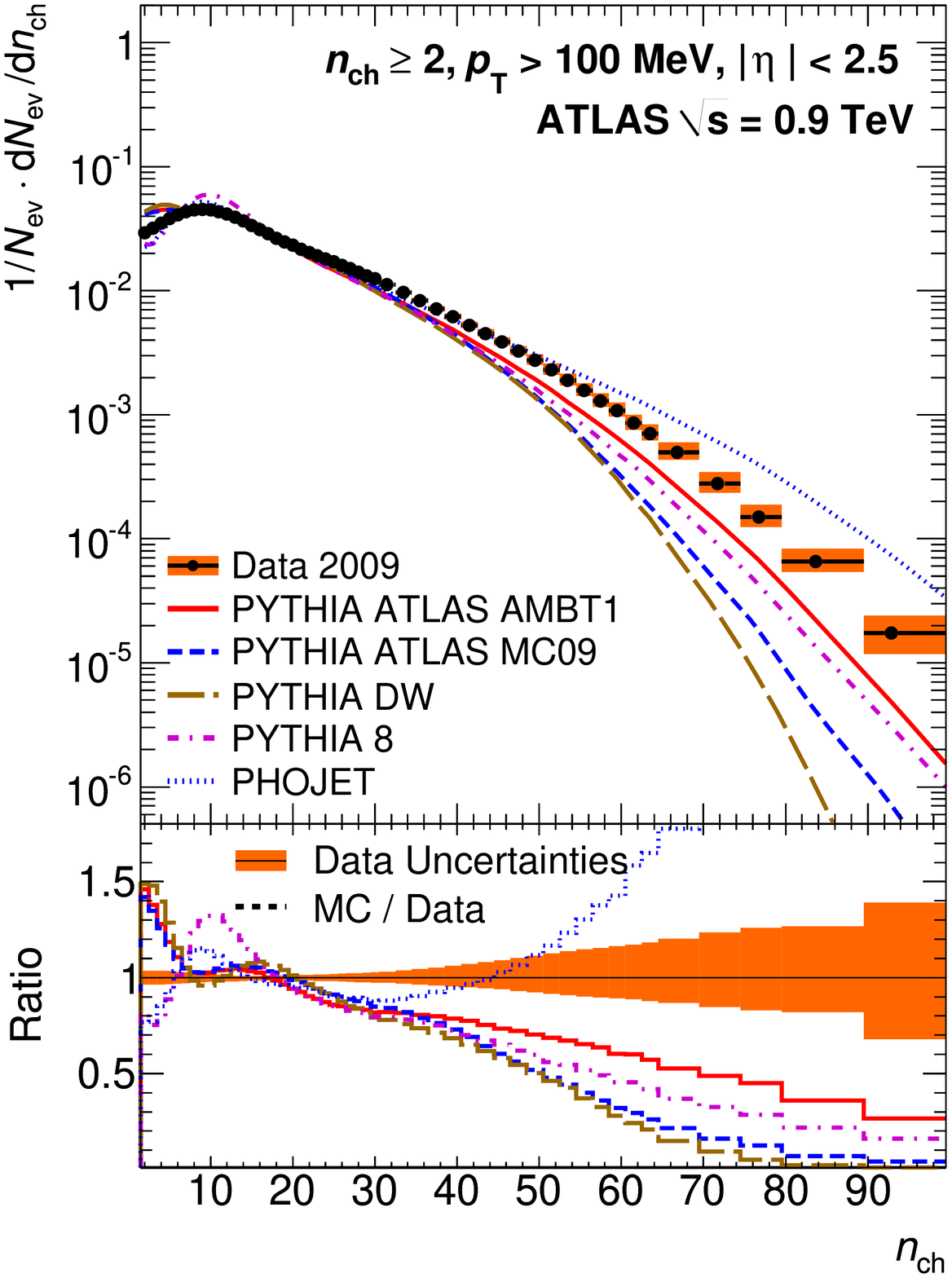}}	
	\subfigure[\label{dndnch_7_pt100}]{\includegraphics[width=0.43\textwidth]{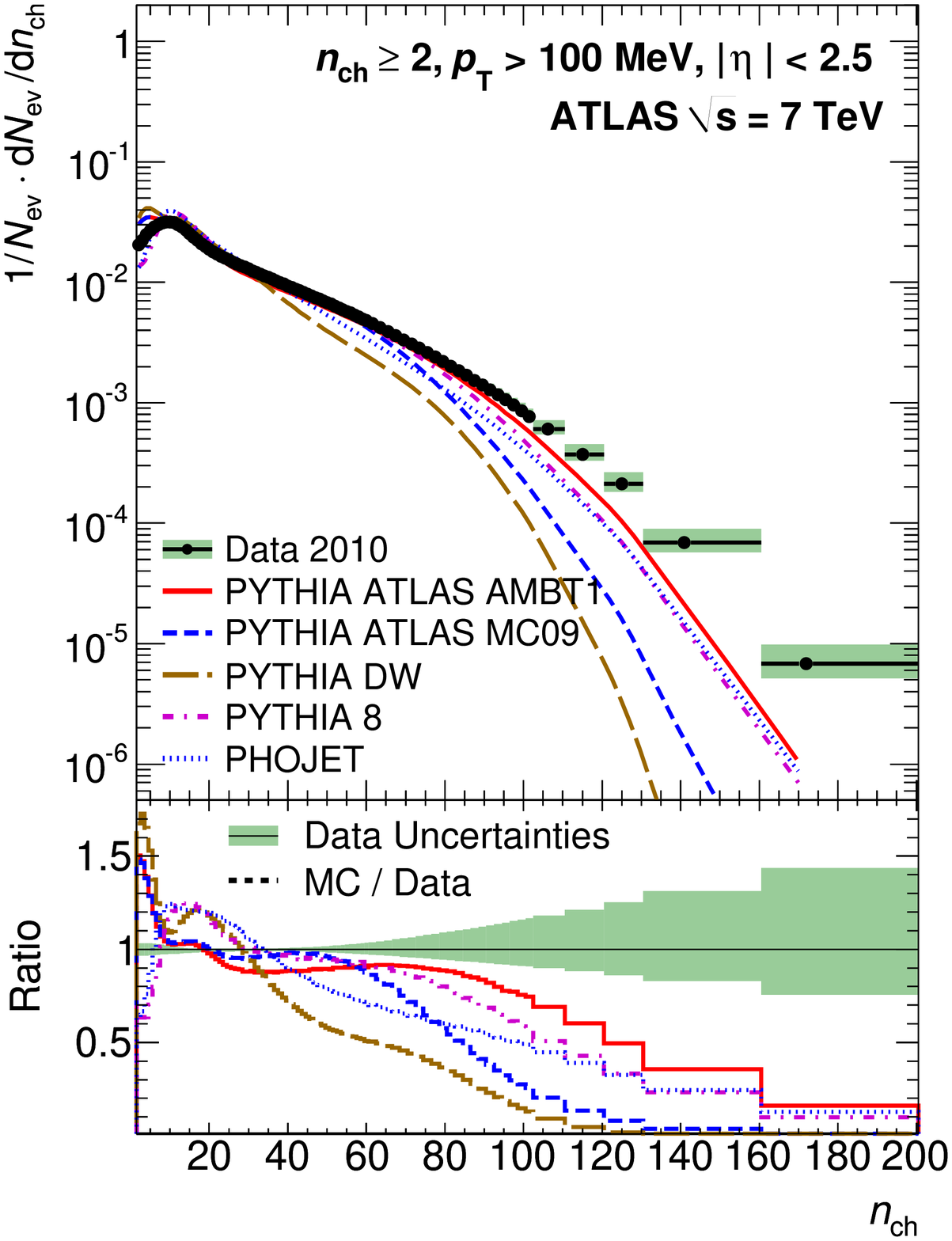}}		
	\subfigure[\label{dndnch_900_nch6}]{\includegraphics[width=0.43\textwidth]{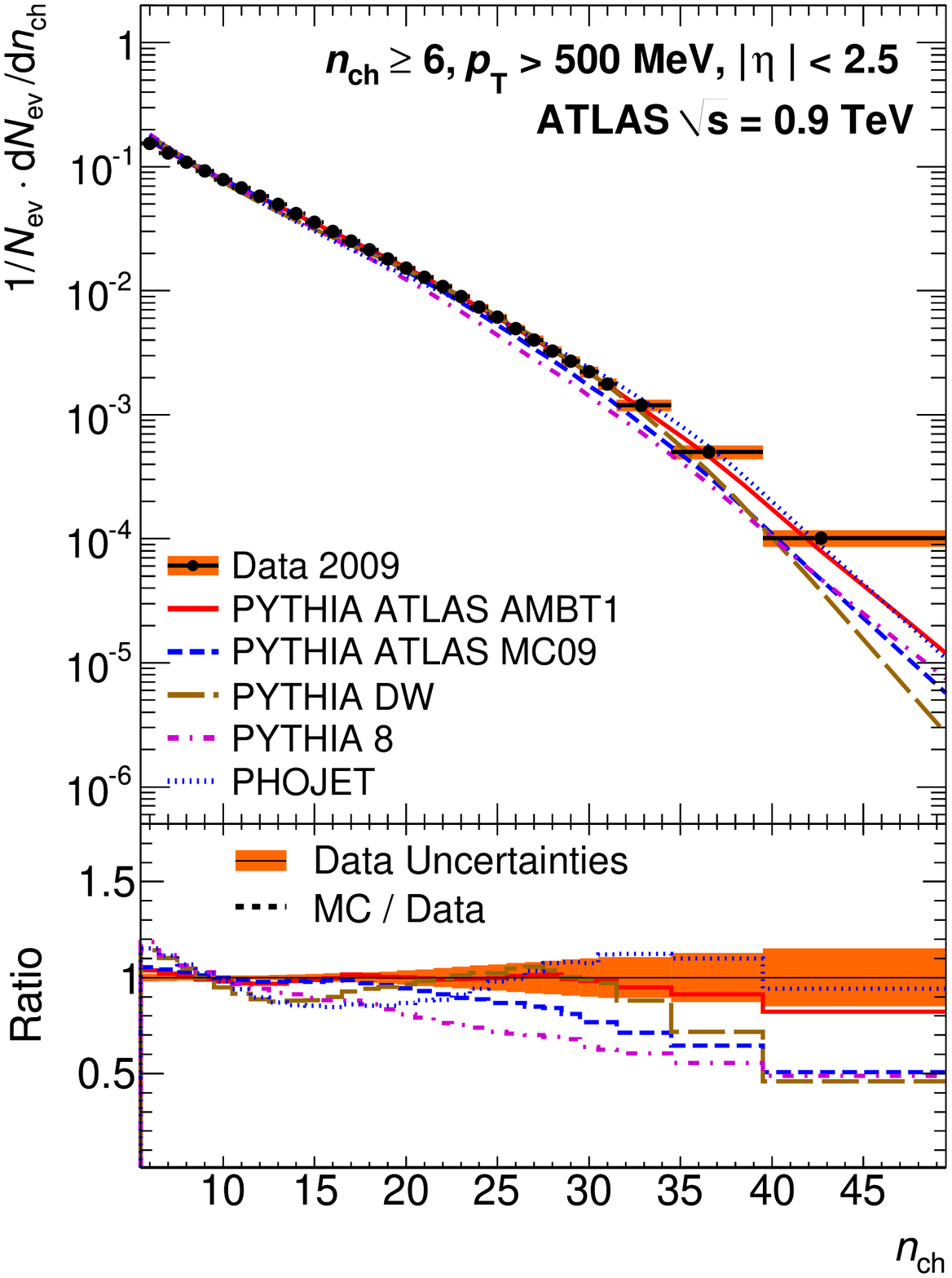}}	
	\subfigure[\label{dndnch_7_nch6}]{\includegraphics[width=0.43\textwidth]{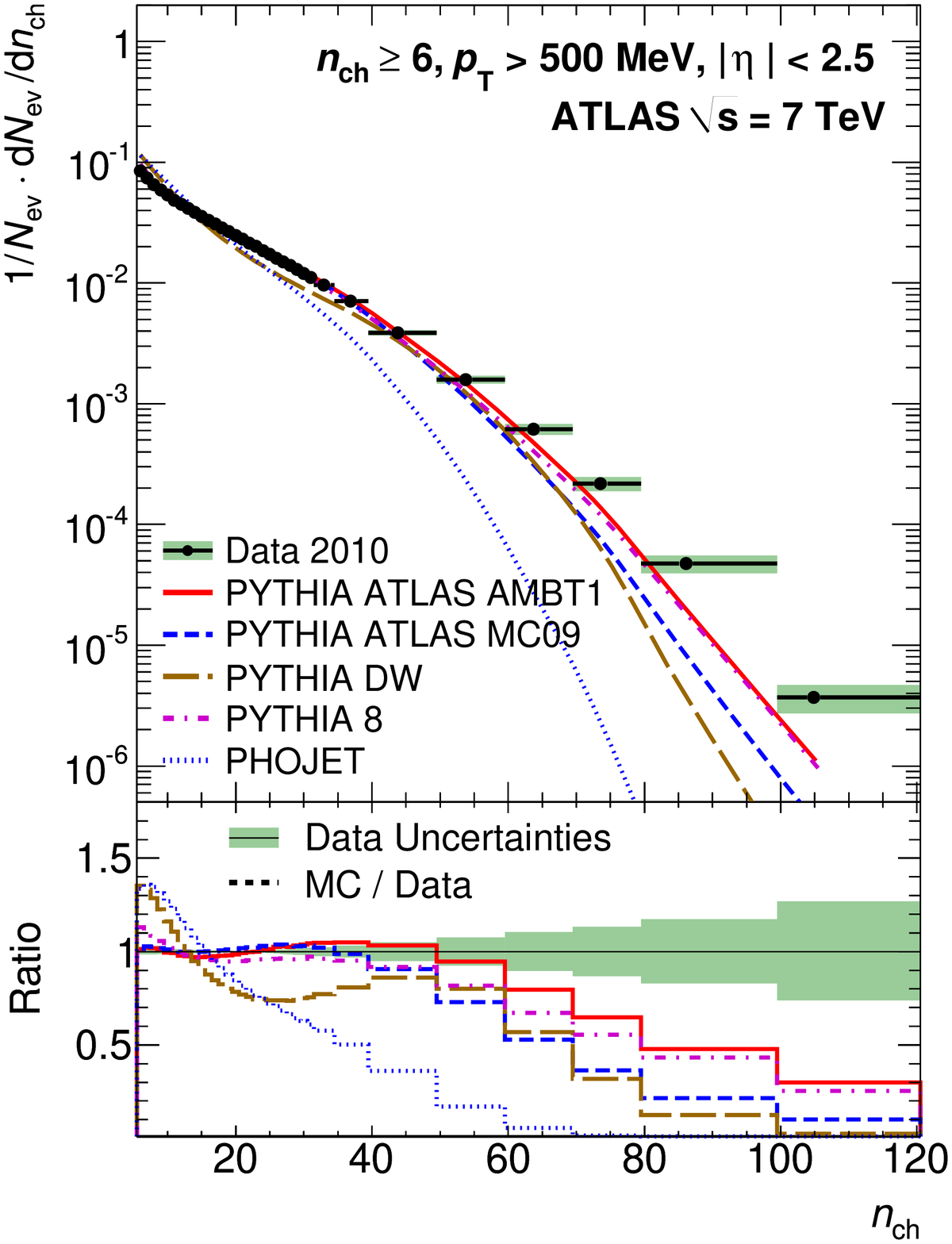}}	
\caption{Charged-particle multiplicity distributions
for events with $\nch~\geq~2$, $\pta~>~100$~MeV (a,b) and $\nch~\geq~6$, $\pta~>~500$~MeV (c,d) and $|\eta|~<~2.5$ at \sqn\ (a,c) and \sqs\ (b,d). 
The dots represent the data and the curves the predictions from different MC models. 
The vertical bars represent the statistical uncertainties,
while the shaded areas show statistical and systematic uncertainties added in quadrature.
The bottom inserts show the ratio of the MC over the data. The values of the ratio histograms refer to the bin centroids.}
\label{fig:dndnch_2}
\end{center}
\end{figure}

Figures~\ref{fig:dndnch_2}a and b show the distribution for the most inclusive phase-space region. 
Here the variations between models at both low and high values of \nch\ are increased and no model predicts the observed spectra. 

Figures~\ref{fig:dndnch_2}c and d show the distribution for the diffraction-reduced phase-space region. 
The distributions are very similar to those in Fig.~\ref{fig:dndnch_1} with a cut at $\nch~\geq~6$; 
only the normalisation is different between the plots. 
The errors are also recomputed as there is a larger cancellation between the numerator and denominator for this phase-space region.

% dN\dnch
% pt > 500

% pt > 100

\subsection{Average Transverse Momentum as a Function of the Number of Charged Particles}
\begin{figure}[htb!]
\begin{center}
	\subfigure[\label{meanpt_900_pt500}]{\includegraphics[width=0.43\textwidth]{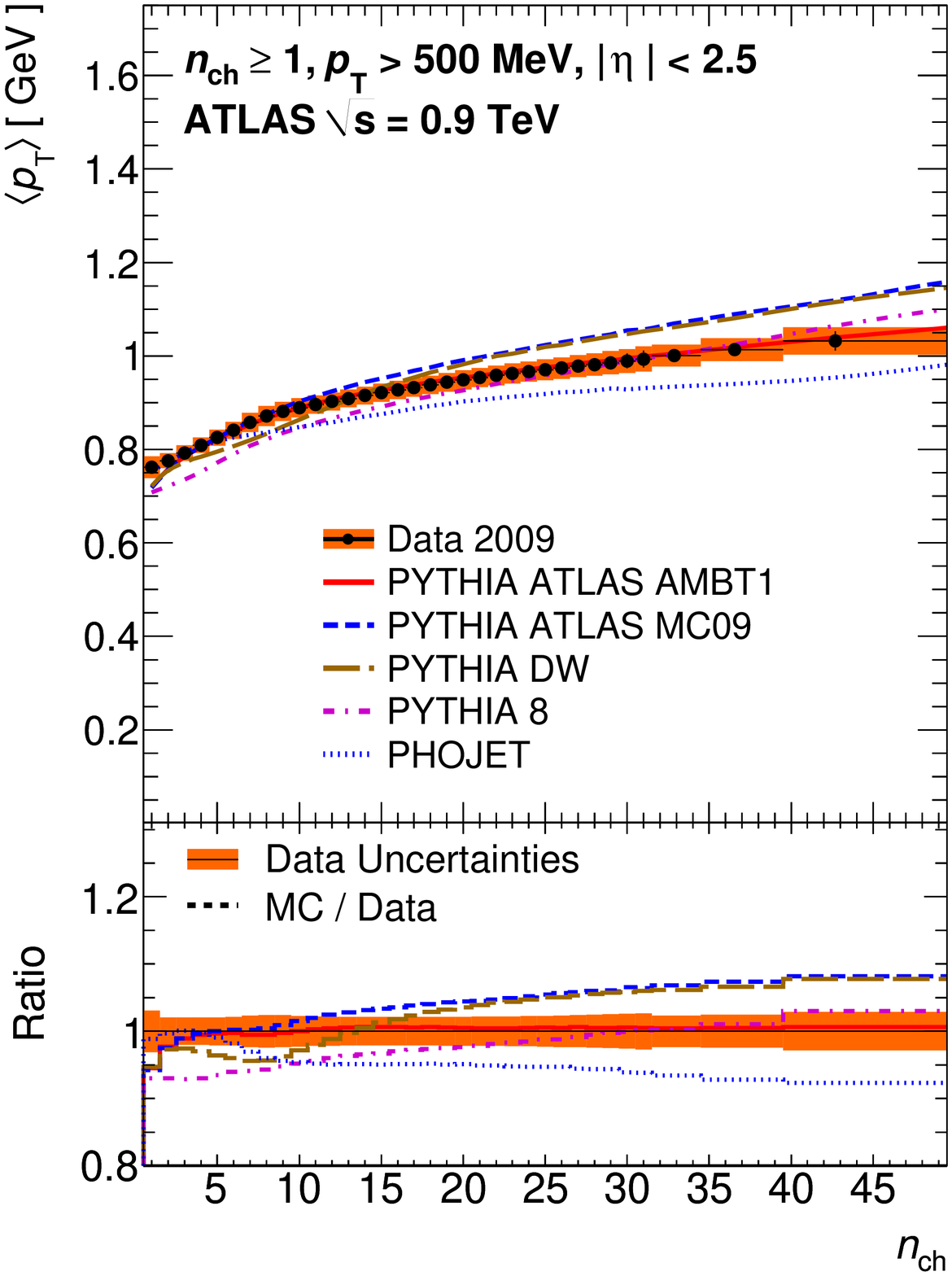}}	
	\subfigure[\label{meanpt_7_pt500}]{\includegraphics[width=0.43\textwidth]{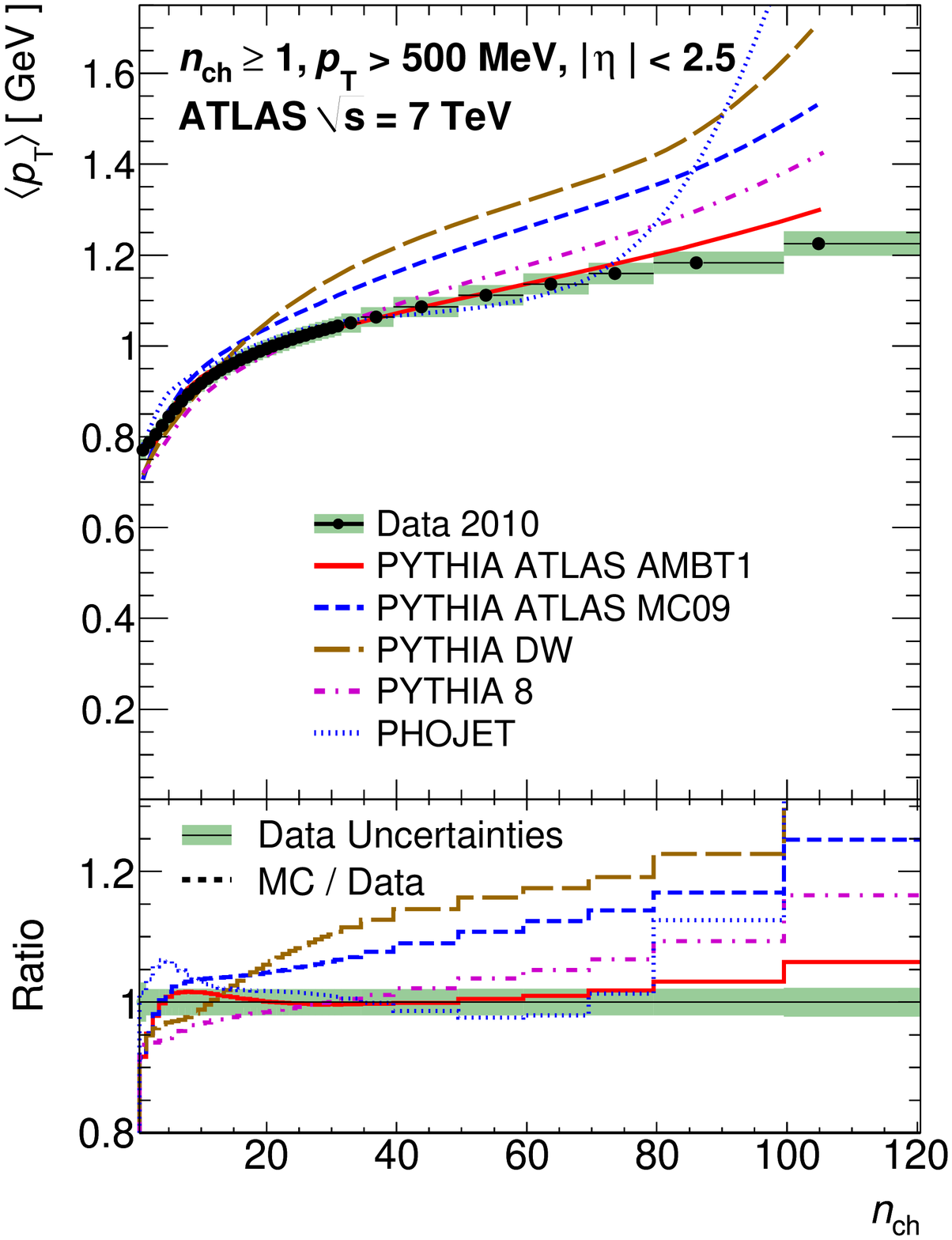}}	

\caption{Average transverse momentum as a function of the number of charged particles in the event 
for events with $\nch~\geq~1$, $\pta~>~500$~MeV and $|\eta|~<~2.5$ at \sqn (a), 
and \sqs (b). 
The dots represent the data and the curves the predictions from different MC models. 
The vertical bars represent the statistical uncertainties,
while the shaded areas show statistical and systematic uncertainties added in quadrature.
The bottom inserts show the ratio of the MC over the data. The values of the ratio histograms refer to the bin centroids.}
\label{fig:meanpt_1}
\end{center}
\end{figure}

\begin{figure}[htb!]
\begin{center}
	\subfigure[\label{meanpt_900_pt100}]{\includegraphics[width=0.43\textwidth]{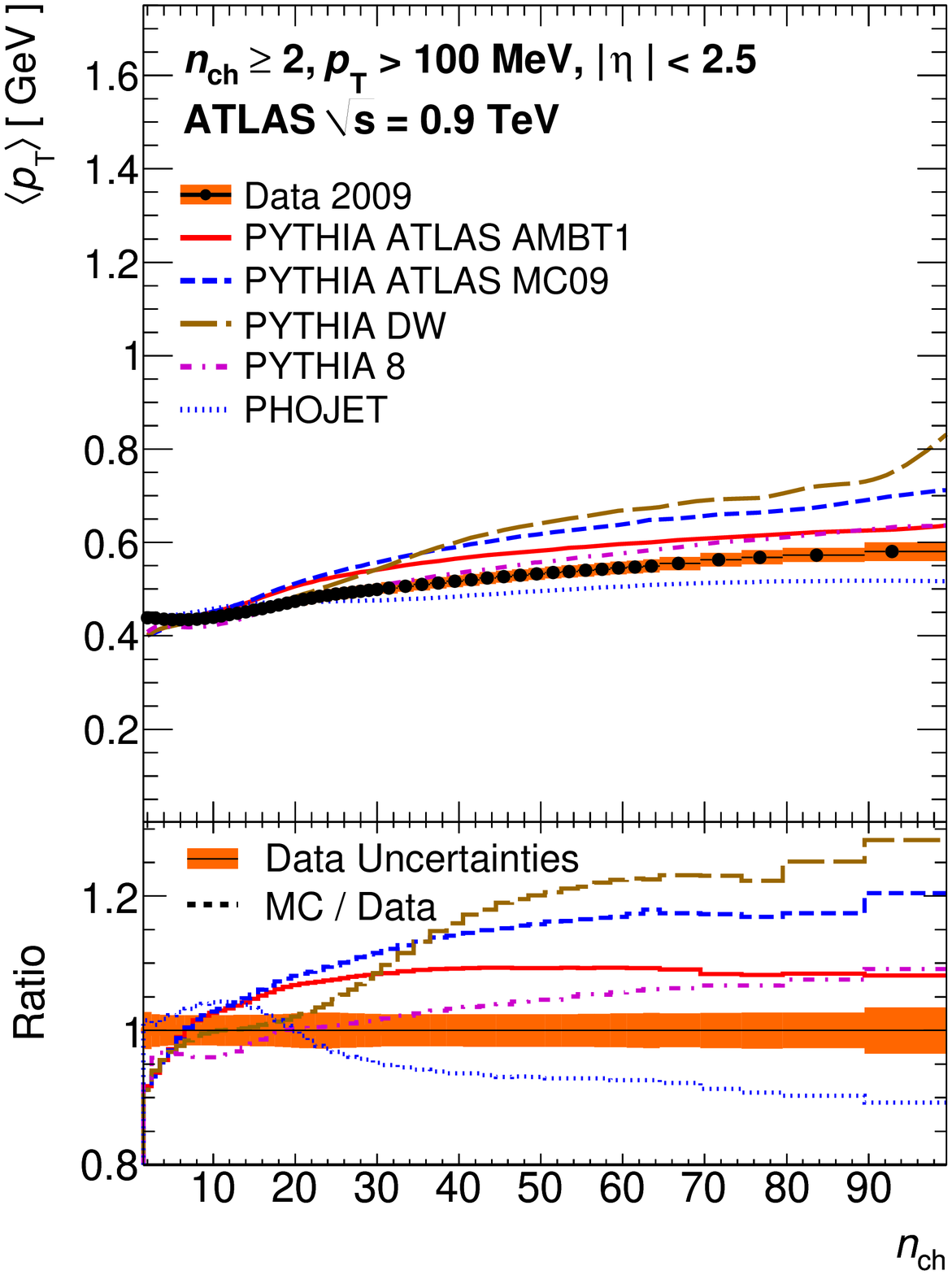}}	
	\subfigure[\label{meanpt_7_pt100}]{\includegraphics[width=0.43\textwidth]{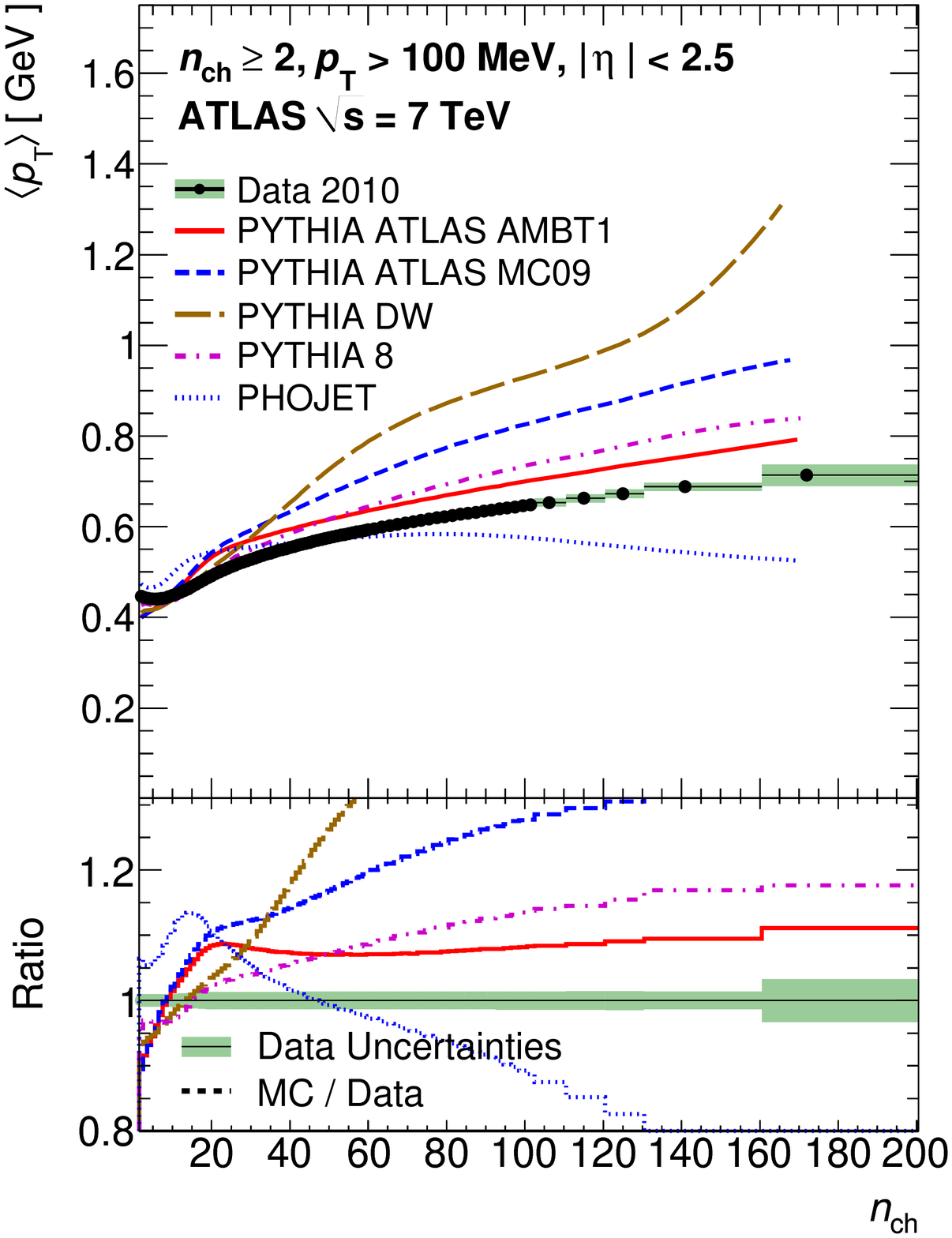}}	

\caption{Average transverse momentum as a function of the number of charged particles in the event 
for events with $\nch~\geq~2$, $\pta~>~100$~MeV 
and $|\eta|~<~2.5$ at \sqn\ (a) and \sqs\ (b). 
The dots represent the data and the curves the predictions from different MC models. 
The vertical bars represent the statistical uncertainties,
while the shaded areas show statistical and systematic uncertainties added in quadrature.
The bottom inserts show the ratio of the MC over the data. The values of the ratio histograms refer to the bin centroids.}
\label{fig:meanpt_2}
\end{center}
\end{figure}

The final set of distributions discussed in the main part of this paper is the average transverse momentum as a function of particle multiplicity.
The measurement of \meanpt\ as a function of charged multiplicity at  \sqt\  is not shown because different track reconstruction methods are used for determining the \pta\ and multiplicity distributions, as discussed in Sec.~\ref{sec:2.36config}.
Figure~\ref{fig:meanpt_1} shows the results for events with $\nch~\geq~1$, $\pta~>~500$~MeV and $|\eta|~<~2.5$.
At 900~GeV the slope vs. \nch\ for high values of \nch\ seems to be well described by most models but the absolute value is best modelled by \py 6 DW. 
At the highest centre-of-mass energy above 20 particles the models vary widely both in slope and in absolute value; at low values of \nch\ none of the models describe the data very well.
In the more inclusive phase-space region, Fig.~\ref{fig:meanpt_2}a and b, the models vary widely, especially at high~$\sqrt{s}$.

% mean pt vs. dnch
% pt > 500

% pt > 100

%------------------------------
\subsection{ $d\nch / d\eta$ at $\eta$ = 0 }

\begin{figure}[htb!]
\begin{center}
\includegraphics[width=0.7\textwidth]{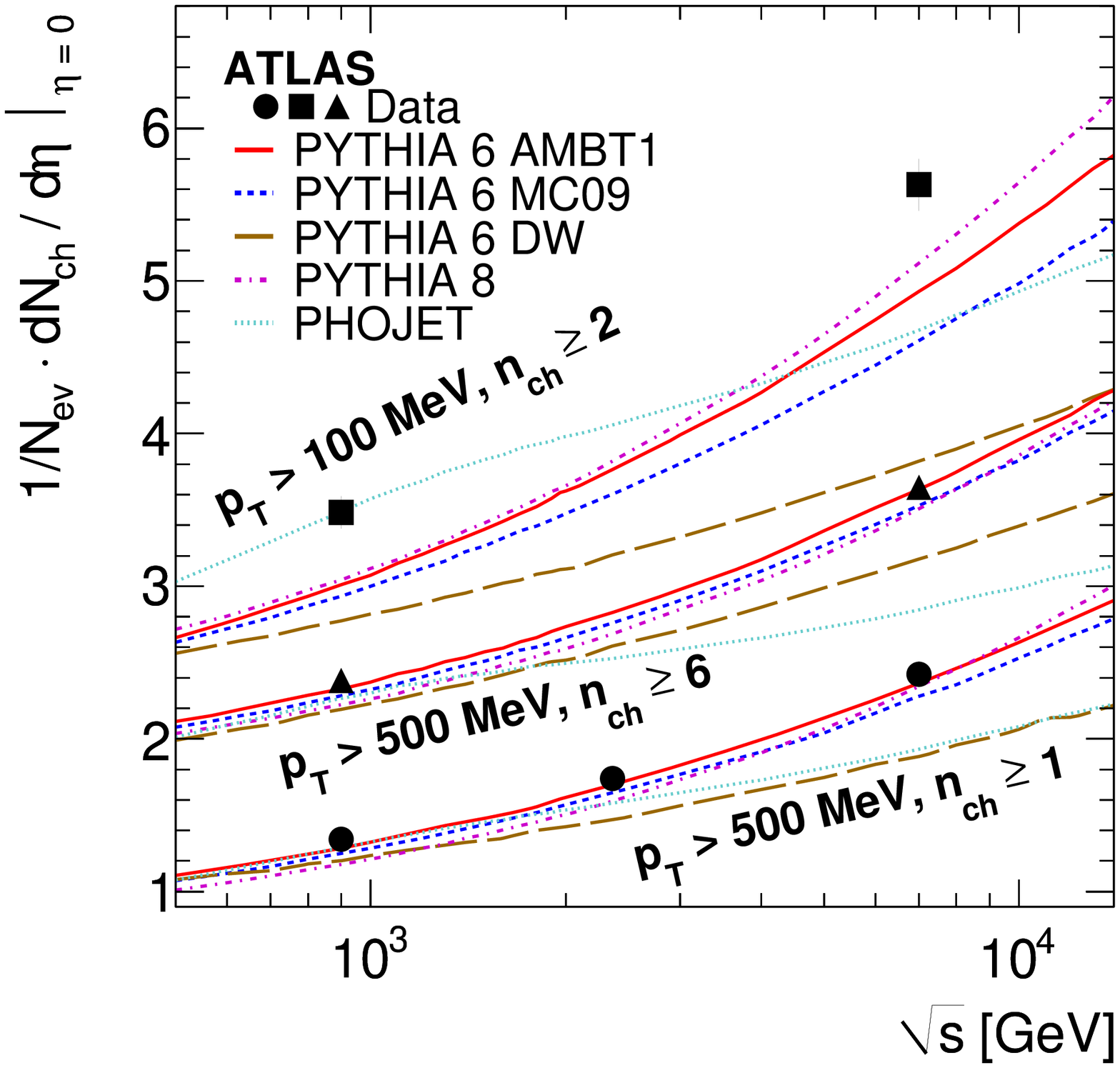}
\caption{The average charged-particle multiplicity per unit of rapidity for $\eta$ = 0 as a function of the centre-of-mass energy. The results with $\nch~\geq~2 $ within the kinematic range $\pta~>~100$~MeV and $|\eta|~<~2.5$ are shown alongside the results with $\nch~\geq~1 $ within the kinematic range $\pta~>~500$~MeV and $|\eta|~<~2.5$ at 0.9, 2.36 and 7~TeV. 
The data are compared to various particle level MC predictions.
The vertical error bars on the data represent the total uncertainty.}
\label{fig:sqrts}
\end{center}
\end{figure}

The mean number of charged particles in the central region is computed by averaging over $|\eta|~<~0.2$. The values for all three phase-space regions and all energies available are shown in Fig.~\ref{fig:sqrts} and in Table~\ref{tab:dndeta_0}. The result quoted at \sqt\ is the value obtained using the Pixel track method.
The phase-space region with largest minimum \pta\ and highest minimum multiplicity ($\pta > 500$~MeV; $\nch \geq 6$), which is the region with the least amount of diffraction, is the one where the models vary the least and the energy extrapolations of most models agree the best with the data. However, in this region the energy extrapolation of \py6 and \pho\ do not agree with the data.
For the most inclusive measurements, none of the models agree with the data and the spread at 7~TeV in the expected values is almost one third of the mean predicted value. 
The observed value is significantly higher at this energy than any of the models.

  \begin{table}[h!]
	\begin{center}
\begin{tabular}{ | c | c | c | c|}
\hline
\hline
Phase-Space Region & Energy & \multicolumn{2}{c|}{$d\nch / d\eta$ at $\eta$ = 0}  \\ 
                        & (TeV) & Measured & \py 6 AMBT1 MC\\ 
\hline
 \multirow{2}{*}{ $\nch~\geq~2$, $\pta~>~100$~MeV} & 0.9 & 3.483 $\pm$ 0.009 (stat) $\pm$ 0.106 (syst)& 3.01 \\ 
                                                                                         & 7    & 5.630 $\pm$ 0.003 (stat) $\pm$ 0.169 (syst)& 4.93 \\
                                                                                                                   
\hline
 \multirow{3}{*}{ $\nch~\geq~1$, $\pta~>~500$~MeV}& 0.9 & 1.343 $\pm$ 0.004 (stat) $\pm$ 0.027 (syst)  &1.28  \\ 
                                                                                        & 2.36 & 1.74 $\pm$ 0.019 (stat) $\pm$ 0.058 (syst) & 1.70  \\
                                                                                        & 7    & 2.423 $\pm$ 0.001 (stat) $\pm$ 0.050 (syst) & 2.36 \\
\hline
 \multirow{2}{*}{ $\nch~\geq~6$, $\pta~>~500$~MeV} & 0.9 & 2.380 $\pm$ 0.009 (stat) $\pm$ 0.027 (syst) & 2.33  \\ 
                                                                                          & 7    &  3.647 $\pm$ 0.002 (stat) $\pm$ 0.052 (syst) & 3.63 \\
\hline
\hline
\end{tabular}
\caption{
$d\nch / d\eta$ at $\eta$ = 0 for the three different phase-space regions considered in this paper for the energies where results are available.
For MC, sufficient statistics were generated such that the statistical uncertainty is smaller than the last digit quoted.
\label{tab:dndeta_0}}
\end{center}
\end{table}

% sqrt(s) plot

%---------------------------------------------------------

\subsection{Extrapolation to \pta\ = 0}
\begin{figure}[htb!]
\begin{center}
\includegraphics[width=0.7\textwidth]{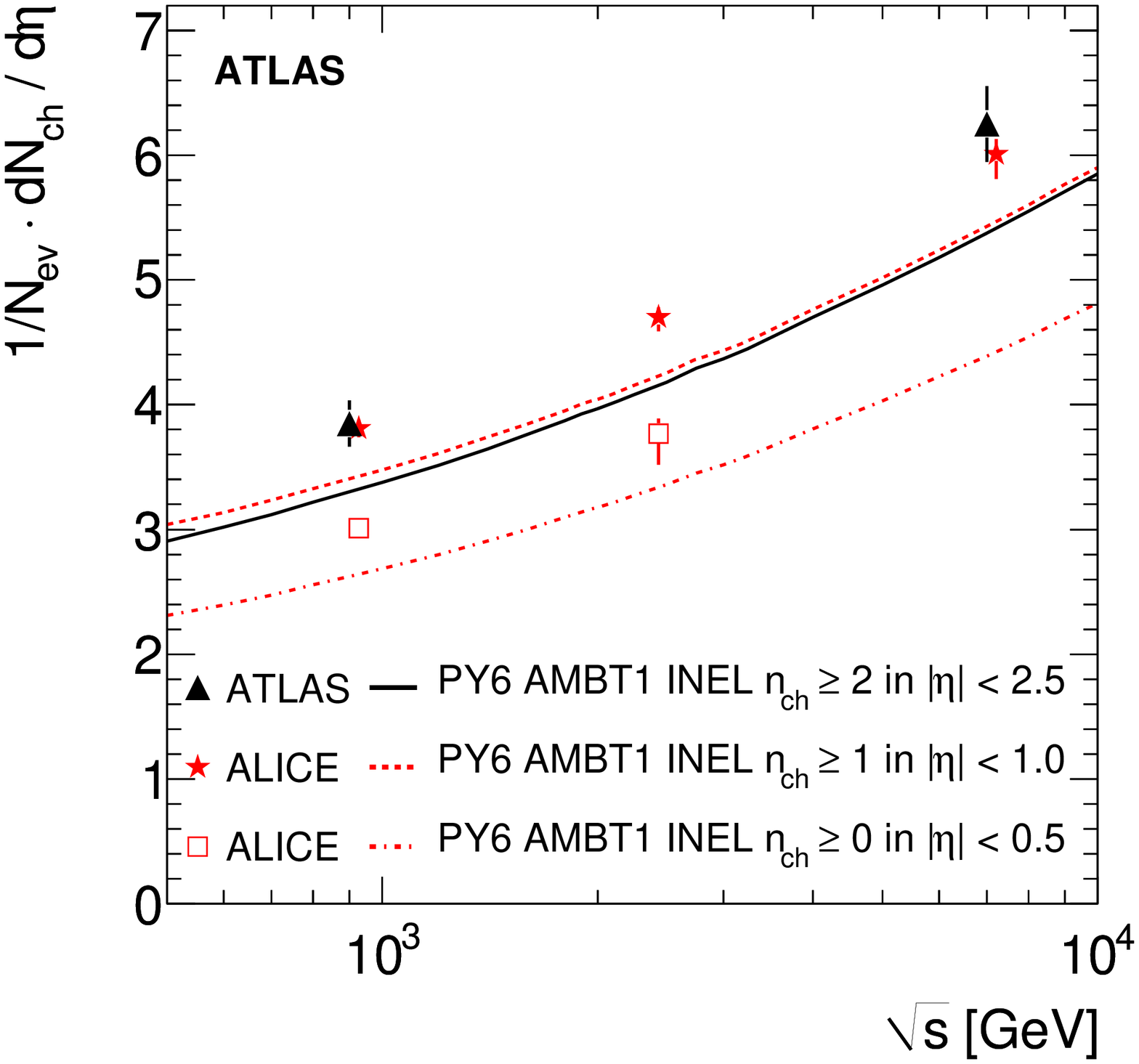}
\caption{The average charged-particle multiplicity per unit of rapidity as a function of the centre-of-mass energy.
The ATLAS results are for $\nch~\geq~2$ in the region $|\eta|~<~2.5$.
For comparison ALICE results for $\nch~\geq~1$ in the region $|\eta|~<~1.0$ and $\nch~\geq~0$ in the region $|\eta|~<~0.5$ are shown.
It should be noted that the ALICE points have been slightly shifted horizontally for clarity.
The data points are compared to \py 6 AMBT1 predictions for the same phase-space regions.
}
\label{fig:alice_comp}
\end{center}
\end{figure}

The mean multiplicities of charged-particles with  $\pta~>~ 100$~MeV within the full $|\eta | < 2.5$ region are computed as the mean of the distributions shown in Fig.~\ref{fig:dndeta_2}a and b.
They are found to be
3.614 $\pm$ 0.006 (stat) $\pm$ 0.170 (syst)
at \sqn\
and
5.881 $\pm$ 0.002 (stat) $\pm$ 0.276 (syst)
at \sqs.
Multiplying these numbers by the model-dependent scale factors obtained in Sec.~\ref{sec:pt_extrap}, 
the averaged inclusive charged-particle multiplicity for events with two or more particles is found to be
3.849 $\pm$ 0.006 (stat) $\pm$ 0.185 (syst) at \sqn\
and
6.252 $\pm$ 0.002 (stat) $\pm$ 0.304 (syst) at \sqs.
This result is interpreted as the average total inelastic multiplicity for events with two or more particles within $| \eta | < 2.5$.
Figure~\ref{fig:alice_comp} compares these results to recently published ALICE results~\cite{alice_mb2,alice_mb} for inclusive inelastic as well as inelastic with more than one particle. The ALICE results are quoted as averages over $|\eta|~<~1.0$ and $|\eta|~<~0.5$, respectively.

%\clearpage
%%%%%%%%%%%%%%%%%%%%%%%%%%%%
\section{Conclusions}\label{sec:concl}

Charged-particle multiplicity measurements with the ATLAS detector using the first collisions delivered by the LHC during 2009 and 2010 are presented.
Based on over three hundred thousand proton-proton inelastic interactions at 900~GeV, just under six thousand at 2.36~TeV and over ten million at 7~TeV, 
the properties of events in three well-defined phase-space regions
 were studied.  
The data were corrected with minimal model dependence to obtain inclusive distributions.
The selected kinematic range and the precision of this analysis highlight clear differences between Monte Carlo models and the measured distributions.  
In all the kinematic regions considered, the particle multiplicities are higher than predicted by the Monte Carlo models.

The three different phase-space regions studied, from the most inclusive to the one with the smallest diffractive contribution, highlight various aspects of the charged-particle spectra.
In general, the agreement between the models and the data is better in the phase-space regions with higher minimum \pta\ cutoff, where diffractive contributions are less significant.

For the \sqn\ measurements with the \pta\ threshold of 500~MeV, these results supersede the results presented in~\cite{MB1}.

\section{Acknowledgements}

We wish to thank CERN for the efficient commissioning and operation of the LHC during this initial high-energy data-taking period as well as the support staff from our institutions without whom ATLAS could not be operated efficiently.

We acknowledge the support of ANPCyT, Argentina; YerPhI, Armenia; ARC, Australia; BMWF, Austria; ANAS, Azerbaijan; SSTC, Belarus; CNPq and FAPESP, Brazil; NSERC, NRC and CFI, Canada; CERN; CONICYT, Chile; CAS, MOST and NSFC, China; COLCIENCIAS, Colombia; MSMT CR, MPO CR and VSC CR, Czech Republic; DNRF, DNSRC and Lundbeck Foundation, Denmark; ARTEMIS, European Union; IN2P3-CNRS, CEA-DSM/IRFU, France; GNAS, Georgia; BMBF, DFG, HGF, MPG and AvH Foundation, Germany; GSRT, Greece; ISF, MINERVA, GIF, DIP and Benoziyo Center, Israel; INFN, Italy; MEXT and JSPS, Japan; CNRST, Morocco; FOM and NWO, Netherlands; RCN, Norway;  MNiSW, Poland; GRICES and FCT, Portugal;  MERYS (MECTS), Romania;  MES of Russia and ROSATOM, Russian Federation; JINR; MSTD, Serbia; MSSR, Slovakia; ARRS and MVZT, Slovenia; DST/NRF, South Africa; MICINN, Spain; SRC and Wallenberg Foundation, Sweden; SER,  SNSF and Cantons of Bern and Geneva, Switzerland;  NSC, Taiwan; TAEK, Turkey; STFC, the Royal Society and Leverhulme Trust, United Kingdom; DOE and NSF, United States of America.  

The crucial computing support from all WLCG partners is acknowledged gratefully, in particular from CERN and the ATLAS Tier-1 facilities at TRIUMF (Canada), NDGF (Denmark, Norway, Sweden), CC-IN2P3 (France), KIT/GridKA (Germany), INFN-CNAF (Italy), NL-T1 (Netherlands), PIC (Spain), ASGC (Taiwan), RAL (UK) and BNL (USA) and in the Tier-2 facilities worldwide.

We thank Peter Skands for useful discussions concerning the AMBT1 tune.

\bibliographystyle{atlasnote}
\bibliography{MinBias2Paper}

\clearpage
 \appendix
 
\section{Distributions Used in AMBT1 Tuning}\label{sec:ambt1_dist}
Table~\ref{tab:datasets_ATLAS} and~\ref{tab:datasets_Tev} show the list of all distributions from ATLAS and the Tevatron, respectively, used in the ATLAS Minimum Bias Tune 1 (AMBT1).
The Analysis column refers to  the event selection used in the particular analysis.
The Tuning range column refers to the portion of the phase-space region that is considered for the tune.

\begin{table}[h!]
\begin{center}
\begin{tabular}{cccc}
\hline
Analysis & Observable & Tuning range \\
\hline
ATLAS $0.9~TeV$, minimum bias, $\nch\ge6$ & $\frac{1}{\nev} \cdot  \frac{\mathrm{d} \Nch}{\mathrm{d} \eta}$ & $-2.5 < \eta < 2.5$ \\
ATLAS $0.9~TeV$, minimum bias, $\nch\ge6$ & $\frac{1}{\nev}\cdot \frac{1}{2 \pi  p_\mathrm{T}} \cdot \frac{\mathrm{d}^2 \Nch}{\mathrm{d} \eta \mathrm{d} p_\mathrm{T}}$ & $\pta \ge5.0$~GeV \\
ATLAS $0.9~TeV$, minimum bias, $\nch\ge6$ & $\frac{1}{\nev} \cdot \frac{\mathrm{d} \nev}{\mathrm{d} \nch}$ & $\nch \ge20$ \\
ATLAS $0.9~TeV$, minimum bias, $\nch\ge6$ & $\langle p_\mathrm{T}\rangle ~ {\mathrm vs.} ~ \nch{\rm}$ & $\nch \ge10$\\
\hline
ATLAS $0.9~TeV$,  UE in minimum bias & $\langle \frac{\mathrm{d}^2N_{\mathrm{ch}}}{\mathrm{d}\eta \mathrm{d}\phi}\rangle$ vs. $\pta^\mathrm{lead}$ (towards) & $\pta^\mathrm{lead}\ge 5.5~GeV$ \\
ATLAS $0.9~TeV$, UE in minimum bias  & $\langle \frac{\mathrm{d}^2N_{\mathrm{ch}}}{\mathrm{d} \eta \mathrm{d}\phi}\rangle$ vs. $\pta^\mathrm{lead}$ (transverse) & $\pta^\mathrm{lead}\ge 5.5~GeV$ \\
ATLAS $0.9~TeV$, UE in minimum bias & $\langle \frac{\mathrm{d}^2N_{\mathrm{ch}}}{\mathrm{d}\eta \mathrm{d}\phi}\rangle$ vs. $\pta^\mathrm{lead}$ (away) & $\pta^\mathrm{lead}\ge 5.5~GeV$ \\
ATLAS $0.9~TeV$, UE in minimum bias & $\langle \frac{\mathrm{d}^2\sum\pta}{\mathrm{d} \eta \mathrm{d}\phi}\rangle$ vs. $\pta^\mathrm{lead}$ (towards) & $\pta^\mathrm{lead}\ge 5.5~GeV$ \\
ATLAS $0.9~TeV$, UE in minimum bias & $\langle \frac{\mathrm{d}^2\sum\pta}{\mathrm{d} \eta \mathrm{d}\phi}\rangle$ vs. $\pta^\mathrm{lead}$ (transverse) & $\pta^\mathrm{lead}\ge 5.5~GeV$ \\
ATLAS $0.9~TeV$, UE in minimum bias & $\langle \frac{\mathrm{d}^2\sum\pta}{\mathrm{d} \eta \mathrm{d}\phi}\rangle$ vs. $\pta^\mathrm{lead}$ (away) & $\pta^\mathrm{lead}\ge 5.5~GeV$ \\ 
\hline
ATLAS $7~TeV$, minimum bias, $\nch\ge6$ & $\frac{1}{\nev} \cdot  \frac{\mathrm{d} \Nch}{\mathrm{d} \eta}$ &$ -2.5 < \eta < 2.5$ \\
ATLAS $7~TeV$, minimum bias, $\nch\ge6$ & $\frac{1}{\nev}\cdot \frac{1}{2 \pi  p_\mathrm{T}} \cdot \frac{\mathrm{d}^2 \Nch}{\mathrm{d} \eta \mathrm{d} p_\mathrm{T}}$ & $\pta \ge5.0$~GeV \\
ATLAS $7~TeV$, minimum bias, $\nch\ge6$ & $\frac{1}{\nev} \cdot \frac{\mathrm{d} \nev}{\mathrm{d} \nch}$ & $\nch \ge40$ \\
ATLAS $7~TeV$, minimum bias, $\nch\ge6$ & $\langle p_\mathrm{T}\rangle ~ {\mathrm vs.} ~ \nch{\rm}$ & $\nch \ge10$\\
\hline
ATLAS $7~TeV$,  UE in minimum bias & $\langle \frac{\mathrm{d}^2N_{\mathrm{ch}}}{\mathrm{d} \eta \mathrm{d}\phi}\rangle$ vs. $\pta^\mathrm{lead}$ (towards) & $\pta^\mathrm{lead}\ge 10~GeV$ \\
ATLAS $7~TeV$, UE in minimum bias & $\langle \frac{\mathrm{d}^2N_{\mathrm{ch}}}{\mathrm{d}\eta \mathrm{d}\phi}\rangle$ vs. $\pta^\mathrm{lead}$ (transverse) & $\pta^\mathrm{lead}\ge 10~GeV$ \\
ATLAS $7~TeV$, UE in minimum bias & $\langle \frac{\mathrm{d}^2N_{\mathrm{ch}}}{\mathrm{d}\eta \mathrm{d}\phi}\rangle$ vs. $\pta^\mathrm{lead}$ (away) & $\pta^\mathrm{lead}\ge 10~GeV$ \\
ATLAS $7~TeV$, UE in minimum bias & $\langle \frac{\mathrm{d}^2\sum\pta}{\mathrm{d}\eta\mathrm{d}\phi}\rangle$ vs. $\pta^\mathrm{lead}$ (towards) & $\pta^\mathrm{lead}\ge 10~GeV$ \\
ATLAS $7~TeV$, UE in minimum bias & $\langle \frac{\mathrm{d}^2\sum\pta}{\mathrm{d}\eta\mathrm{d}\phi}\rangle$ vs. $\pta^\mathrm{lead}$ (transverse) & $\pta^\mathrm{lead}\ge 10~GeV$ \\
ATLAS $7~TeV$, UE in minimum bias & $\langle \frac{\mathrm{d}^2\sum\pta}{\mathrm{d}\eta\mathrm{d}\phi}\rangle$ vs. $\pta^\mathrm{lead}$ (away) & $\pta^\mathrm{lead}\ge 10~GeV$ \\
\hline
 \end{tabular}
\end{center}
\caption{ATLAS observables and ranges of distributions used in the AMBT1 tuning.}
\label{tab:datasets_ATLAS}
\end{table}

\begin{table}[h!]
  \begin{center}
  \begin{tabular}{lc}
  \hline
  Observables \\
\hline
  \multicolumn{1}{l}{\textit{CDF Run I underlying event in dijet events}\cite{leadingjets2} (leading jet analysis)} \\
  \Nch\  density vs. leading jet \pta\ (transverse), JET20   \\
  \Nch\ density vs. leading jet \pta\  (toward), JET20        \\
   \Nch\ density vs. leading jet \pta\  (away), JET20         \\
  $\sum \pta$ density vs. leading jet \pta\  (transverse), JET20   \\
  $\sum \pta$ density vs. leading jet \pta\  (toward), JET20       \\
  $\sum \pta$ density vs. leading jet \pta\  (away), JET20         \\
  \Nch\ density vs. leading jet \pta\  (transverse), min bias \\
  \Nch\ density vs. leading jet \pta\  (toward), min bias     \\
  \Nch\ density vs. leading jet \pta\  (away), min bias       \\
  $\sum \pta$ density vs. leading jet \pta\  (transverse), min bias \\
  $\sum \pta$ density vs. leading jet \pta\  (toward), min bias    \\
  $\sum \pta$ density vs. leading jet \pta\ (away), min bias      \\
  \pta\ distribution (transverse), leading $\pta > 5$ GeV \\
  \pta\ distribution (transverse), leading $\pta > 30$ GeV \\

\hline
  \multicolumn{2}{l}{\textit{CDF Run I underlying event in MIN/MAX-cones}\cite{maxmincone} (``MIN-MAX'' analysis)}  \\ 
   $\langle p_T^\mathrm{max} \rangle$ vs. $E_T^\mathrm{lead}$, $\sqrt{s}=1800$~GeV \\
  $\langle p_T^\mathrm{min} \rangle$ vs. $E_T^\mathrm{lead}$, $\sqrt{s}=1800$~GeV \\
  $\langle p_T^\mathrm{diff} \rangle$ vs. $E_T^\mathrm{lead}$, $\sqrt{s}=1800$~GeV \\
  $\langle N_\mathrm{max}\rangle$ vs. $E_T^\mathrm{lead}$, $\sqrt{s}=1800$~GeV \\
  $\langle N_\mathrm{min}\rangle$ vs. $E_T^\mathrm{lead}$, $\sqrt{s}=1800$~GeV \\
  Swiss Cheese $p_T^\mathrm{sum}$ vs. $E_T^\mathrm{lead}$ (2 jets), $\sqrt{s}=1800$~GeV \\
   $\langle p_T^\mathrm{max} \rangle$ vs. $E_T^\mathrm{lead}$, $\sqrt{s}=630$~GeV \\
  $\langle p_T^\mathrm{min} \rangle$ vs. $E_T^\mathrm{lead}$, $\sqrt{s}=630$~GeV \\
  $\langle p_T^\mathrm{diff} \rangle$ vs. $E_T^\mathrm{lead}$, $\sqrt{s}=630$~GeV \\
  Swiss Cheese $\pta^\mathrm{sum}$ vs. $E_T^\mathrm{lead}$ (2 jets), $\sqrt{s}=630$~GeV \\
\hline
  \multicolumn{1}{l}{\textit{D0~Run II dijet angular correlations}\cite{d0dijets}} \\
  Dijet azimuthal angle, $p_T^\mathrm{max} \in $ [75, 100] ~GeV   \\
  Dijet azimuthal angle, $p_T^\mathrm{max} \in $ [100, 130]~GeV  \\
  Dijet azimuthal angle, $p_T^\mathrm{max} \in $ [130, 180] ~GeV  \\
  Dijet azimuthal angle, $p_T^\mathrm{max} > $180~GeV           \\
\hline
  \multicolumn{1}{l}{\textit{CDF Run II minimum bias}\protect\cite{minbias2}} \\ 
  $\langle\pta\rangle$ of charged particles vs.~\Nch, $\sqrt{s}=1960$~GeV  \\

\hline
  \multicolumn{2}{l}{\textit{CDF Run I Z \pta}\protect\cite{Zpt}} \\ 
  $\frac{d\sigma}{d\pta^Z}$ , $\sqrt{s}=1800$~GeV  \\
 
   \hline
\end{tabular}
\end{center}
\caption{Tevatron datasets used in the AMBT1 tuning. No specific cuts on the tuning ranges were made.}
\label{tab:datasets_Tev}
\end{table}

\clearpage
%%%%%%%%%%%%%%%%%%%%%%%%%%%%%%%%

\section{Additional Phase-Space Regions} \label{sec:more_phase_spaces}

Two additional phase-space regions are considered in this appendix:
\begin{itemize}
\item at least twenty charged particles in the kinematic range $|\eta|<$~2.5 and $\pta> 100$~MeV,
\item at least one charged particle in the kinematic range $|\eta|<$~2.5 and $\pta> 2.5$~GeV.
\end{itemize}
The correction procedures as well as methods used to extract the systematic uncertainties are identical to the three phase-space regions presented in the main part of the paper.
The first phase-space region is chosen to be compared with the other diffraction-reduced phase-space region with six particles above 500~MeV and allows the study of the interplay between the number of particles and the \pta , in particular for the study of diffraction models.
The second additional phase-space region is chosen so as to be less influenced by non-perturbative parts of the non-diffractive modeling and to be useful for predicting high-\pta\ particle rates, for example for trigger studies.

\begin{figure}[htb!]
\begin{center}
	\subfigure[\label{dndeta_900_nch20}]{\includegraphics[width=0.43\textwidth]{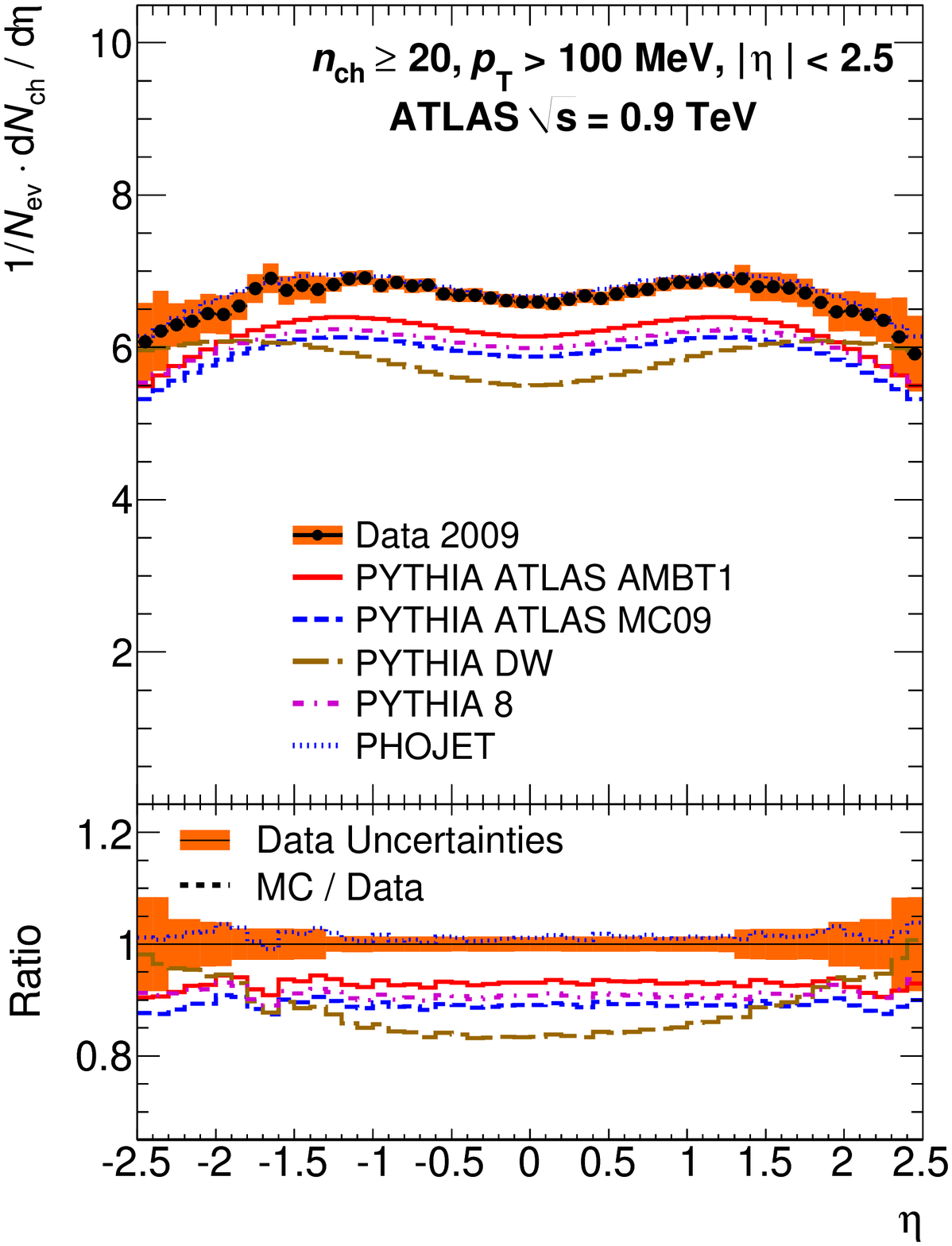}}	
	\subfigure[\label{dndeta_7_nch20}]{\includegraphics[width=0.43\textwidth]{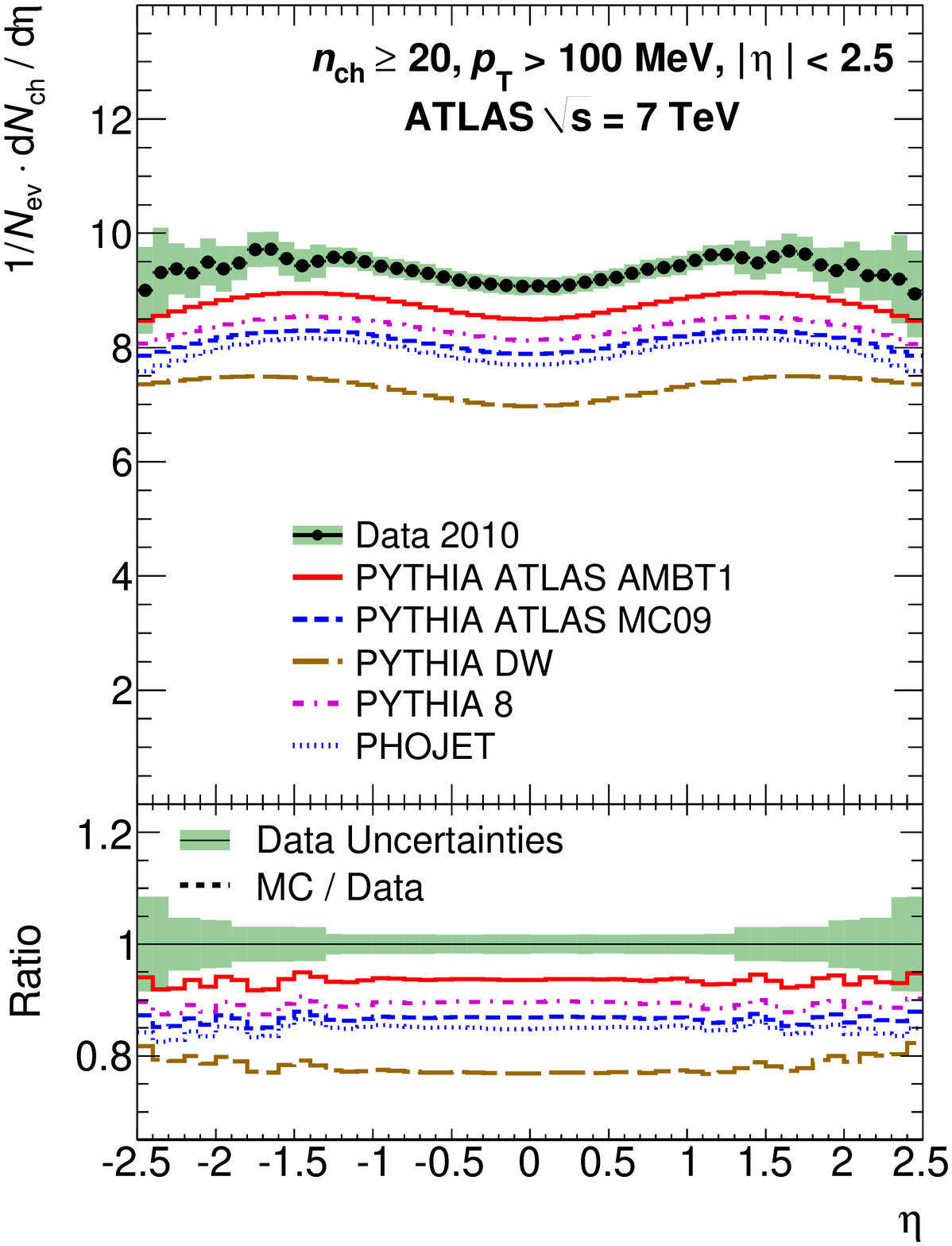}}
	\subfigure[\label{dndeta_900_pt2500}]{\includegraphics[width=0.43\textwidth]{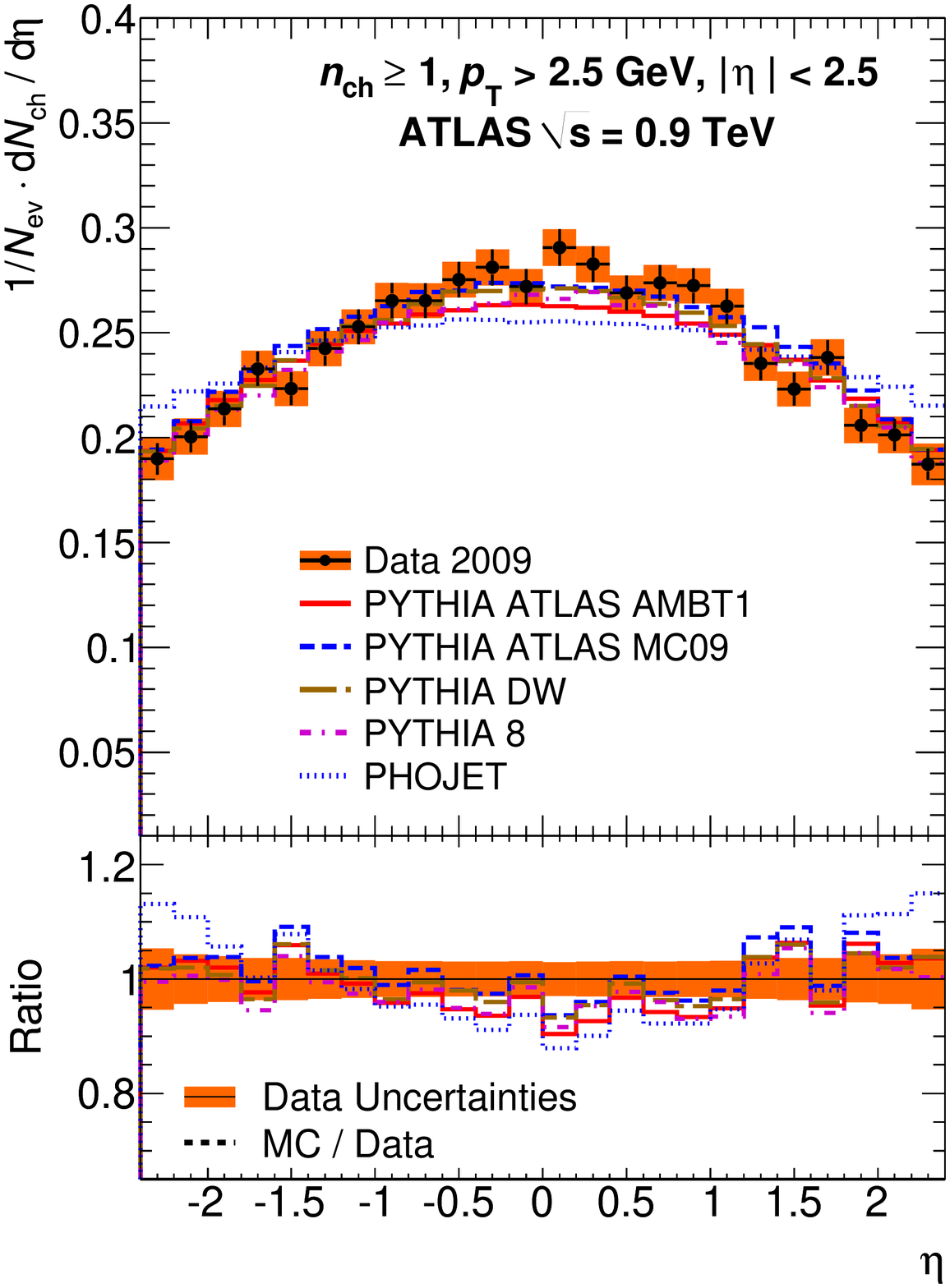}}	
	\subfigure[\label{dndeta_7_pt2500}]{\includegraphics[width=0.43\textwidth]{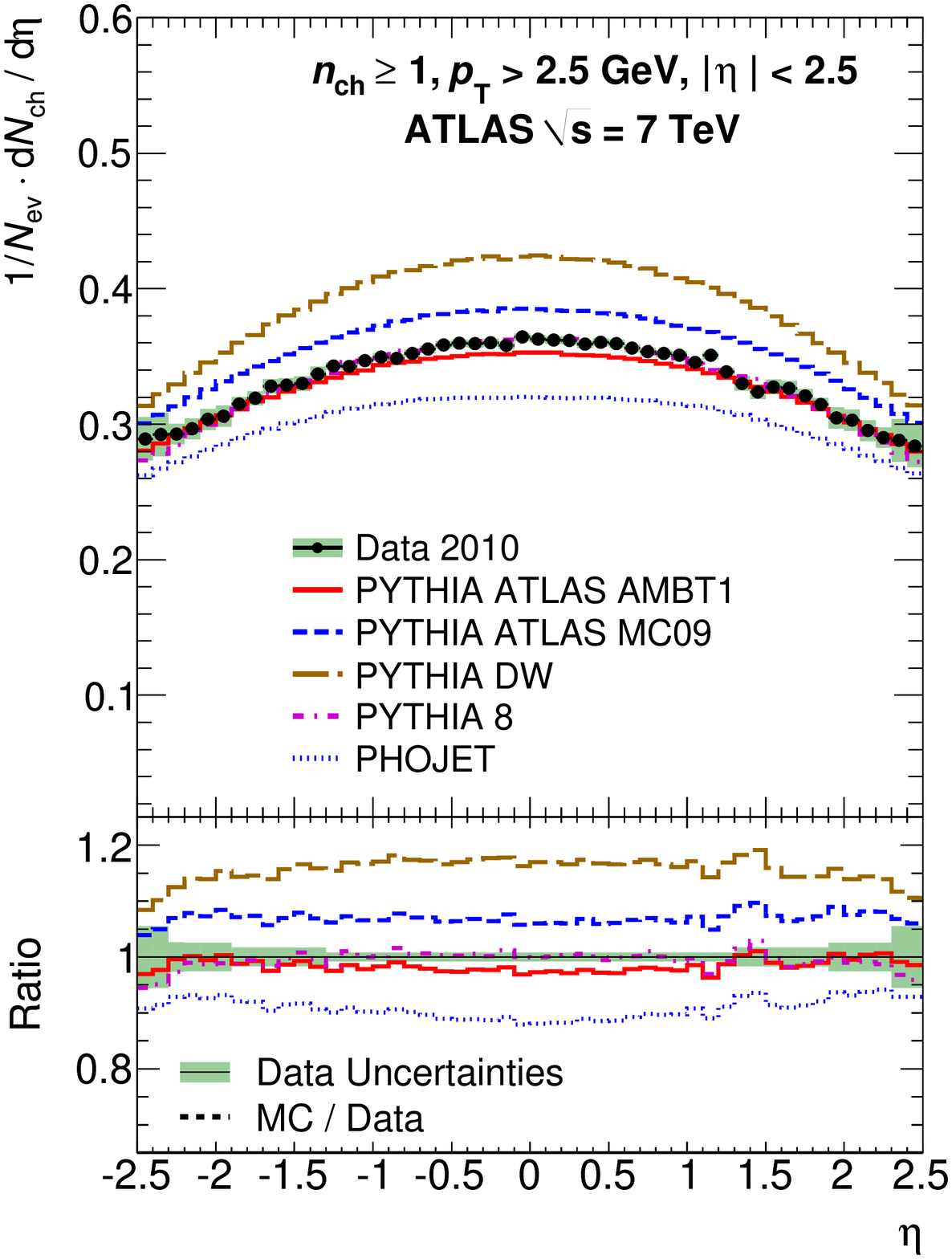}}

\caption{Charged-particle multiplicities as a function of the pseudorapidity
for events with $\nch \geq 20$, $\pta > 100$ MeV (a,b) and $\nch \geq 1$, $\pta > 2.5$~GeV (c,d) and $|\eta| < 2.5$ at \sqn\ (a,c) and \sqs\ (b,d). 
The dots represent the data and the curves the predictions from different MC models. The vertical bars represent the statistical uncertainties,
while the shaded areas show statistical and systematic uncertainties added in quadrature.
The bottom inserts show the ratio of the MC over the data. The values of the ratio histograms refer to the bin centroids.}\label{fig:dndeta_more}
\end{center}
\end{figure}

\begin{figure}[htb!]
\begin{center}
	\subfigure[\label{dndpt_900_nch20}]{\includegraphics[width=0.43\textwidth]{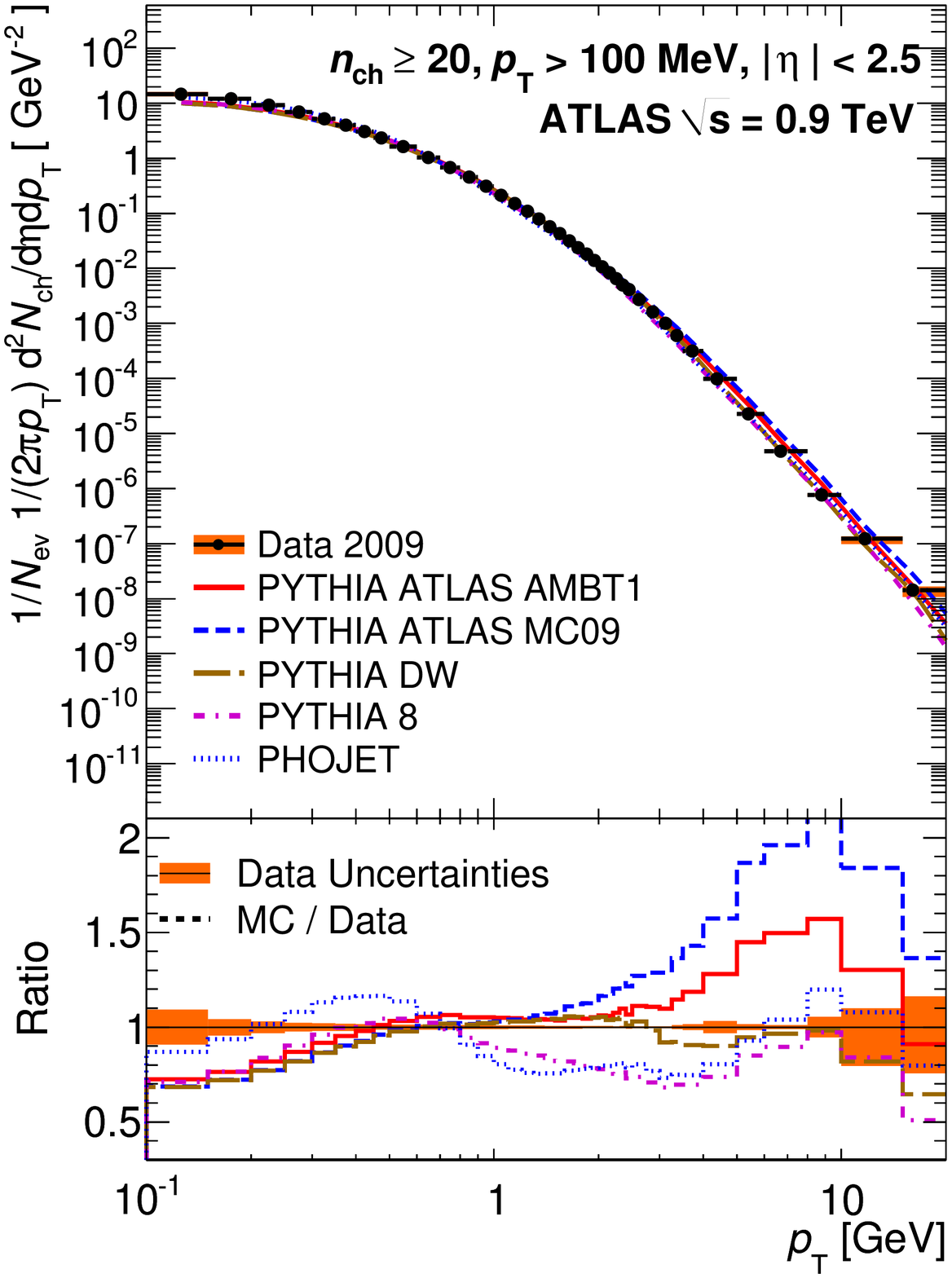}}	
	\subfigure[\label{dndpt_7_nch20}]{\includegraphics[width=0.43\textwidth]{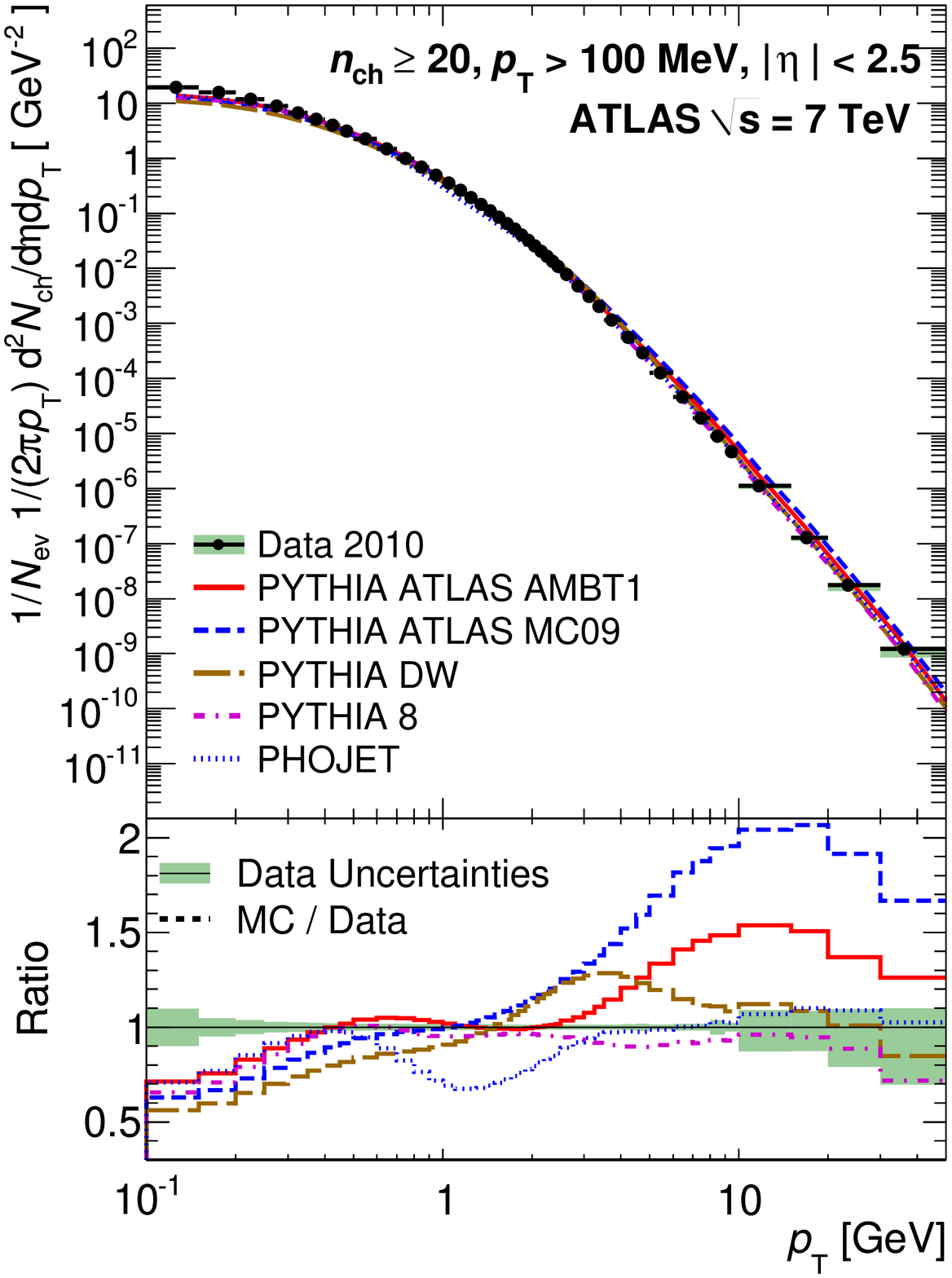}}	
	\subfigure[\label{dndpt_900_pt2500}]{\includegraphics[width=0.43\textwidth]{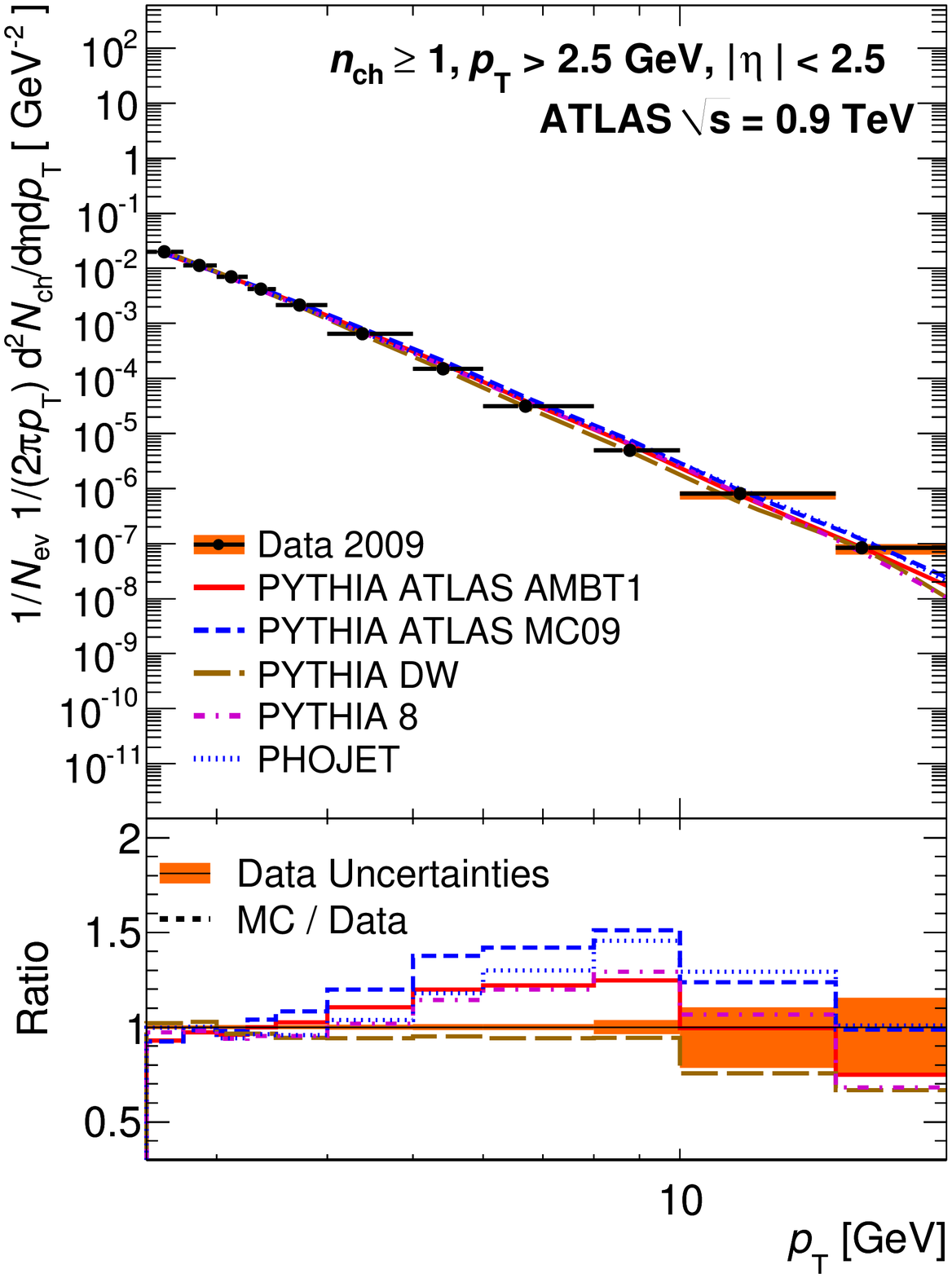}}	
	\subfigure[\label{dndpt_7_pt2500}]{\includegraphics[width=0.43\textwidth]{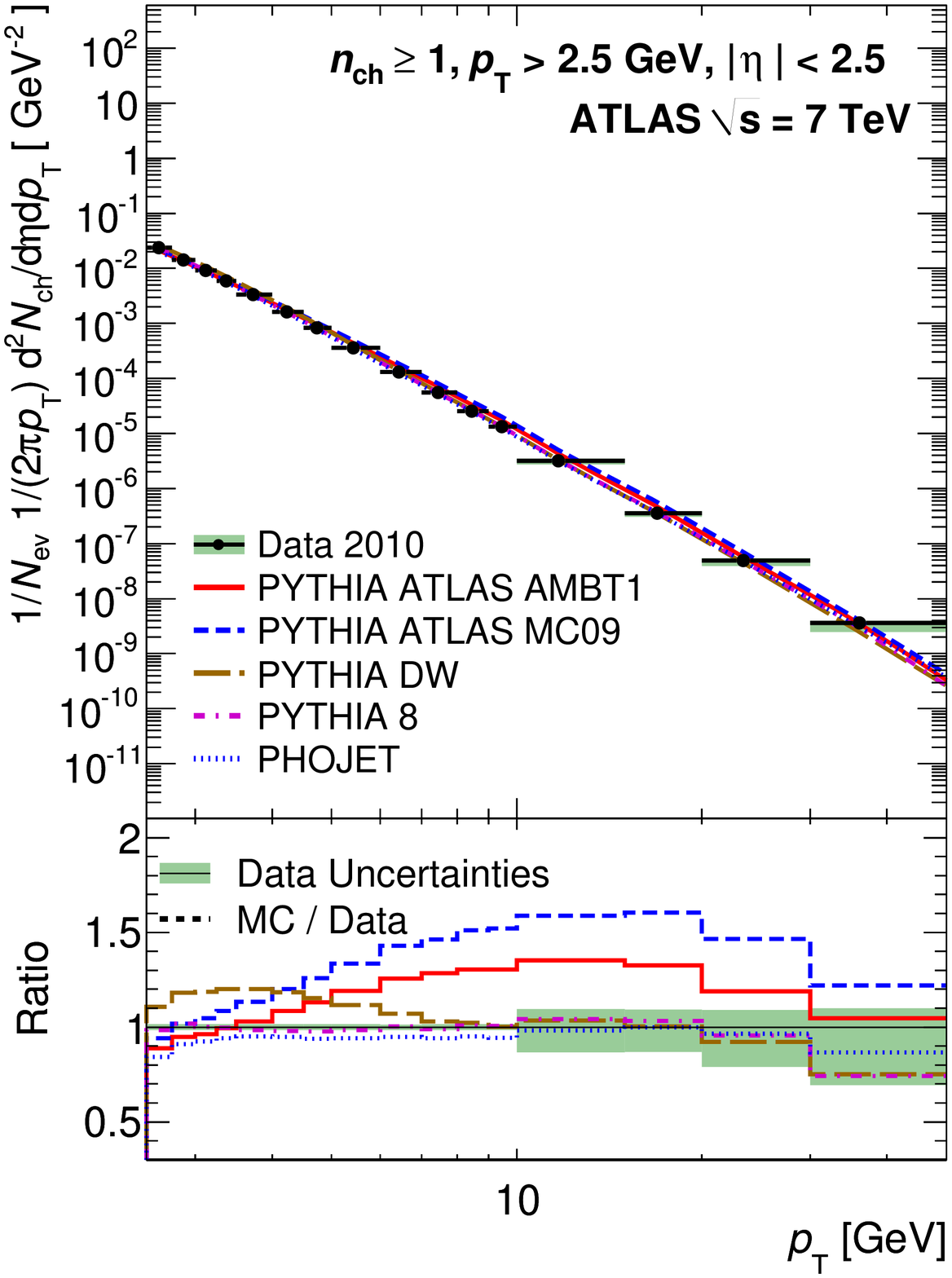}}

\caption{Charged-particle multiplicities as a function of the transverse momentum
for events with $\nch \geq 20$, $\pta > 100$~MeV (a,b) and $\nch \geq 1$, $\pta > 2.5$~GeV (c,d) and $|\eta| < 2.5$ at \sqn\ (a,c) and \sqs\ (b,d). 
The dots represent the data and the curves the predictions from different MC models. The vertical bars represent the statistical uncertainties,
while the shaded areas show statistical and systematic uncertainties added in quadrature.
The bottom inserts show the ratio of the MC over the data. The values of the ratio histograms refer to the bin centroids.}\label{fig:dndpt_more}
\end{center}
\end{figure}

\begin{figure}[htb!]
\begin{center}

	\subfigure[\label{dndnch_900_nch20}]{\includegraphics[width=0.43\textwidth]{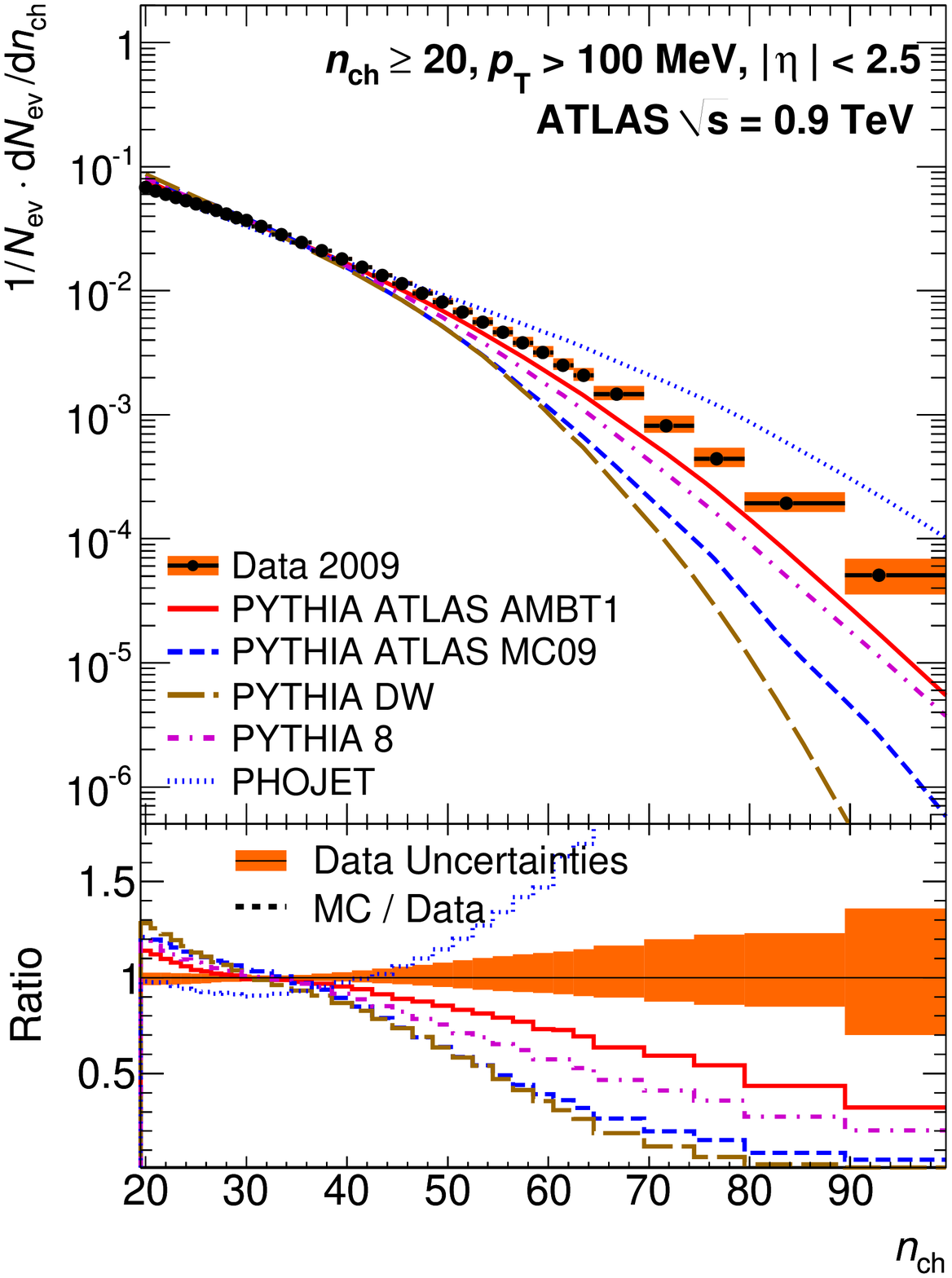}}	
	\subfigure[\label{dndnch_7_nch20}]{\includegraphics[width=0.43\textwidth]{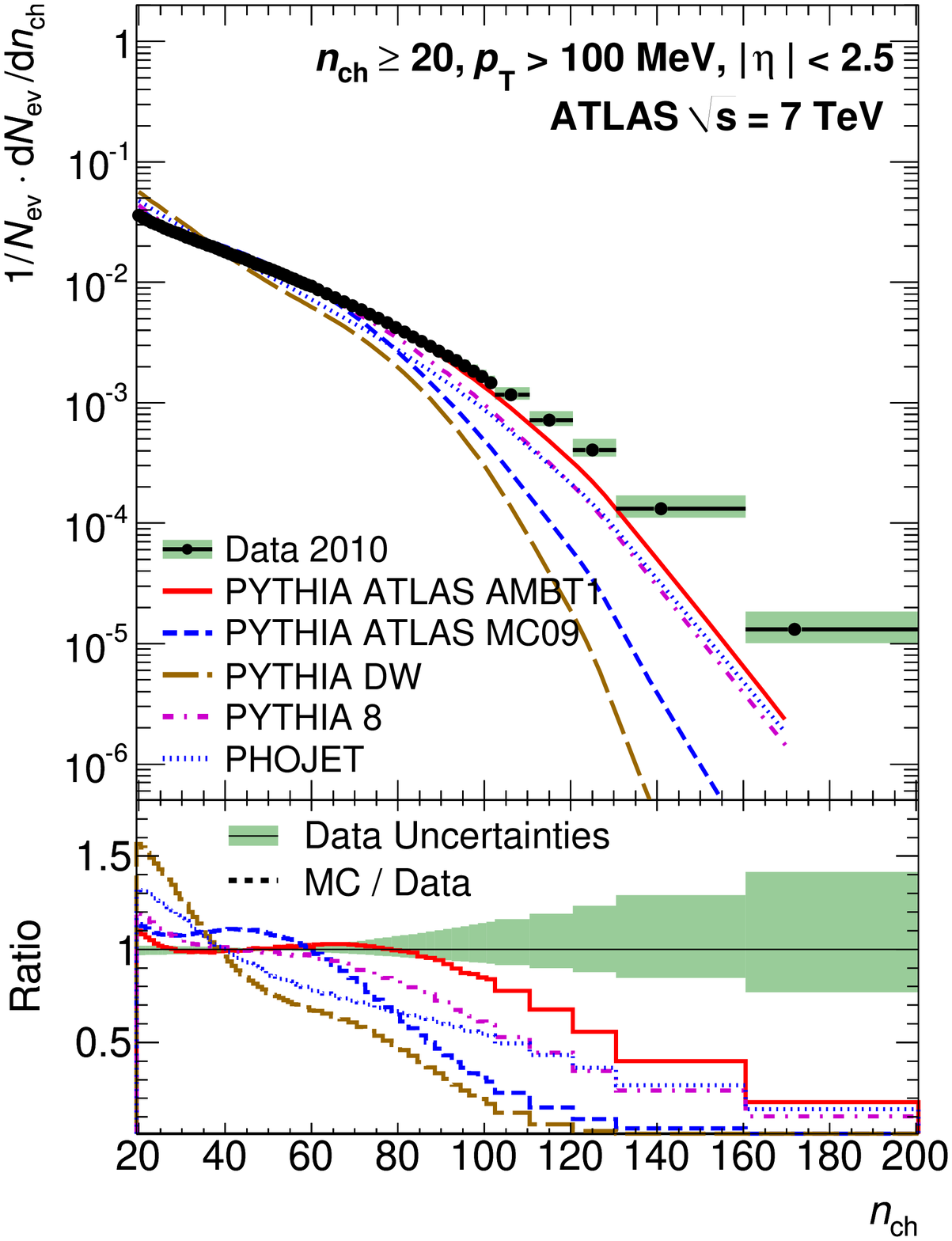}}
	\subfigure[\label{dndnch_900_pt2500}]{\includegraphics[width=0.43\textwidth]{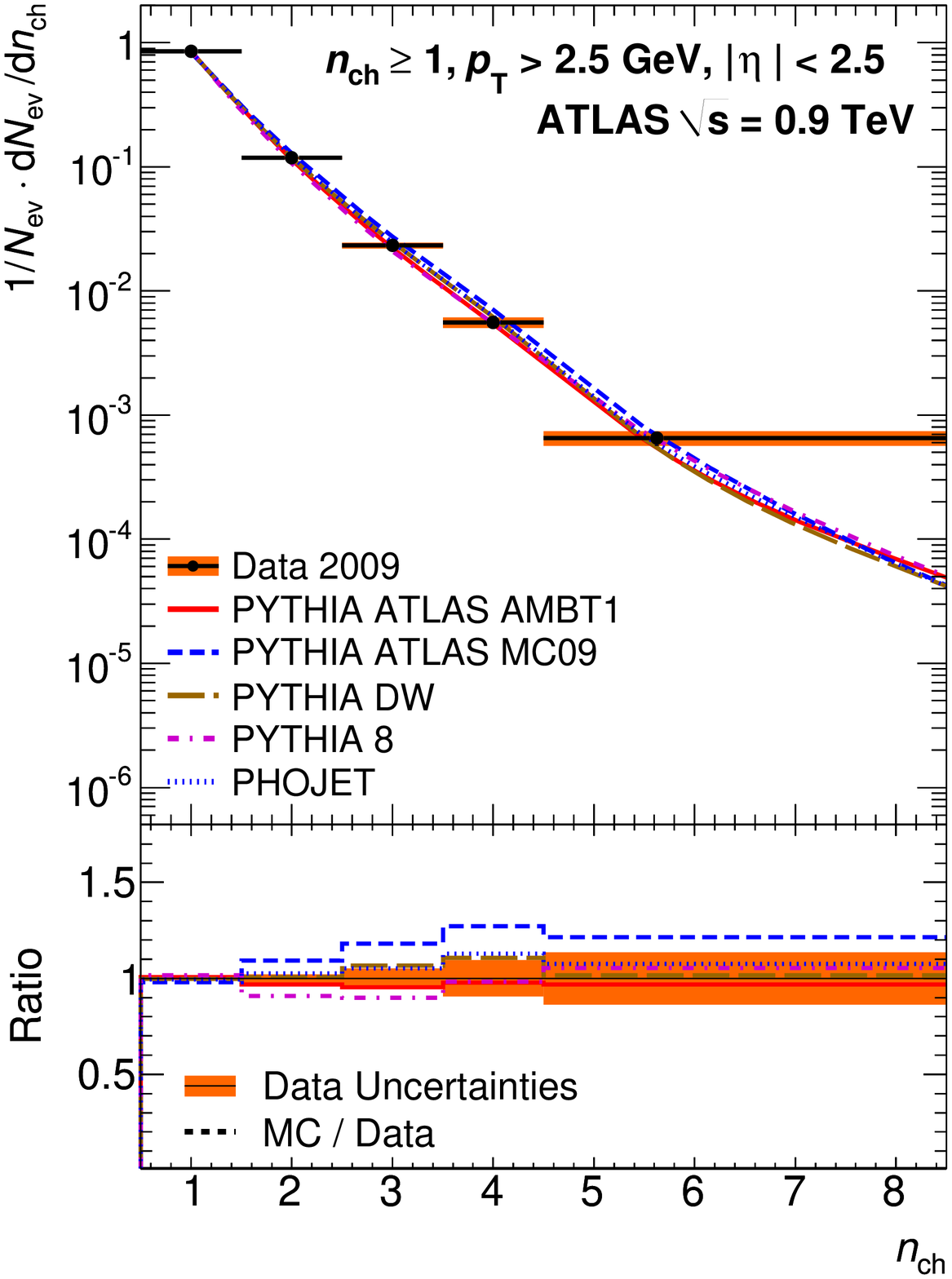}}	
	\subfigure[\label{dndnch_7_pt2500}]{\includegraphics[width=0.43\textwidth]{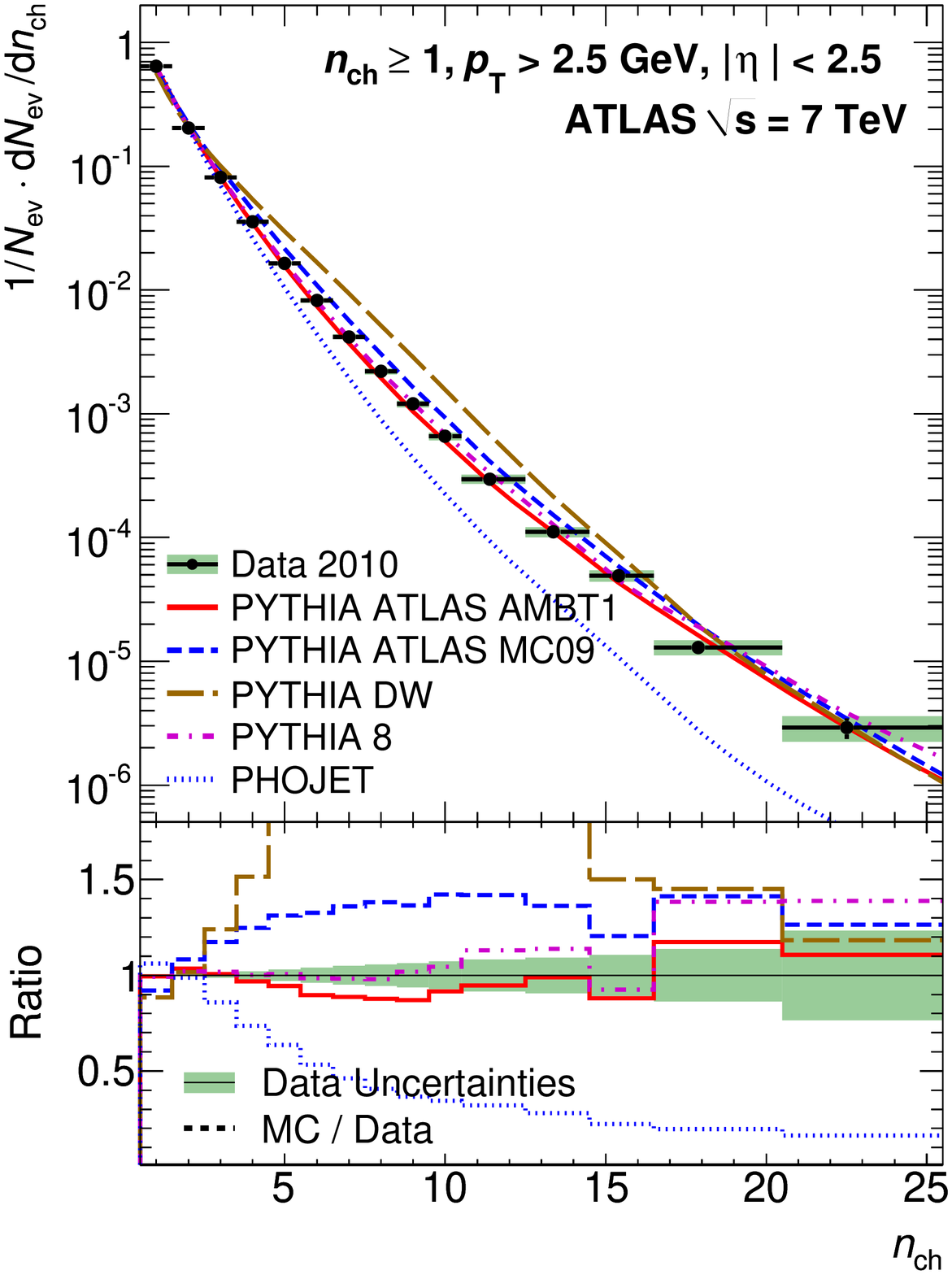}}
\caption{Charged-particle multiplicity distributions
for events with $\nch \geq 20$, $\pta > 100$~MeV (a,b) and $\nch \geq 1$, $\pta > 2.5$~GeV (c,d) and $|\eta| < 2.5$ at \sqn\ (a,c) and \sqs\ (b,d). 
The dots represent the data and the curves the predictions from different MC models. 
The vertical bars represent the statistical uncertainties,
while the shaded areas show statistical and systematic uncertainties added in quadrature.
The bottom inserts show the ratio of the MC over the data. The values of the ratio histograms refer to the bin centroids.}\label{fig:dndnch_more}
\end{center}
\end{figure}

\begin{figure}[htb!]
\begin{center}
	\subfigure[\label{meanpt_900_pt2500}]{\includegraphics[width=0.43\textwidth]{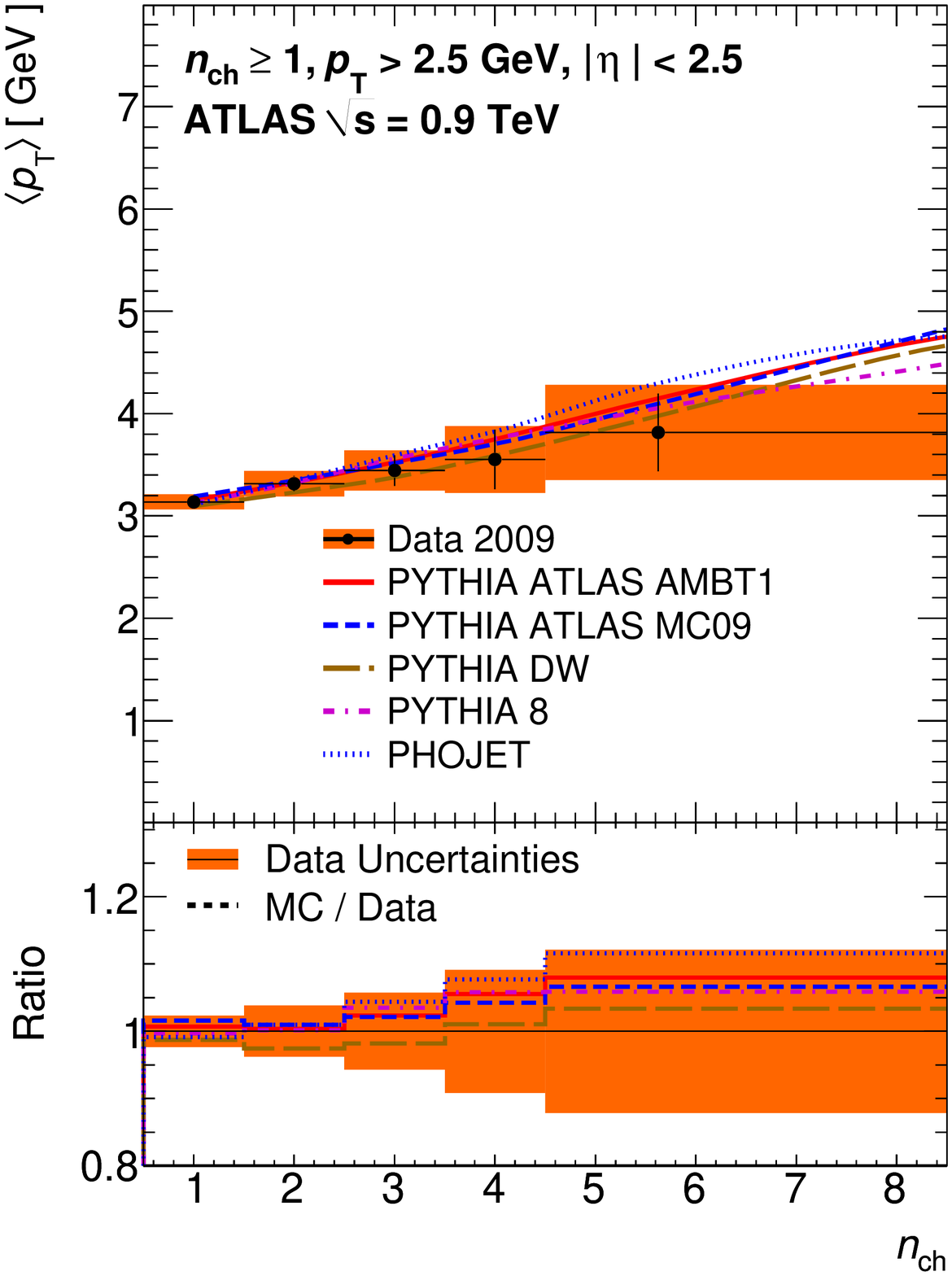}}	
	\subfigure[\label{meanpt_7_pt2500}]{\includegraphics[width=0.43\textwidth]{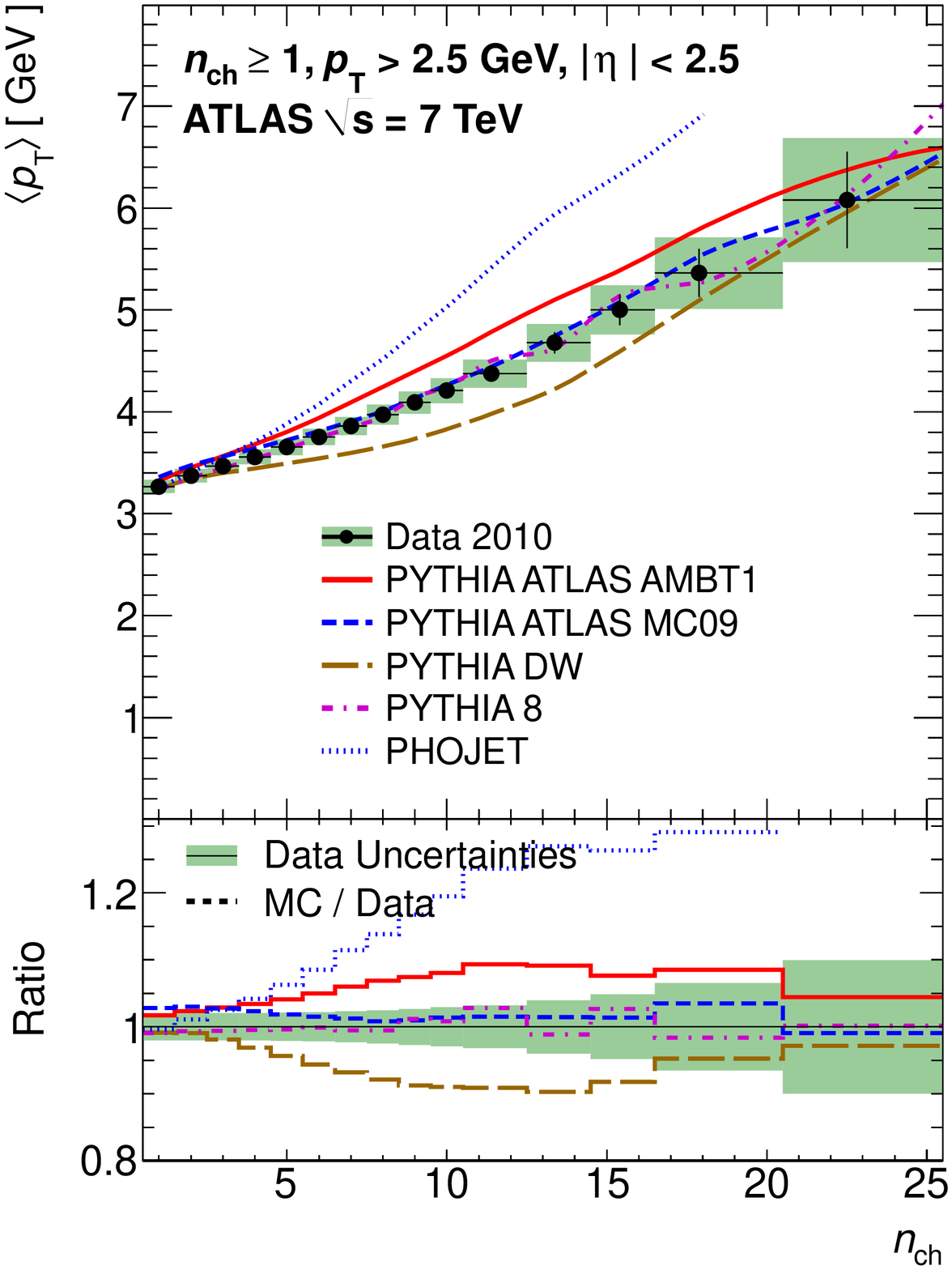}}

\caption{Average transverse momentum as a function of the number of charged particles in the event 
for events with $\nch \geq 1$, $\pta > 2.5$~GeV and $|\eta| < 2.5$ at \sqn\ (a) and \sqs\ (b). 
The dots represent the data and the curves the predictions from different MC models. 
The vertical bars represent the statistical uncertainties,
while the shaded areas show statistical and systematic uncertainties added in quadrature.
The bottom inserts show the ratio of the MC over the data. The values of the ratio histograms refer to the bin centroids.}\label{fig:meanpt_more}
\end{center}
\end{figure}

\begin{figure}[htb!]
\begin{center}
\includegraphics[width=0.9\textwidth]{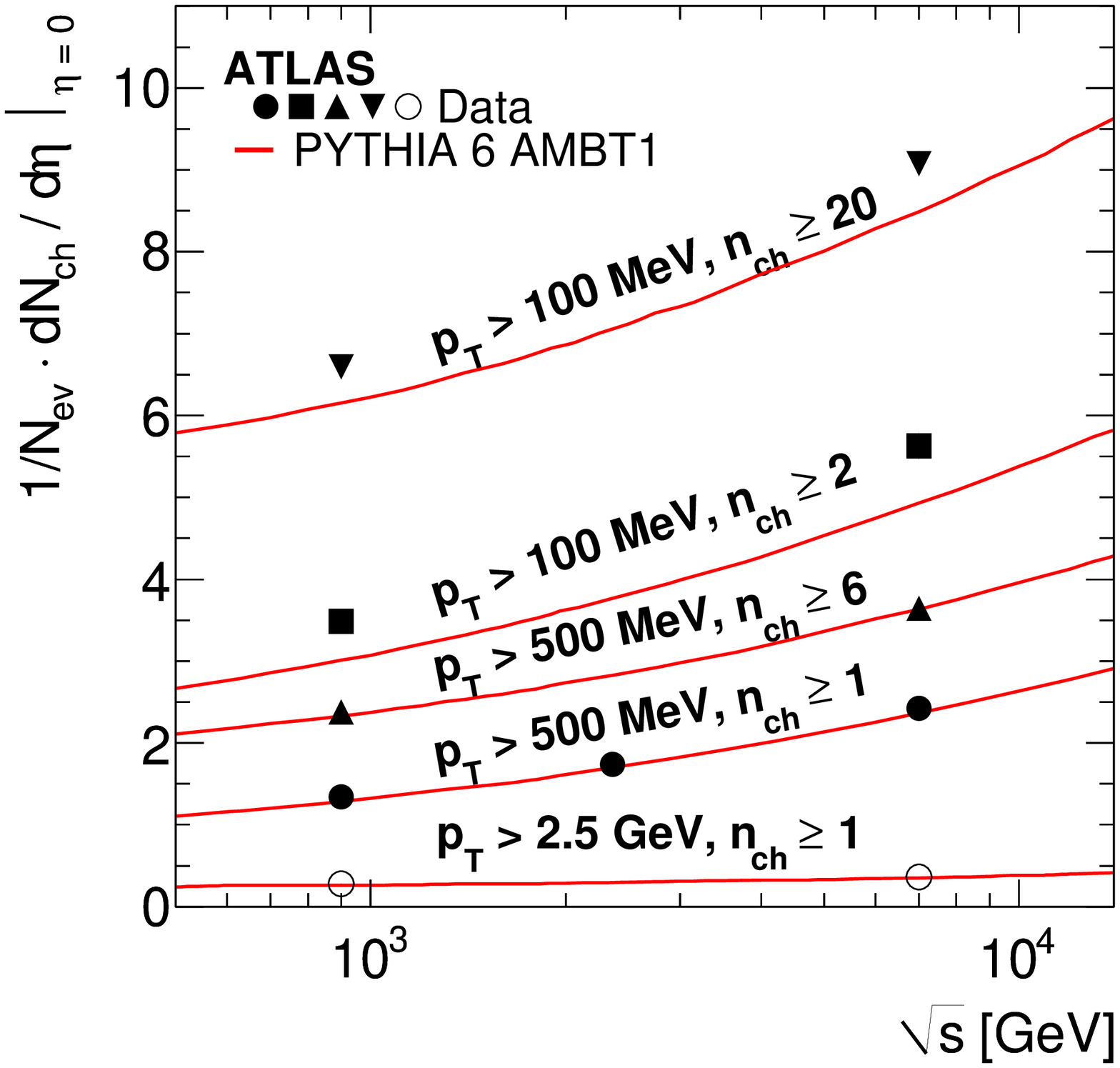}
\caption{
The average charged particle multiplicity per unit of rapidity for $\eta$ = 0 as a function of the centre-of-mass energy. 
All the measured phase-space regions and energies are shown as triangles and compared to predictions from \py6 AMBT1 tune.
The phase-space region label is above the corresponding curves and points.
Combined statistical and systematic uncertainties are approximately equal to or smaller than the data points.
}
\label{fig:sqrts_all}
\end{center}
\end{figure}

Table~\ref{tab:nevents_more} shows the number of selected events and tracks for these two additional phase-space regions at both \sqn\ and \sqs.
Figures~\ref{fig:dndeta_more} to \ref{fig:meanpt_more} show the four kinematic distributions.
Table~\ref{tab:dndeta_0_more} shows the results for the mean track multiplicity at central eta (obtained as the average between $-0.2<\eta<0.2$). Figure~\ref{fig:sqrts_all} shows the mean track multiplicity at central rapidity for all centre-of-mass energies and phase-space regions presented in this paper, along with predictions from \py6 AMBT1.

\begin{table}[h!]
 \begin{center}
   \begin{tabular}{|c | c | c | c | c | c | c | }
   \hline\hline
   \multicolumn{2}{|c}{ Phase-Space Region} &   \multicolumn{2}{|c}{ \sqn} & \multicolumn{2}{|c|}{ {\sqs}}  \\
\nch & min \pta & Events & Tracks & Events & Tracks \\ \hline
20  & 100  MeV & 69,833 & 1,966,059
		& 4,029,563 &  153,553,344\\
1  & 2.5 GeV & 19,016 & 22,233
		& 1,715,637 & 2,690,534 \\
   \hline\hline
   \end{tabular}
   \caption{\label{tab:nevents_more} Number of events and tracks in the two additional phase-space regions and energies considered in this appendix.
}
 \end{center}
\end{table}

% diffraction limited phase-space
% dn/deta

% dn/dpt2

% dN/dnch

% mean pt vs. nch

  \begin{table}[h!]
	\begin{center}
\begin{tabular}{ | c | c | c |}
\hline
\hline
Phase-Space Region & Energy& $d\nch / d\eta$ at $\eta$ = 0 \\ 
                        & (TeV) & Measured \\ 
\hline
 \multirow{2}{*}{ $\nch \geq 20$, $\pta > 100$~MeV} & 0.9 &  6.596 $\pm$ 0.025 (stat) $\pm$ 0.080 (syst)\\ 
                                                                                         & 7    &  9.077 $\pm$ 0.005 (stat) $\pm$ 0.157 (syst) \\
                                                                                                                   
\hline
 \multirow{2}{*}{ $\nch \geq 1$, $\pta > 2.5$~GeV}& 0.9 & 0.281 $\pm$ 0.006 (stat) $\pm$ 0.0005 (syst)   \\ 
                                                                                        & 7    &  0.362 $\pm$ 0.001 (stat) $\pm$ 0.002 (syst) \\
\hline
\hline
\end{tabular}
\caption{
$d\nch / d\eta$ at $\eta$ = 0 for the additional two different phase-space regions considered in this paper for \sqn\ and \sqs.
\label{tab:dndeta_0_more}}
\end{center}
\end{table}

\clearpage
\begin{flushleft}
{\Large The ATLAS Collaboration}

\bigskip

G.~Aad$^{\rm 48}$,
B.~Abbott$^{\rm 111}$,
J.~Abdallah$^{\rm 11}$,
A.A.~Abdelalim$^{\rm 49}$,
A.~Abdesselam$^{\rm 118}$,
O.~Abdinov$^{\rm 10}$,
B.~Abi$^{\rm 112}$,
M.~Abolins$^{\rm 88}$,
H.~Abramowicz$^{\rm 153}$,
H.~Abreu$^{\rm 115}$
% ...
\bigskip

% ...
$^{10}$ Institute of Physics, Azerbaijan Academy of Sciences, H. Javid Avenue 33, AZ 143 Baku, Azerbaijan\\
$^{11}$ Institut de F\'isica d'Altes Energies, IFAE, Edifici Cn, Universitat Aut\`onoma  de Barcelona,  ES - 08193 Bellaterra (Barcelona), Spain\\
% ...
$^{48}$ Albert-Ludwigs-Universit\"{a}t, Fakult\"{a}t f\"{u}r Mathematik und Physik, Hermann-Herder Str. 3, D - 79104 Freiburg i.Br., Germany\\
$^{49}$ Universit\'e de Gen\`eve, Section de Physique, 24 rue Ernest Ansermet, CH - 1211 Geneve 4, Switzerland\\
% ...
$^{88}$ Michigan State University, Department of Physics and Astronomy, High Energy Physics Group, East Lansing, MI 48824-2320, United States of America\\
% ...
$^{111}$ University of Oklahoma, Homer L. Dodge Department of Physics and Astronomy, 440 West Brooks, Room 100, Norman, OK 73019-0225, United States of America\\
$^{112}$ Oklahoma State University, Department of Physics, 145 Physical Sciences Building, Stillwater, OK 74078-3072, United States of America\\
% ...
$^{115}$ LAL, Univ. Paris-Sud, IN2P3/CNRS, Orsay, France\\
% ...
$^{118}$ Oxford University, Department of Physics, Denys Wilkinson Building, Keble Road, Oxford OX1 3RH, United Kingdom\\
% ...
$^{153}$ Tel Aviv University, Raymond and Beverly Sackler School of Physics and Astronomy, Ramat Aviv, IL - Tel Aviv 69978, Israel\\
% ...
\end{flushleft}

\end{document}